\documentclass[10pt,acmsmall,screen]{acmart}
\usepackage{booktabs}   
\usepackage{subcaption} 

\def\finalversion{}
\def\techreportversion{} 

\usepackage{amsfonts}

\usepackage{mathpartir}
\usepackage{pftools}  

\usepackage{bm}
\usepackage{bussproofs} 

\usepackage{iris}

\usepackage[T1]{fontenc} 

\usepackage[final]{listings}
\usepackage{tikz}
\usetikzlibrary{matrix,arrows,positioning,calc,fit,backgrounds, decorations.pathreplacing, calligraphy}

\usepackage[shortlabels, inline]{enumitem}
\makeatletter
\newcommand{\myitem}[1][]{
  \protected@edef\@currentlabel{#1}%
\item[#1]
}
\makeatother

\usepackage{mathtools}

\usepackage{etoolbox}

\usepackage{array}
\newcolumntype{L}{>$l<$}

\usepackage{stmaryrd}

\usepackage{dashbox}

\usepackage{scalerel}

\usepackage{mathpartir}

\usepackage{xspace}
\usepackage{multirow}

\usepackage{breakcites}

\usepackage{verbatim}

\newcommand{\moreless}[2]{\ifdefined\techreportversion#1\else#2\fi}
\newcommand{\techreport}[1]{\moreless{#1}{}}

\newcommand{\defFunc}[2]{\newcommand{#1}{\mathsf{#2}}}

\renewcommand{\le}{\leqslant}
\renewcommand{\ge}{\geqslant}

\newcommand{\Nat}{\mathbb{N}}

\newcommand{\QQ}{\mathbb{Q}}

\newcommand{\impl}{\Rightarrow}
\renewcommand{\ra}{\rightarrow}

\renewcommand{\provesIff}{\mathrel{\dashv\vdash}}

\renewcommand{\dom}{\operatorname{\mathsf{dom}}}

\newcommand{\pto}{\rightharpoonup}

\newcommand{\defeq}{\coloneqq}

\newcommand{\paren} [1] {\ensuremath{ \left( {#1} \right) }}

\newcommand{\setcomp}[2]{\ensuremath{\left\{#1\;\middle|\;#2\right\}}}

\newcommand{\refSec}[1]{\S\ref{#1}}
\newcommand{\refFig}[1]{Figure~\ref{#1}}

\newcommand{\refTab}[1]{Table~\ref{#1}}

\newcommand{\refEqn}[1]{\ensuremath{(\ref{#1})}}
\newcommand{\refRule}[1]{\ruleref{#1}}
\newcommand{\refApp}[1]{\moreless{Appendix~\ref{#1}}{\cite[Appendix~\ref{#1}]{oopsla21-techreport}}}

\newcommand{\code}[1]{\textnormal{\texttt{#1}}}
\newcommand{\grasshopper}{GRASShopper\xspace}

\newcommand{\helperFn}[1]{\textcolor{purple}{\code{#1}}}

\newcommand{\invariant}[2]{\boxed{#2}^{#1}}
\newcommand{\invName}{\mathcal{N}}
\newcommand{\inv}[1]{\invariant{\invName}{#1}}
\newcommand{\sinv}[1]{\invariant{}{#1}}

\newcommand{\ghostState}[2]{\dbox{\ensuremath{#2}}^{\gname_{#1}}}

\newcommand{\atomicUpdate}{\mathsf{AU}}

\makeatletter%
\@ifundefined{dplus}{%
\newcommand\dplus{\mathbin{+\kern-1.0ex+}}
}{}
\makeatother%

\newcommand{\raOp}{\cdot}

\newcommand{\setm}{\textmon{Set}}
\newcommand{\authFrag}{\circ}
\newcommand{\authAuth}{\bullet}

\newcommand{\fracHalf}{\text{\textonehalf}}

\newcommand{\heapRep}{\mathsf{node}}

\newcommand{\result}{\mathit{res}}

\newcommand{\some}{\mathsf{Some}}
\newcommand{\none}{\mathsf{None}}

\newcommand{\inFootprint}{\mathsf{inFP}}

\newcommand{\globalInv}{\mathsf{G}}
\newcommand{\lockR}{\mathsf{L}}

\defFunc{\inset}{ins}
\defFunc{\linkset}{lnks}
\defFunc{\outset}{outs}
\defFunc{\edgeset}{es}
\defFunc{\reachset}{rs}
\defFunc{\inreach}{inr}

\newcommand{\nodeDom}{\mathfrak{N}}

\newcommand{\mDom}{M}
\newcommand{\mOp}{+}
\newcommand{\mBigOp}{\sum}

\newcommand{\mZero}{0}

\newcommand{\edgeFn}{\mathit{e}}

\newcommand{\flow}{\mathit{fl}}

\defFunc{\inflowFn}{inf}
\defFunc{\outflowFn}{outf}

\newcommand{\outflow}{\mathit{out}}

\newcommand{\interface}{I}
\newcommand{\ointerface}{J}
\newcommand{\rinterface}{R}
\defFunc{\interfaceFn}{int}
\defFunc{\footprintFn}{ffp}

\newcommand{\intComp}{\oplus}
\newcommand{\intBigComp}{\bigoplus}

\newcommand{\inflowOfInt}[1]{{#1}.\inflow}

\newcommand{\outflowOfInt}[1]{{#1}.\outflow}

\defFunc{\init}{HInit}
\defFunc{\hunique}{HUnique}
\defFunc{\hcont}{HContents}
\defFunc{\capacity}{cap}
\defFunc{\capacityAux}{capAux}
\defFunc{\flowFn}{flow}
\defFunc{\flowmapFn}{flm}
\defFunc{\userEdgeFn}{edges}

\defFunc{\goodCondition}{\nu}

\defFunc{\pathCount}{pc}

\newcommand{\KS}{\mathsf{KS}}
\newcommand{\VS}{\mathsf{V}}
\newcommand{\inflow}{\mathit{in}}

\defFunc{\lock}{lock}
\defFunc{\keyset}{ks}
\defFunc{\pathset}{path}
\defFunc{\inflows}{In}

\defFunc{\composition}{comp}
\defFunc{\projection}{proj}

\defFunc{\abstractionFn}{edge}
\defFunc{\hrepSpatial}{spatialRep}

\newcommand{\mkblue}[1]{\textcolor{blue}{#1}}

\newcommand{\annot}[1]{\mkblue{\left\{\begin{aligned}#1\end{aligned}\right\}}}

\newcommand{\annotAtom}[1]{\mkblue{\left\langle\,\begin{aligned}#1\end{aligned}\,\right\rangle}}

\newcommand{\hoareTriple}[3]{\annot{#1} \; #2 \; \annot{#3}}
\newcommand{\atomicTriple}[3]{\annotAtom{#1} \; #2 \; \annotAtom{#3}}

\DeclareMathOperator*{\Sep}{\scalerel*{\ast}{\sum}}
\newcommand{\magicwand}{\mathrel{-\mkern-6mu*}}

\newcommand{\ite}[3]{\paren{#1 \;?\; #2 : #3}}

\newcommand{\mcsprotoupdate}{\mathsf{Updatable}}
\newcommand{\mcssearchspec}{\mathsf{SearchSpec}}
\newcommand{\mcsupsertspec}{\mathsf{UpsertSpec}}
\newcommand{\tid}{\mathit{tid}}
\newcommand{\mcslsearch}{\overline{\code{search}}}
\newcommand{\mcscont}{\mathsf{SR}}
\newcommand{\mcslinv}{\mathsf{Inv}}
\newcommand{\mcsinv}{\mcslinv_{\mathit{tpl}}}
\newcommand{\lsminv}{\mcslinv_{\mathit{LSM}}}

\newcommand{\mcsstate}{\mathsf{MCS}}
\newcommand{\mcslstate}{\overline{\mcsstate}}

\newcommand{\maxH}[1][H]{\bar{#1}}
\newcommand{\maxTS}{\mathsf{HClock}}
\newcommand{\ts}{\mathsf{ts}}
\newcommand{\valof}{\mathsf{val}}
\newcommand{\cir}{C_{\mathit{ir}}}
\newcommand{\Cir}[1]{B_{#1}}

\newcommand{\mcbtable}[8]{\begin{tabular}{ |c|c| } 
  \hline
  $#1$ & $#5$ \\
  \hline 
  $#2$ & $#6$ \\
  \hline 
  $#3$ & $#7$ \\ 
  \hline
  $#4$ & $#8$ \\ 
  \hline
  \end{tabular}}

\newcommand{\mctable}[6]{\begin{tabular}{ |c|c| } 
  \hline
  $#1$ & $#4$ \\
  \hline 
  $#2$ & $#5$ \\
  \hline 
  $#3$ & $#6$ \\ 
  \hline
  \end{tabular}}

\newcommand{\mctabletworows}[4]{\begin{tabular}{ |c|c| } 
  \hline
  $#1$ & $#3$ \\
  \hline
  $#2$ & $#4$ \\
  \hline
  \end{tabular}}
\newcommand{\mcclrtable}[7]{{\color{#1} \mctable{#2}{#3}{#4}{#5}{#6}{#7}}}
\newcommand{\mcredtable}{\mcclrtable{red}}

\newcommand{\closed}{\mathsf{closed}}
\newcommand{\resolve}[2]{\code{Resolve}\; {#1}\; \code{to}\; {#2}}
\newcommand{\newproph}{\code{NewProph}}
\newcommand{\proph}{\mathsf{Proph}}

\newcommand{\reg}{\mathsf{Reg}}
\newcommand{\helpingstate}{\mathsf{State}}
\newcommand{\proto}{\mathsf{Prot}}
\newcommand{\helpingproto}{\mathsf{Prot}_{\mathit{help}}}
\newcommand{\linpending}{\mathsf{Pending}}
\newcommand{\lindone}{\mathsf{Done}}
\newcommand{\token}{\mathsf{Tok}}

\newcommand{\pendingstate}{\mathsf{Pending}}

\newcommand{\true}{\mathit{true}}
\newcommand{\false}{\mathit{false}}

\newcommand{\nodePred}{\mathsf{N}_\mathsf{L}}
\newcommand{\nodeShar}{\mathsf{N}_\mathsf{S}}

\lstdefinelanguage{SPL}{
  morekeywords={acc, method, struct,if,else,returns,procedure,requires,ensures,:=,var,
    new,old,free,implicit,modifies,call,locals,assume,assert,choose,havoc,ghost,
    predicate,function,invariant,while, return,atomic, split, type, field, result,
    mark, unmark, define, datatype, domain, axiom},
  deletekeywords={union,int},
  numbers=left,
  xleftmargin=2em,
  escapeinside={@}{@},
  numberstyle=\tiny,
  basicstyle=\footnotesize\ttfamily,
  columns=flexible,
  morecomment=*[s][\color{green!60!black}]{/*}{*/},
  morecomment=*[l][\color{green!60!black}]{//},
  moredelim=**[is][\color{purple}]{|<}{>|},
  mathescape=true,
}

\tikzset{%
  array/.style={matrix of nodes,nodes={draw, minimum size=5mm, anchor=center},column sep=-\pgflinewidth, row sep=-\pgflinewidth, nodes in empty cells,anchor=center},
  ptr/.style={*->, shorten <=-(1.8pt+1.4\pgflinewidth)},
  edge/.style={->},
  dedge/.style={<->, dashed},
  fedge/.style={->, dashed},
  unode/.style={circle, draw=black, thick, minimum size=1cm},
  usnode/.style={draw=black, thick, minimum size=1cm},
  mnode/.style={circle, draw=black, thick, fill=gray!20, minimum size=1cm},
  stackVar/.style={circle, fill=none, inner sep=0pt, minimum size=1cm, font=\normalsize},
  hnode/.style={circle, fill=none, inner sep=0pt, minimum size=1mm, font=\normalsize},
  gnode/.style={circle, draw=black, thick, minimum size=1cm},
  pnode/.style={circle, draw=black, thick, minimum size=8mm},
  rnode/.style={draw=black, thick, minimum size=8mm},
  prio/.style={circle, fill=none, inner sep=0pt, minimum size=8mm, font=\footnotesize},
  dnode/.style={circle, draw=black, thick, dotted, minimum size=1cm},
  inflow/.style={circle, fill=none, inner sep=0pt, minimum size=5mm, font=\normalsize},
  phantomNode/.style={circle, fill=none, inner sep=0pt, minimum
    size=0pt}
}

\moreless{
\acmArticle{1}
\settopmatter{printacmref=false}
\setcopyright{none}
\acmConference{}{}{}
\acmISBN{} 
\acmPrice{}
\acmDOI{} 
\startPage{1}
}{
\setcopyright{acmlicensed}
\acmPrice{}
\acmDOI{10.1145/3485490}
\acmYear{2021}
\copyrightyear{2021}
\acmSubmissionID{oopsla21main-p70-p}
\acmJournal{PACMPL}
\acmVolume{5}
\acmNumber{OOPSLA}
\acmArticle{113}
\acmMonth{10}
}

\usepackage[normalem]{ulem}

\newcommand{\defineAuthor}[3]{
  \expandafter\newcommand\csname #1\endcsname[1]{%
    \ifdefined\finalversion{##1}%
    \else{\ifdefined\monochrome{\color{green!35!black}{##1}}%
      \else{\color{#3}##1}\fi}%
    \fi}
  \expandafter\newcommand\csname #1out\endcsname[1]{%
    \ifdefined\finalversion{}%
    \else{\ifdefined\monochrome{}%
      \else{\color{#3}{\sout{##1}}}%
      \fi}%
    \fi}
  \expandafter\newcommand\csname #1footnote\endcsname[1]{%
    \ifdefined\finalversion{}%
    \else{\ifdefined\monochrome{}%
      \else{\csname #1\endcsname{\footnote{\csname #1\endcsname{#2: ##1}}}}%
      \fi}%
    \fi}
}
\defineAuthor{tw}{THOMAS}{blue!80!black}
\defineAuthor{sk}{SID}{green!40!black}
\defineAuthor{des}{DENNIS}{purple}
\defineAuthor{np}{NISARG}{brown!80!black}
\defineAuthor{ed}{LIZ}{orange!80!black}

\moreless{
  \includeonly{intro,motivation,multicopy,template,proof,maintenance,evaluation,related,conclusion,iris-proofs}
}{
  \includeonly{intro,motivation,multicopy,template,proof,maintenance,evaluation,related,conclusion}}

\begin{document}

\title{Verifying Concurrent Multicopy Search Structures}         


\author{Nisarg Patel}
\affiliation{
  \institution{New York University}            
  \country{USA}
}
\email{nisarg@nyu.edu}          

\author{Siddharth Krishna}
\affiliation{
  \institution{Microsoft Research}            
  \city{Cambridge}
  \country{UK}
}
\email{siddharth@cs.nyu.edu}          

\author{Dennis Shasha}
\affiliation{
  \institution{New York University}            
  \country{USA}
}
\email{shasha@cims.nyu.edu}         

\author{Thomas Wies}
\affiliation{
  \institution{New York University}            
  \country{USA}
}
\email{wies@cs.nyu.edu}         

\begin{abstract}
Multicopy search structures such as log-structured merge (LSM) trees 
are optimized for high insert/update/delete (collectively known as upsert) performance. In such data structures, an upsert on key $k$, which adds $(k,v)$ where $v$ can be a value or a tombstone, is added to the root node even if $k$ is already present in other nodes. Thus there may be multiple copies of $k$ in the search structure. A search on $k$ aims to return the value associated with the most recent upsert.
We present a general framework for verifying linearizability of concurrent multicopy search structures that abstracts from the underlying representation of the data structure in memory, enabling proof-reuse across diverse implementations. Based on our framework, we propose template algorithms for a) LSM structures forming arbitrary directed acyclic graphs and b) differential file structures, and formally verify these templates in the concurrent separation logic Iris. We also instantiate the LSM template to obtain the first verified \tw{concurrent in-memory} LSM tree implementation.
\end{abstract}






\maketitle

\begingroup
\let\clearpage\relax
\section{Introduction}
\label{sec-intro}

\citet{DBLP:conf/pldi/KrishnaPSW20} demonstrated how to simplify the verification of concurrent search structure algorithms by abstracting implementations of diverse data structures such as B-trees, lists, and hash tables into templates that can be verified once and for all.
The template algorithms considered in~\cite{DBLP:journals/tods/ShashaG88,DBLP:conf/pldi/KrishnaPSW20} handle only search structures that perform all operations on keys \emph{in-place}. That is, an operation on key $k$ searches for the unique node containing $k$ in the structure and then performs any necessary modifications on that node. Since every key occurs at most once in the data structure at any given moment, we refer to these structures as \emph{single-copy (search) structures}.

Single-copy structures achieve high performance for reads.  However, some applications, such as event logging, require high write performance, possibly at the cost of decreased read speed and increased memory overhead. This demand is met by data structures that store upserts (inserts, deletes or updates) to a key $k$ \emph{out-of-place}  at a new node instead of overwriting a previous copy of $k$ that was already present in some other node. Performing out-of-place upserts  can be done in constant time (e.g., always at the head of a list). A consequence of this design is that the same key $k$ can now be present multiple times simultaneously in the data structure. Hence, we refer to these structures as \emph{multicopy (search) structures}.


Examples of multicopy structures include the differential file structure~\cite{DBLP:journals/tods/SeveranceL76}, log-structured merge (LSM) tree~\cite{DBLP:journals/acta/ONeilCGO96}, and the Bw-tree~\cite{DBLP:conf/icde/LevandoskiLS13a}. These concurrent data structures are widely used in practice, including in state-of-the-art database systems such as Apache Cassandra~\cite{apache-cassandra} and Google LevelDB~\cite{level-db}.

Like the verification method proposed by~\citet{DBLP:conf/pldi/KrishnaPSW20}, we aim to prove that the concurrent search structure of interest is linearizable~\cite{DBLP:conf/crypto/HerlihyT87}, i.e., each of its operations appears to take effect atomically at a \emph{linearization point} and behaves according to a sequential specification. \tw{For multicopy structures, the sequential specification is that of a (partial) mathematical map that maps a key to the last value that was upserted for that key.}
The framework proposed in~\cite{DBLP:journals/tods/ShashaG88,DBLP:conf/pldi/KrishnaPSW20}
does not extend to multicopy structures as it critically relies on the fact that every key is present in at most one node of the data structure at a time. Moreover, searches in multicopy structures exhibit dynamic non-local linearization points (i.e., the linearization point of a search is determined by and may be present during the execution of concurrently executing upserts).
This introduces a technical challenge that is not addressed by this prior work. We discuss further related work in \refSec{sec-related}.

\paragraph{Contributions.}

This paper presents a framework for constructing linearizability proofs of concurrent multicopy structures with the goal of enabling proof reuse across data structures.
\refFig{fig-proof-structure} provides an overview of our work. The paper starts by describing the basic intuition behind the correctness proof of any multicopy structure (\refSec{sec-motivation}). 
We then derive an abstract notion of multicopy structures similar to the abstract single-copy structures in the edgeset framework~\cite{DBLP:journals/tods/ShashaG88} (\refSec{sec-mcs}).
By introducing this intermediate abstraction level ("Template Level" in the figure) at which we can verify concurrent multicopy structure template algorithms, we aid proof reuse in two ways. First, the template algorithms abstract from the concrete representation of the data structure, allowing their proofs to be reused across diverse template instantiations. Second, the specification against which the templates are verified ("Search Recency") admits simpler linearizability proofs than the standard client-level specification of a search structure. The proof relating the client-level and template-level specification~(\refSec{sec-client-template-refinement}) can be reused across all templates.

\begin{figure}
  \centering
  \includegraphics[scale=.43]{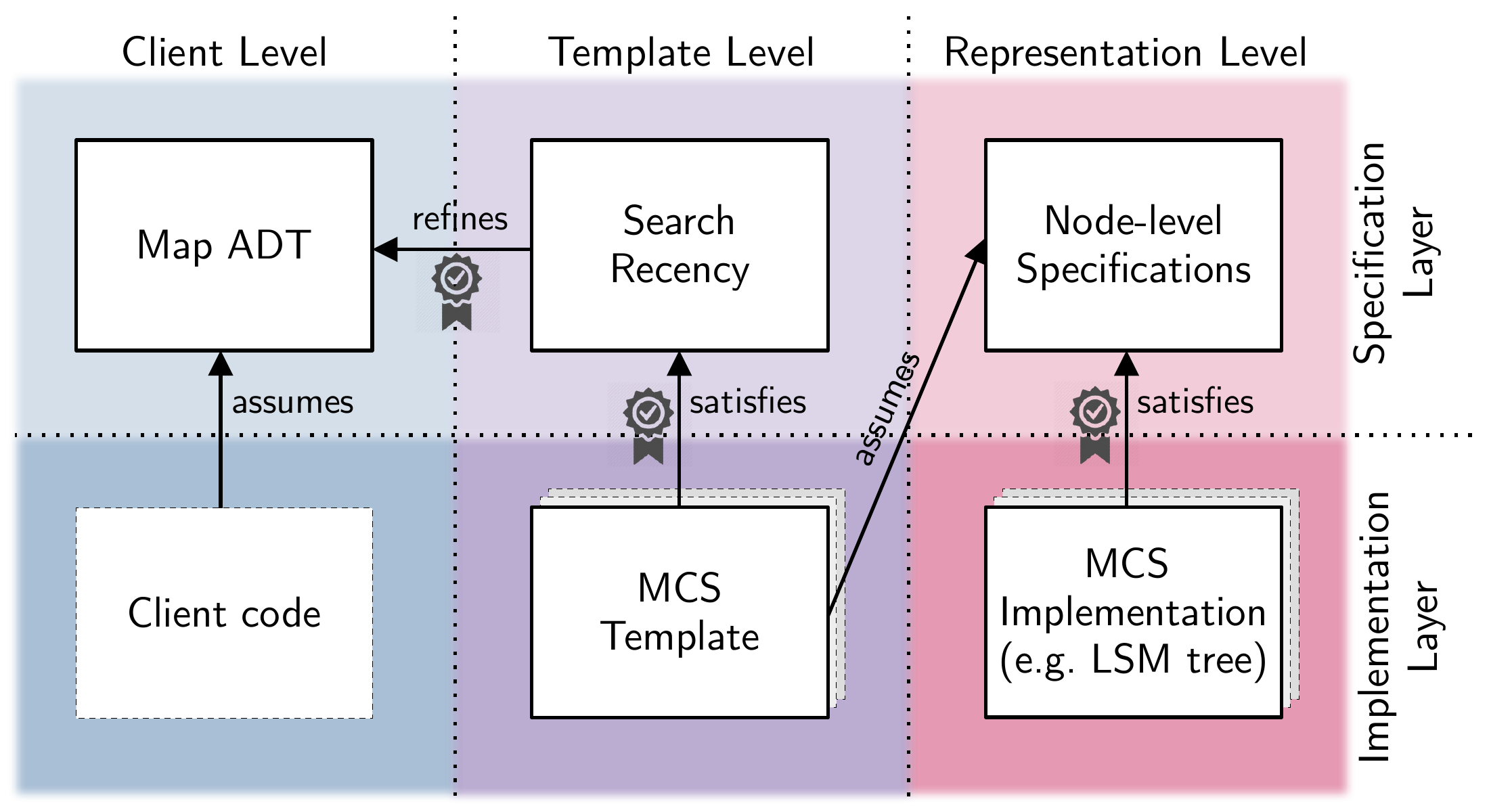}
  \caption{The structure of our verification effort. MCS stands for multicopy structure.\label{fig-proof-structure}}
\end{figure}

We demonstrate our framework by developing and verifying concurrent multicopy templates for a) LSM structures and b) differential file structures. The LSM template applies to existing LSM trees as well as to structures that form arbitrary directed acyclic graphs (DAGs) (\refSec{sec-template}, \refSec{setc-template-proof}, and \refSec{sec-maintenance}). The template and its proof
support implementations based on different heap representations such as lists, arrays, and B-link trees.
Verifying an instantiation of one of the two templates for a specific implementation  involves only sequential reasoning about node-level operations.

We have mechanized both the proof relating client-level and template-level specifications as well as the verification of our template algorithms in the Coq-based interactive proof mode of the concurrent separation logic Iris~\cite{DBLP:journals/jfp/JungKJBBD18,DBLP:conf/popl/KrebbersTB17,DBLP:journals/pacmpl/KrebbersJ0TKTCD18}. Similar to~\cite{DBLP:conf/pldi/KrishnaPSW20}, our formalization uses the flow framework~\cite{DBLP:conf/esop/KrishnaSW20,DBLP:journals/pacmpl/KrishnaSW18} to enable local reasoning about inductive invariants of a general multicopy structure graph.
In order to obtain a concrete multicopy structure, we have instantiated the node-level operations assumed by the LSM template for the LSM tree and verified their implementations in the automated separation logic verifier \grasshopper~\cite{DBLP:conf/tacas/PiskacWZ14} (\refSec{sec-evaluation}).
The result is the first formally-verified \tw{concurrent in-memory} LSM tree implementation.

\techreport{
  This is an extended version of a conference paper~\cite{multicopy-oopsla-paper}. All additional materials are provided in the appendix.
}

\section{Motivation and Overview}
\label{sec-motivation}

From a client's perspective, a multicopy structure implements a partial mathematical map $M \colon \KS \pto \VS$ of keys $k \in \KS$ to values $v \in \VS$. We refer to $M$ as the \emph{logical contents} of the structure.
The data structure supports insertions and deletions of key/value pairs on $M$ and searches for the value $M(k)$ associated with a given key $k$.

The insert and delete operations are implemented by a single generic operation referred to as an {\em upsert}.
The sequential specification of upsert is as follows.
The operation takes a key-value pair $(k, v)$ and updates $M$ to $M[k \rightarrowtail v]$, associating $k$ with the given value $v$.
To delete a key $k$ from the structure, one upserts the pair $(k, \square)$ where $\square$ is a dedicated \emph{tombstone} value used to indicate that $k$ has been deleted. 
The sequential specification of a search for a key $k$ is then as
expected: it returns $M(k)$ if $M$ is defined for $k$ and $\square$ otherwise.

Multicopy structures are commonly used in scenarios where the nodes representing the data structure's logical contents $M$ are spread over multiple media such as memory, solid-state drives, and hard disk drives. Each node therefore contains its own data structure that is designed for the particular characteristics of the underlying medium, typically an unsorted array at the root to allow upserts to perform fast appends and a classical single-copy search structure (e.g., a hash structure or arrays with bloom filters) for non-root nodes. The non-root nodes are typically read-only, so concurrency at the node level is not an issue. In this paper, we consider the multicopy data structure as a graph of nodes. We study template algorithms on that graph.



\subsection{A Library Analogy to Multicopy Search Structures}
\label{sec-library-analogy}

To train your intuition about multicopy structures, consider a library of books in which new editions of the same book arrive over time. Thus the first edition of book $k$ can enter and later the second edition, then the third and so on. A patron of this library who enters the library at time $t$ and is looking for book $k$ should find an edition that is either current at time $t$ or one that arrives in the library after $t$. We call this normative property \emph{search recency}.

Now suppose the library is organized as a sequence of rooms. All new books are put in the first room (near the entrance). When a new edition $v$ of a book arrives in the first room, any previous editions of that book in that room are thrown out.  When the first room becomes full,   the books in that room are moved to the second room. If a previous edition of some book is already in the second room, that previous edition is thrown out. When the second room becomes full,  its books are moved to the third room using the same throwing out rule, and so on. This procedure maintains the time-ordering invariant that the editions of the same book are ordered from most recent (at or nearer to the first room) to least recent (farther away from the first room) in the sequence of rooms.

A patron's search for $k$ starting at time $t$ begins in the first room. If the search finds any edition of $k$ in that room, the patron takes a photocopy of that edition. If not, the search proceeds to the second room and so on.

Now suppose that the latest edition at time $t$ is edition $v$ and there is a previous edition $v'$. Because of the time-ordering invariant and the fact that the search begins at the first room, the  search will encounter $v$ before it encounters $v'$. The search may also encounter an even newer edition of $k$, but will never encounter an older one before returning. That  establishes the search recency property.


Any concurrent execution of inserts and searches is equivalent to a serial execution in which (i) each insert is placed in its relative order of entering the root node with respect to other inserts and (iia) a search $s$ is placed after the insert whose edition $s$ copies if that insert occurred after $s$ began or (iib) a search $s$ is placed at the point when $s$ began, if the edition that $s$ copies was inserted before $s$ began (or if $s$ returns no edition at all).

Because the searches satisfy the search recency property, the concurrent execution is \emph{linearizable}~\cite{DBLP:journals/toplas/HerlihyW90}, which is our ultimate correctness goal.

\des{Note that the analogy as written has treated only inserts and searches. However, updates and deletions can be implemented as inserts: an update to  book $k$ can be implemented  as the insertion of a new edition; a delete of  book $k$ can be implemented as the insertion of an edition whose value is a ``tombstone'' which is an indication that book $k$ has been deleted.}


\subsection{Log-Structured Merge Trees}
\label{sec-lsm}

A prominent example of a multicopy structure is the LSM tree, which closely corresponds to the library analogy described above. The data structure consists of a root node $r$ stored in memory (the first room in the library), and a linked list of nodes $n_1,n_2,\dots,n_l$ stored on disk (the remaining rooms).  \refFig{fig-lsm-tree} (a) shows an example.

\begin{figure}[t]
  \centering
  \tikzset{
    unode/.style={circle, draw=black, thick, minimum size=4.5em, inner sep=0pt},
  }
  \begin{minipage}{.29\linewidth}
  \begin{tikzpicture}[>=stealth, every node/.style={scale=0.8}, font=\footnotesize]

    \def\xsep{2}
    \def\ysep{-1.8}
    \def\xshift{2.5}
    \def\s{.5}
    \def\t{.8}

    \node[unode, label = left:{$r$}] (r) {\begin{tabular}{c} $(k_2, d)$ \end{tabular}};
    \node[hnode, draw=white] (r') at ($(r) - (0,\ysep/2)$) {};
    \node[unode, label = left:{$n_1$}] (c1) at ($(r) + (0, \ysep)$) {\begin{tabular}{c} $(k_1, \square)$ \\ $(k_2, b)$ \end{tabular}};
    \node[unode, scale = 1, label = left:{$n_2$}] (c2) at ($(c1) + (0, \ysep)$) {\begin{tabular}{c} $(k_2, a)$ \\ $(k_3, c)$ \end{tabular}};
    \node[unode, scale = 1, label = left:{$n_3$}] (c3) at ($(c2) + (0, \ysep)$) {\begin{tabular}{c} $(k_1, c)$ \\ $(k_3, b)$ \end{tabular}};

    \draw[edge,dashed,-] ($(r) + (-1.9,\ysep/2)$) to ($(r) + (1.3,\ysep/2)$);
    \draw[edge] (r') to (r);
    \draw[edge] (r) to (c1);
    \draw[edge] (c1) to (c2);
    \draw[edge] (c2) to (c3);

   \node[stackVar, draw=white] (rc) at ($(r) + (-\xsep/1.5, \ysep/2.5)$) {memory};
   \node[stackVar, draw=white] (dc) at ($(c1) + (-\xsep/1.28, -\ysep/2.6)$) {disk};

   \node[hnode] (step) at ($(r) + (-\xsep/1.2,0)$) {\bf(a)};
    
 \end{tikzpicture}
  \end{minipage}%
  \begin{minipage}{.30\linewidth}
  \begin{tikzpicture}[>=stealth, every node/.style={scale=.8}, font=\footnotesize]

    \def\xsep{2.0}
    \def\ysep{-1.8}
    \def\xshift{2.5}
    \def\s{.5}
    \def\t{.8}


    \node[unode, label = left:{$r$}, minimum size=4.5em] (r) {};
    \node[hnode, draw=white] (r') at ($(r) - (0, \ysep/2)$) {};
    \node[unode, label = left:{$n_1$}] (c1) at ($(r) + (0, \ysep)$) {\begin{tabular}{c} $(k_1, \square)$ \\ $(k_2, d)$ \end{tabular}};
    \node[unode, scale = 1, label = left:{$n_2$}] (c2) at ($(c1) + (0, \ysep)$) {\begin{tabular}{c} $(k_2, a)$ \\ $(k_3, c)$ \end{tabular}};
    \node[unode, scale = 1, label = left:{$n_3$}] (c3) at ($(c2) + (0,\ysep)$) {\begin{tabular}{c} $(k_1, c)$ \\ $(k_3, b)$ \end{tabular}};

    \draw[edge,dashed,-] ($(r) + (-1.3,\ysep/2)$) to ($(r) + (1.3,\ysep/2)$);

    \draw[edge] (r') to (r);
    \draw[edge] (r) to (c1);
    \draw[edge] (c1) to (c2);
    \draw[edge] (c2) to (c3);


    \node[hnode] (step) at ($(r) + (-\xsep/1.2,0)$) {\bf(b)};
   \draw[edge,-] ($(r) + (-1.1*\xsep,-\ysep/2)$) to ($(c3) + (-1.1*\xsep,\ysep/3)$);
    
 \end{tikzpicture}
\end{minipage}%
  \begin{minipage}{.3\linewidth}
  \begin{tikzpicture}[>=stealth, every node/.style={scale=0.8}, font=\footnotesize]

    \def\xsep{2}
    \def\ysep{-1.8}
    \def\xshift{2.5}
    \def\s{.5}
    \def\t{.8}

    \node[unode, label = left:{$r$}] (r) {\begin{tabular}{c} $(k_2, d)$ \end{tabular}};
    \node[hnode, draw=white] (r') at ($(r) - (0,\ysep/2)$) {};
    \node[unode, label = left:{$n_1$}] (c1) at ($(r) + (0,\ysep)$) {};
    \node[unode, scale = 1, label = left:{$n_2$}] (c2) at ($(c1) + (0, \ysep)$) 
      {\begin{tabular}{c} $(k_1,
        \square)$ \\
        $(k_2, b)$ \\
        $(k_3, c)$
      \end{tabular}};
    \node[unode, scale = 1, label = left:{$n_3$}] (c3) at ($(c2) + (0, \ysep)$) {\begin{tabular}{c} $(k_1, c)$ \\ $(k_3, b)$ \end{tabular}};


    \draw[edge,dashed,-] ($(r) + (-1.3,\ysep/2)$) to ($(r) + (1.3,\ysep/2)$);

    \draw[edge] (r') to (r);
    \draw[edge] (r) to (c1);
    \draw[edge] (c1) to (c2);
    \draw[edge] (c2) to (c3);


   \node[hnode] (step) at ($(r) + (-\xsep/1.2,0)$) {\bf(c)};
   \draw[edge,-] ($(r) + (-1.1*\xsep,-\ysep/2)$) to ($(c3) + (-1.1*\xsep,\ysep/3)$);

 \end{tikzpicture}
\end{minipage}
\caption{\textbf{(a)} High-level structure of an LSM tree. \textbf{(b)} LSM tree obtained from (a) after flushing node $r$ to disk. \textbf{(c)} LSM tree obtained (a) after compacting nodes $n_1$ and $n_2$.}
  \label{fig-lsm-tree}
\end{figure}
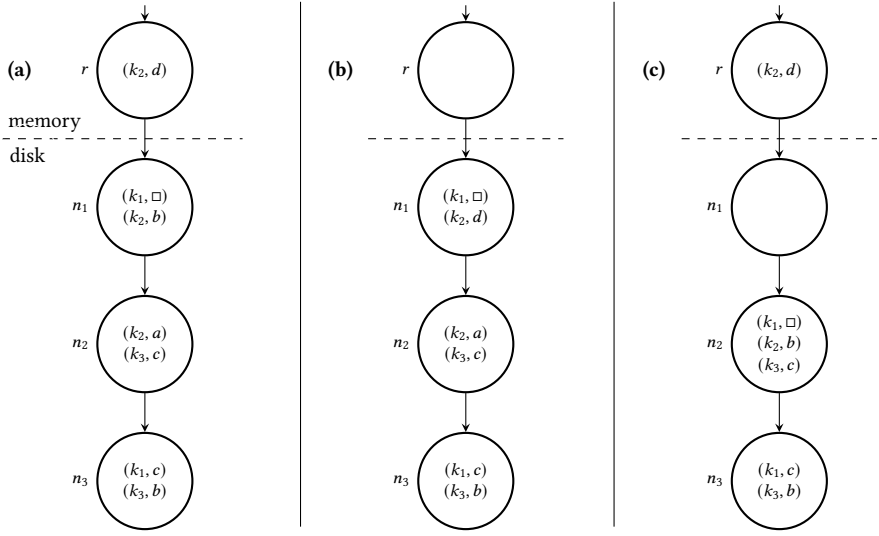

The LSM tree operations essentially behave as outlined in the library analogy. The upsert operation takes place at the root node $r$.  
A search for a key $k$ traverses the list starting from the root node and retrieves the  value associated with the first copy of $k$ that is encountered. If the retrieved value is $\square$ or if no entry for $k$ has been found after traversing the entire list, then the search determines that $k$ is not present in the data structure. Otherwise, it returns the retrieved value. For instance, a search for key $k_1$ on the LSM tree depicted in \refFig{fig-lsm-tree} (a) would determine that this key is not present since the retrieved value is $\square$ from node $n_1$. Similarly, $k_4$ is not present since there is no entry for this key. On the other hand, a search for $k_2$ would return $d$ and a search for $k_3$ would return $c$.




To prevent the root node from growing too large, the LSM tree performs {\em flushing}. As the name suggests, the flushing operation flushes the data from the root node to the disk by moving its contents to the first disk node. \refFig{fig-lsm-tree} (b) shows the LSM tree obtained from \refFig{fig-lsm-tree} (a) after flushing the contents of $r$ to the disk node $n_1$.

Similar to flushing, a \emph{compaction} operation moves data from full nodes on disk to their successor. In case there is no successor, then a new node is created at the end of the structure. During the merge, if a key is present in both nodes, then the most recent (closer-to-the-root) copy is kept, while older copies are discarded. \refFig{fig-lsm-tree} (c) shows the LSM tree obtained from \refFig{fig-lsm-tree} (a) after compacting nodes $n_1$ and $n_2$. Here, the copy of $k_2$ in $n_2$ has been discarded. In practice, the length of the data structure is bounded by letting the size of newly created nodes grow exponentially.

The net effect of all these operations is that the data structure satisfies the time-ordering invariant and searches achieve search recency.

The LSM tree can be tuned by implementing workload- and
hardware-specific data structures at the node level. In addition,
research has been directed towards optimizing the layout of nodes and
developing different strategies for the maintenance
operations
used to reorganize these data structures. This has resulted in a variety of implementations today (e.g.~\cite{DBLP:conf/sosp/RajuKCA17,DBLP:conf/sigmod/DayanI18,DBLP:conf/usenix/WuXSJ15,DBLP:conf/icde/Thonangi017, DBLP:journals/vldb/LuoC20}). Despite the differences between these implementations, they generally follow the same high-level algorithms for the core search structure operations.

We construct template algorithms for concurrent multicopy structures from the high-level descriptions of their operations and then prove the correctness of these operations. Notably our LSM DAG template generalizes the LSM tree so that the outer data structure can be a DAG rather than just a list. 
A number of existing LSM structures are based on trees (e.g. \cite{DBLP:conf/sigmod/SearsR12,DBLP:conf/usenix/WuXSJ15}). Practical implementations of tree-based concurrent search structures often have additional pointer structures layered on top of the tree that make them DAGs. For instance, many implementations use the \emph{link technique} to increase performance. Here, when a maintenance operation relocates a key $k$ from one node to another, it adds a pointer linking the two nodes, which ensures that $k$ remains reachable via the old search path. A concurrent thread searching for $k$ that arrives at the old node can then follow the link, avoiding a restart of the search from the root. Our verified templates can be instantiated to lock-based implementations of this technique.



\section{Multicopy Search Structure Framework}
\label{sec-mcs}


We build our formal framework of multicopy structures on the concurrent separation logic Iris~\cite{DBLP:journals/jfp/JungKJBBD18}. A detailed introduction to Iris is beyond the scope of this paper. We  therefore introduce only the relevant features of the logic as we use them.

\subsection{Multicopy Search Structures}

We abstract away from the data organization within the nodes, and treat the data structure as consisting of nodes in a mathematical directed acyclic graph. 
%

Since  copies of a single key $k$ can be present in different nodes simultaneously, we need a mechanism to differentiate between these copies. \tw{To that end, we augment each entry $(k, v)$ stored in a node with the unique timestamp $t$ identifying the point in time when $(k,v)$ was upserted: $(k, (v, t))$.} The timestamp plays the role of the book edition in the library analogy from the last section. \tw{For example in \refFig{fig-multicopy}, $(k_3, c)$ was upserted after $(k_2,a)$, which was upserted after $(k_3, b)$.} To generate these timestamps, we use a single global clock\tw{, which we initialize to $1$}. Note that the timestamp associated with an upserted value is auxiliary, or \emph{ghost}, data that we use in our proofs to track the temporal ordering of the copies present in the structure at any point. Implementations do not need to explicitly store this timestamp information.


Formally, let $\KS$ be the set of all keys \tw{and $\VS$ a set of values with a dedicated tombstone value $\square \in \VS$}. A multicopy (search) structure is a directed acyclic graph $G=(N,E)$ with nodes $N$ and edges $E \subseteq N \times N$. We assume that there is a dedicated \emph{root node} $r \in N$ which uniquely identifies the structure. 
\tw{Each node $n$ of the graph is labeled by its contents $C_n \colon \KS \pto \VS \times \Nat$, which is a partial map from keys to pairs of values and timestamps. For a node $n$ and its contents $C_n$, we say $(k,(v,t))$ is in the contents of $n$ if $C_n(k) = (v,t)$.} We denote the absence of \tw{an entry for a key} $k$ in $n$ by $C_n(k) = \bot$ and let $\dom(C_n)\defeq\setcomp{k}{C_n(k) \neq \bot}$. \tw{We further write $\valof(C_n) \colon \KS \pto \VS$ for the partial function that strips off the timestamp information from the contents of a node, $\valof(C_n) \defeq \lambda k.\, \ite{\exists v.\, C_n(k)=(v,\_)}{v}{\bot}$.}
  
For each edge $(n,n')\in E$ in the graph, the \emph{edgeset} $\edgeset(n,n')$ is the set of keys $k$ for which an operation arriving at a node $n$ would traverse $(n,n')$ if $k \notin \dom(C_n)$.
We require that the edgesets of all outgoing edges of a node $n$ are pairwise disjoint. \refFig{fig-multicopy} shows a potential abstract multicopy structure graph consistent with the LSM tree depicted in \refFig{fig-lsm-tree} (a). Here, all edges have edgeset $\KS$.

 
\begin{figure}[t]
  \centering
  \tikzset{
    unode/.style={circle, draw=black, thick, minimum size=4.5em, inner sep=2pt},
  }
  \begin{tikzpicture}[>=stealth,  every node/.style={scale=.8}, font=\footnotesize]

    \def\xsep{2.2}
    \def\ysep{-7}
    \def\xshift{2.5}
    \def\s{.5}
    \def\t{.8}

    \node[usnode, label = below:{$r$}] (r) {\mctable{k_1}{k_2}{k_3}{\bot}{(d,7)}{\bot}};
    \node[hnode, draw=white] (r') at ($(r) - (\xsep/2,0)$) {};
    \node[usnode, label = below:{$n_1$}] (n1) at ($(r) + (\xsep, 0)$) {\mctable{k_1}{k_2}{k_3}{(\square,6)}{(b,5)}{\bot}};
    \node[usnode, label = below:{$n_2$}] (n2) at ($(n1) + (\xsep, 0)$) {\mctable{k_1}{k_2}{k_3}{\bot}{(a,3)}{(c,4)}};
    \node[usnode, label = below:{$n_3$}] (n3) at ($(n2) + (\xsep, 0)$){\mctable{k_1}{k_2}{k_3}{(c,2)}{\bot}{(b,1)}};

    \draw[edge] (r') to (r);
    \draw[edge] (r) to (n1);
    \draw[edge] (n1) to (n2);
    \draw[edge] (n2) to (n3);

  \end{tikzpicture}
  \caption{Abstract multicopy data structure graph for the LSM tree in \refFig{fig-lsm-tree} (a).}
  \label{fig-multicopy}
\end{figure}
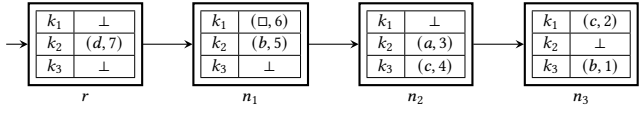

\subsection{Client-Level Specification}

Our goal is to prove the linearizability of concurrent multicopy structure templates with respect to their desired sequential client-level specification. 
As discussed earlier, the sequential specification is that of a map ADT, i.e., the \emph{logical contents} of the data structure is a mathematical map from keys to \tw{values, $M \colon \KS \to \VS$. The map $M$ associates every key $k$ with the most recently upserted value $v$ for $k$, respectively, $\square$ if $k$ has not yet been upserted:
\[ M(k) \defeq
  \begin{cases}
    v & \text{if } \exists n\, t.\; C_n(k)=(v,t) \land t = \max \setcomp{t'}{\exists n'\, v'.\; C_{n'}(k) = (v',t')}\\
    \square & \text{otherwise}
  \end{cases}
  \]
We call $M(k)$ the \emph{logical value} of key $k$.}

  Linearizability of a data structure is defined in terms of the concurrent execution histories of the data structure's operations~\cite{DBLP:conf/crypto/HerlihyT87,DBLP:journals/toplas/HerlihyW90}. Hoare logics like Iris emphasize proof decomposition, which means, in particular, that they strive to reason only about a single data structure operation at a time. It is therefore difficult to specify linearizability directly in such logics. Instead, we specify the intended behavior of each data structure operation in terms of an \emph{atomic triple}~\cite{DBLP:conf/popl/JacobsP11, DBLP:conf/ecoop/PintoDG14, DBLP:conf/popl/JungSSSTBD15, DBLP:conf/lics/FruminKB18, DBLP:journals/pacmpl/JungLPRTDJ20}. Atomic triples can be thought of as the concurrent counterparts of sequential Hoare triples. They formalize the intuition that a linearizable operation appears to take effect atomically at a single point in time, the operation's \emph{linearization point}.

More precisely, an atomic triple $\atomicTriple{\vec{x}.\;P}{e}{\val.\; Q}$ is made up of a precondition $P$, which may refer to the variables $\vec{x}$, a postcondition $Q$, which relates the variables $\vec{x}$ and the return value $\val$, and a program $e$. The triple states that $e$ may assume that for each of its atomic steps up to its linearization point, the shared state satisfies $P$ for possibly different values of $\vec{x}$ in each step. At the linearization point, $e$ then changes the shared state to one that satisfies $Q$ in one atomic step. Afterwards, $e$ no longer access resources in $P$ or $Q$. Intuitively, concurrently executing threads may interfere with $e$ by modifying the shared state but they are required to maintain $P$ as an invariant.

\tw{Now suppose that $\mcslstate(r, M)$ is a \emph{representation predicate} that provides the client view of a multicopy structure with root $r$, abstracting its shared state by the logical contents $M$. We then require that the \code{search} and \code{upsert} methods respect the following client-level atomic specifications:
\begin{align}
& \atomicTriple{M.\, \mcslstate(r,M)}{\code{upsert}\;r\;k\;v}{\mcslstate(r, M[k \rightarrowtail v])} \label{eq:upsert-seq-spec}\\
  & \atomicTriple{M.\,\mcslstate(r,M)}{\code{search}\;r\;k}{\code{$v$}.\; \mcslstate(r, M) * M(k)=v} \label{eq:search-seq-spec}
\end{align}
The specification of \code{upsert} updates the logical value of $k$ to $v$. Thus \code{upsert} performs the ``insert'' of the library analogy. The \code{search} specification states that \code{search} returns the logical value $M(k)=v$ of its query key $k$.} 



\subsection{Template-Level Specification: Search Recency}
\label{sec-atomic-spec-search-recency}

The verification of multicopy structures requires reasoning about the dynamic non-local linearization points of \code{search}, which are determined by the concurrently executing \code{upsert}s.
We want to avoid having to do this reasoning each time we verify a new template for a multicopy structure implementation. Our strategy is to provide  an alternative template-level specification that uses a more detailed abstraction of the computation history rather than just the logical contents. This alternative specification will then have fixed local linearization points,  simplifying the verification.

We say that \code{search} satisfies \emph{search recency} if each concurrent invocation \code{search$\,r\,k$} either returns the logical \tw{value} associated with $k$ at the point when the search started, or any other copy of $k$ that was upserted between the search's start time and the search's end time.

We will show that if \code{search}es satisfy search recency and \code{upsert}s take effect in a single atomic step that changes the logical contents $M$ according to (\ref{eq:upsert-seq-spec}), then the multicopy structure is linearizable. 

\tw{\skout{Define}\sk{We start by defining} the \emph{upsert history}
$H \subseteq \KS \times (\VS \times \Nat)$ of a multicopy data structure as the set of all copies $(k, (v,t))$ that have been upserted thus far.
In particular, we require that any multicopy structure will maintain the following predicates concerning $H$ and the global clock $t$:
\begin{align*}
    \init(H) \defeq {}
    & \forall k.\; (k,(\square,0)) \in H
    \\[.5em]
    \hunique(H) \defeq {}
    & \forall k\, t'\, v_1\, v_2.\; (k,(v_1,t')) \in H \land (k,(v_2,t')) \in H \Rightarrow v_1 = v_2
    \\[.5em]
    \maxTS(t,H) \defeq {}
    & \forall (k,(\_,t')) \in H.\; t' < t
\end{align*}
The predicate $\hunique(H)$ ensures that we can lift the total order $t_1 \le t_2$ on timestamps to a total order $(v_1, t_1) \le (v_2, t_2)$ on the pairs of values and timestamps occurring in $H$. The lifted order simply ignores the value component. Together with $\init(H)$, this ensures that the following function is well-defined:
\[\maxH \defeq \lambda k.\, \max \setcomp{(v,t)}{(k,(v,t)) \in H}.\]
The latest copy of a key \npout{that was upserted} will always be contained in some node $n$ of the data structure. If the data structure implementation maintains the additional invariant, $H \supseteq \bigcup_{n \in N} C_n$, then this guarantees that $\maxH$ is consistent with the logical contents $M$, i.e., for all keys $k$, $\maxH(k)=(M(k),\_)$. Finally, the predicate $\maxTS(t,H)$ guarantees that $\hunique(H)$ is preserved when a new entry $(k,(v,t))$ is added to $H$ for the current value of the global clock $t$.}

Assume that, similar to $\mcslstate(r, M)$, we are given a template-level representation predicate $\mcsstate(r, t, H)$ that abstracts the state of a multicopy structure by its upsert history $H$ and the current value $t$ of the global clock. The desired template-level specification of \code{upsert} in terms of the new abstraction is simply:
\tw{
\begin{equation}
  \label{eq:upsert-spec}
  \atomicTriple{t\, H.\; \mcsstate(r, t, H)}{\code{upsert}\;r\;k\;v}{\mcsstate(r, t + 1, H \cup (k,(v,t)))}\end{equation}
It states that \code{upsert} advances the value of the global clock from $t$ to $t+1$ and adds a new copy $(k,(v,t))$ to the upsert history $H$.}

\tw{The postcondition of \code{search} needs to express two properties. First, we must necessarily have $(k,(v,t')) \in H$, where $v$ is the value returned by \code{search}, $t'$ is $v$'s associated timestamp, and $H$ is the value of the upsert history at the linearization point. Moreover, let $H_0$ be the value of the upsert history at the start of the search and define $(v_0,t_0) \defeq \maxH[H_0](k)$. Then either $v$ is the logical value of $k$ at that point (i.e. $v = v_0$) or $t'$ is the timestamp of an upsert for $k$ that happened after the search started, i.e., $t_0 < t'$.} This is equivalent to demanding that for all $t_0'$ \tw{and $v'$} such that \tw{$(k,(v',t_0')) \in H_0$}, the returned timestamp $t'$ satisfies $t_0' \le t'$. We define the auxiliary abstract predicate \tw{$\mcscont(k, v, t)$ to mean that $(k,(v,t)) \in H$} for the value $H$ of the upsert history at the time point when the predicate is evaluated.\footnote{In \refSec{sec-ghost-state} we will express $\mcscont(k,v,t)$ using appropriate Iris ghost state that keeps track of the upsert history.}
Using this predicate, the template-level specification of \code{search} is then expressed as follows:
\tw{\begin{equation}
  \label{eq:search-spec}
  \begin{array}{l}
  \forall v_0'\, t_0'.\; \mcscont(k, v_0', t_0') \magicwand\\
  \quad \atomicTriple{t\, H.\; \mcsstate(r, t, H)}{\code{search}\;r\;k}{v.\; \exists t'.\, \mcsstate(r, t, H) \,*\, t_0'\!\le\!t' \,*\, (k,(v,t'))\!\in\!H}
\end{array}
\end{equation}}
Here, we use the \emph{magic wand} connective $\magicwand$ to express that the auxiliary local precondition $\mcscont(k, v_0', t_0')$ must be satisfied at the time point when \code{search} is invoked.


In the next section, we deal with the complexity of non-local dynamic linearization points of searches once and for all by proving that any multicopy structure that satisfies the template-level specification also satisfies the desired client-level specification.
To prove the correctness of a given concurrent multicopy structure, it then suffices to show that \code{upsert} satisfies its corresponding template-level specification \refEqn{eq:upsert-spec} and \code{search} satisfies  \refEqn{eq:search-spec}. When proving the validity of the template-level atomic triple of \code{search} for a particular implementation (or template), one can now always \emph{commit} the atomic triple (i.e., declare a linearization point) at the point when the return value \tw{$v$} of the search is determined, i.e., when \tw{$(k,(v,t')) \in H$} is established. This linearization point is now independent of  concurrently executing \code{upsert}s.

\section{Relating the Client-Level and Template-Level Specifications}
\label{sec-client-template-refinement}

We next prove that any concurrent execution of \code{upsert} and \code{search} operations that satisfy the template-level specifications~\refEqn{eq:upsert-spec} and~\refEqn{eq:search-spec} can be linearized to an equivalent sequential execution that satisfies the client-level specifications. Intuitively, this can be done by letting the \code{upsert}s in the equivalent sequential execution occur in the same order as their atomic commit points in the concurrent execution, and by letting each \code{search$\,r\,k$} occur at the earliest time after \tw{the timestamp $t'$ associated with the returned value $v$} of $k$. That is, if \tw{$v = v_0$ (recall that $(v_0,t_0) = \maxH(k)$ where $H$ is the upsert history at the start of the search on $k$)}, then the search occurs right after it was invoked in the concurrent execution. Otherwise we must have $t' > t_0$ and the search occurs after the upsert at time $t'$.
The fact that such an upsert must exist follows from the template-level specifications.

The intuitive proof argument above relies on explicit reasoning about execution histories. Instead, we aim for a thread-modular proof that reasons about individual \code{search}es and \code{upsert}s in isolation, so that we can mechanize the proof in a Hoare logic like Iris. The proof we present below takes inspiration from that of the RDCSS data structure by~\citet{DBLP:journals/pacmpl/JungLPRTDJ20}.

\subsection{Challenges and Proof Outline}

Iris prophecies~\cite{DBLP:journals/pacmpl/JungLPRTDJ20},  based on the idea first introduced by~\citet{DBLP:conf/lics/AbadiL88}, allow a thread to predict what will happen in the future. In particular, one can use prophecies to predict future events in order to reason about non-fixed linearization points~\cite{DBLP:phd/ethos/Vafeiadis08, DBLP:conf/tamc/ZhangFFSL12}. In our case, a thread executing \code{search} can use a prophecy to predict, at the beginning of the search, the \tw{value $v$} that it will eventually return. In a thread-modular correctness proof, one can then decide on how to linearize the operation based on the predicted value.

The linearization point of a \code{search} operation  occurs when an instruction of a concurrent \code{upsert} is executed. One can view this as a form of \emph{helping}~\cite{DBLP:conf/pldi/LiangF13}: when an \code{upsert} operation commits and adds \tw{$(k,(v,t'))$} to the upsert history $H$, it also commits all the (unboundedly many) concurrently executing \code{search} operations for $k$ \tw{that will return $v$}. We encode this \emph{helping protocol} in the predicate $\mcslstate(r,M)$ that captures the shared (ghost) state of the data structure, by taking advantage of Iris's support for higher-order ghost state.

At a high level, the proof then works as follows. We augment \code{search} with auxiliary ghost code that creates and resolves the relevant prophecies. We do this by defining the wrapper function $\mcslsearch$ given in \refFig{fig-search'-algorithm}. The right side shows the specifications of the two functions related to manipulating (one-shot) prophecies in Iris. The function $\newproph$ returns a fresh prophecy $p$ that predicts the value $v_p$. This fact is captured by the resource $\proph(p, v_p)$ in the postcondition of the Hoare triple specifying $\newproph$. The resource $\proph(p, v_p)$ can be owned by a thread as well as transferred between threads via shared resources such as the representation predicate $\mcsstate(r,t,H)$ (as is usual in concurrent separation logics). The resource is also exclusive, meaning it cannot be duplicated.

The function $\mcslsearch$ uses $\newproph$ to create two prophecies, which it binds to $\tid$ and $p$. The prophecy $p$ predicts the value $v$ that will eventually be returned by \code{search}. The value $\tid$ predicted by the second prophecy will be used later as a unique identifier of the thread performing the search when we encode the helping protocol, taking advantage of the fact that each prophecy returned by $\newproph$ is fresh. Freshness of prophecies also ensures that each prophecy can  be resolved only once, which is done using $\resolve{p}{v}$. This operation consumes the resource $\proph(p, v_p)$ and yields the proposition $v_p = v$. It is used on line~\ref{line-mc-search'-resolve} of $\mcslsearch$ to express that the value predicted by $p$ is indeed the value \tw{$v$} returned by \code{search}.

\tw{If $v$ is equal to the current logical value $v_0$} of $k$ at the start of $\mcslsearch$, then the proof commits the client-level atomic triple right away. \tw{If instead $v_0 \neq v$}, then the proof \textit{registers} the thread's client-level atomic triple in the shared predicate $\mcslstate(r,M)$. The registered atomic triple serves as an obligation to commit the atomic triple. This obligation will be discharged by the \code{upsert} operation adding \tw{$(k,(v,t'))$} to $H$. The proof of $\mcslsearch$ then uses the template-level specification of \code{search} to conclude that it can collect the committed triple from the shared predicate after \code{search} has returned.

\begin{figure}
  \centering
\begin{minipage}[t]{.48\textwidth}
\begin{lstlisting}[aboveskip=0pt,belowskip=0pt]
let $\mcslsearch$ $r$ $k$ =
  let $\tid$ = $\newproph$ in @\label{line-mc-search'-proph-tid}@
  let $p$ = $\newproph$ in @\label{line-mc-search'-proph}@
  let $v$ = search $r$ $k$ in @\label{line-mc-search'-search}@
  $\resolve{p}{v}$; $v$ @\label{line-mc-search'-resolve}@
\end{lstlisting}
\end{minipage}\hfill
\begin{minipage}[t]{.48\textwidth}
  \strut\vspace*{-\baselineskip}\newline
\begin{mathpar}
  \inferH{one-shot-prophecy-creation}{}
  {\hoareTriple{\mathsf{True}}{\newproph}{p.\; \exists v_p.\; \proph(p, v_p)}}
  \\
  \inferH{one-shot-prophecy-resolution}{}
  {\hoareTriple{\proph(p, v_p)}{\resolve{p}{v}}{v_p = v}}
\end{mathpar}
\end{minipage}
\caption{Wrapper augmenting \code{search} with prophecy-related ghost code, whose specification is on the right.
\label{fig-search'-algorithm}}
\end{figure}

Relating the high-level and low-level specification of \code{upsert} is straightforward. However, the proof of \code{upsert} also needs to do its part of the helping protocol by 
scanning over all the searches that are currently registered in the shared predicate $\mcslstate(r,M)$ and committing those that return the copy of $k$ added by the \code{upsert}.

In the remainder of this section we explain this helping proof in more detail.

\subsection{Keeping Track of the Upsert History}
\label{sec-ghost-state}

Our thread-modular proof exploits the observation that the upsert history $H$ only increases over time. Thus, assertions  such as \tw{$(k,(v,t)) \in H$}, as used in our specification of search recency, are stable under interference. This style of reasoning follows the classic idea of establishing lower bounds on monotonically evolving state (see e.g.~\cite{DBLP:conf/ifip/Jones83, fahndrich2003heap, DBLP:conf/esop/JensenB12}). We formalize this in Iris using user-defined ghost state.

Iris expresses ownership of ghost state by the proposition $\ghostState{}{a}$ which asserts ownership of a piece $a$ of the ghost location $\gname$.
It is the ghost analogue of the points-to predicate $x \mapsto \val$ in separation logic, except that $\ghostState{}{a}$  asserts only that $\gname$ contains a value \emph{one of whose parts} is $a$.
This means ghost state can be split and combined according to the rules of the \emph{camera}, the algebraic structure from which the values (like $a$) are drawn. Cameras generalize  partial commutative monoids, which are commonly used to give semantics to separation logics.
\tw{
A camera comes equipped with a set $\monoid$ and a binary composition operation $(\mtimes) \colon \monoid \times \monoid \to \monoid$ that form a commutative monoid. The composition operation gives meaning to the separating conjunction of predicates that express fragmental ownership of ghost state at a ghost location $\gamma$ via the rule: $\ghostState{}{a} * \ghostState{}{b} \provesIff \ghostState{}{a \mtimes b}$.
A simple example of a camera is $\setm(X)$, where $X$ is some set. Here, $M=2^X$ and $(\mtimes)$ is set union. Another example is the \emph{heap camera} of standard separation logic, which consists of mappings from heap locations to values that can be composed by disjoint set union.

Iris also provides generic ``functors'' for constructing new cameras from existing ones.
One example that we will be using in our proofs is the \emph{authoritative} camera $\authm(\monoid)$, which can be constructed from any other camera $\monoid$.
It is used to model situations where threads share an authoritative element $\melt$ of $\monoid$ via a representation predicate and individual threads own fragments $\meltB$ of $\melt$. We denote an authoritative element by $\authAuth \melt$ and a fragment by $\authFrag \meltB$. The composition $\authAuth \melt \cdot \authFrag \meltB$ expresses ownership of the authoritative element $\melt$ and, in addition, $\exists \meltC.\, \melt = \meltB \cdot \meltC$.

For instance, we use the \emph{authoritative set} camera $\authm(\setm(\KS \times (\VS \times \nat)))$ to keep track of the upsert history $H$ at a ghost location $\gamma_s$.} The proposition $\ghostState{s}{\authAuth H}$ states that $H$ is the current authoritative version of the upsert history. This ghost resource is kept in the representation predicate $\mcsstate(r,t,H)$, which is shared among all threads operating on the data structure. The camera can also express lower bounds $H' \subseteq H$ on the authoritative set $H$ using propositions of the form $\ghostState{s}{\authFrag H'}$. That is, the proposition $\ghostState{s}{\authAuth H} * \ghostState{s}{\authFrag H'}$ asserts ownership of the current upsert history $H$ and, in addition, $H' \subseteq H$. We can then define the predicate $\mcscont(k,v,t)$, which expresses that \tw{$\maxH(k)=(v,t)$} was true at some point in the past, as \tw{$\mcscont(k,v,t) \defeq \ghostState{s}{\authFrag \set{(k,(v,t))}}$}.



Iris allows \emph{frame-preserving updates} of ghost state, denoted by the \emph{view shift} connective $\vsR$.
For instance, the following rules capture some frame-preserving updates of authoritative sets:
\begin{mathpar}
  \inferH{auth-set-upd}{H \subseteq H'}{\ghostState{}{\authAuth H} \vsR \ghostState{}{\authAuth H'}}

  \inferH{auth-set-snap}{}
  {\ghostState{}{\authAuth H} \vsR \ghostState{}{\authAuth H} * \ghostState{}{\authFrag H}}

  \inferH{auth-set-frag}
  {}{\ghostState{}{\authfrag H} * \ghostState{}{\authfrag H'} \vsE \ghostState{}{\authfrag (H \cup H')}}
%
\end{mathpar}
The rule \refRule{auth-set-upd} is the only way to update the authoritative element, because, intuitively, it must maintain the validity of all lower bounds.
The authoritative set camera thus implicitly enforces the invariant that the upsert history can only increase.
We use this rule to update the authoritative version of the upsert history at the linearization point of $\code{upsert}$.


The rule \refRule{auth-set-snap} allows us to take a ``snapshot'' of the current authoritative set. We use this rule together with the rule \refRule{auth-set-frag} at the call to \code{search} in $\mcslsearch$ to establish the thread-local precondition $\mcscont(k,v_0',t_0')$ of the specification (\ref{eq:search-spec}) from the shared resource $\ghostState{s}{\authAuth H}$. To this end, we choose \tw{$(v_0',t_0') \defeq \maxH(k)$}, which gives us \tw{$H = H \cup \set{(k,(v_0',t_0'))}$} and thus \tw{$\ghostState{s}{\authFrag \set{(k,(v_0',t_0'))}}$}.




\subsection{The Helping Protocol}
\label{sec-helping-protocol}

\tw{
Before we discuss the details of our encoding of the helping protocol in terms of Iris ghost state, let us recall the
basic structure of a proof of an atomic triple
$\atomicTriple{x.\;P}{e}{\Ret\val. Q}$.
The proof proceeds by proving a standard Hoare triple of the form $\All \Phi. \hoareTriple{\atomicUpdate_{x. P, Q}(\Phi)}{e}{\Ret\val. \Phi(\val)}$. Here, $\atomicUpdate_{x. P, Q}(\Phi)$ is the \emph{atomic update token}, which gives us the \emph{right} to use the resources in the precondition $P$ when executing atomic instructions up to the linearization point. The token also records our \emph{obligation} to preserve $P$ up to the linearization point, where $P$ must be transformed to $Q$ in one atomic step. This step consumes the update token. The universally quantified proposition $\Phi$ can be thought of as the precondition for the continuation of the client of the atomic triple. At the linearization point, when the atomic update token is consumed, the corresponding proof rule produces $\Phi(v)$ as a receipt that the obligation has been fulfilled. This receipt is necessary to complete the proof of the Hoare triple.}


\begin{figure}[t!]
  \tw{
  \begin{align*}
    \mcslstate(r,M) \defeq {}
    & \exists\, t\,H.\;
    \mcsstate(r,t,H) \,*\, \forall k.\; (M(k),\_)\!=\!\maxH(k) \\%
    \mcsstate(r,t,H) \defeq {}
    & \ghostState{s}{\authAuth\; H} \,*\, \init(H) \,*\, \hunique(H) \,*\, \maxTS(t,H) 
    \\[.5em]
    &  \,*\, \mcsinv(t,H) \,*\, \helpingproto(H)
    \\[.5em]
    \helpingproto(H) \defeq {}
    & \exists\, R.\; \ghostState{r}{\authAuth\; R}
      * \Sep_{\tid \in R}        
      \; \exists\, k\, v_p\, t_0\, \Phi\, \token.\;
      \proph(\tid, \_)
      * \helpingstate(H, k, v_p, t_0, \Phi, \token)
    \\[.5em]
    \helpingstate(H, k, v_p, t_0, \Phi, \token) \defeq {}
    & \linpending(H, k, v_p, t_0, \Phi) \lor \lindone(H,k,v_p,t_0, \Phi, \token)
    \\[.5em]
    \linpending(H, k, v_p, t_0, \Phi) \defeq {}
    & 
    \atomicUpdate(\Phi)
    * (\forall t.\; (k, (v_p, t)) \in H \Rightarrow t < t_0)
    \\[.5em]
    \lindone(H,k,v_p,t_0,\Phi, \token) \defeq {}
    &  (\Phi(v_p)\; \lor\; \token) * (\exists t.\; (k, (v_p, t)) \in H \land t \ge t_0)
  \end{align*}}
  \caption{Definition of client-level representation predicate and invariants of helping protocol.\label{fig-helping-protocol-inv}}
\end{figure}


\noindent
\refFig{fig-helping-protocol-inv} shows a simplified definition of \tw{$\mcslstate(r,M)$} and the invariant that encodes the helping protocol.\footnote{\tw{For presentation purposes, the proof outline presented here abstracts from some technical details of the actual proof done in Iris. For a more detailed presentation of our Iris development, we refer the interested reader to \refApp{sec-iris-proofs}.}}
The definitions are implicitly parameterized by a proposition $\mcsinv(r,t,H)$, which abstracts from the resources needed for proving that a specific multicopy structure template satisfies the template-level specifications.
In particular, this invariant will store \npout{the authoritative version of the global clock $t$ and} \tw{the resources needed to represent the node-level contents $C_n$ for each node $n \in N$. It also ties the $C_n$ to $H$, capturing the invariant $H \supseteq \bigcup_{n \in N} C_n$.}

The predicate \tw{$\mcslstate(r,M)$} contains the predicate $\mcsstate(r,t,H)$, used in the template-level atomic triples, and defines \tw{$M$ in terms of $\maxH$}.
The predicate $\mcsstate(r,t,H)$ owns all (ghost) resources associated with the data structure.
In particular, this predicate stores the ghost resource $\ghostState{s}{\authAuth H}$, holding the authoritative version of the current upsert history, the abstract template-level invariant $\mcsinv(r,t,H)$, and the helping protocol predicate $\helpingproto(H)$, described below.
$\mcsstate(r,t,H)$ \tw{also states the three invariants $\init(H)$, $\hunique(t,H)$, and $\maxTS(H)$ discussed earlier, which are needed to prove the atomic triple of $\mcslsearch$.}

The helping protocol predicate $\helpingproto$ contains a \emph{registry} $\ghostState{r}{\authAuth\; R}$ of $\mcslsearch$ thread IDs that require helping from \code{upsert} threads.
For each thread ID $\tid$ in the registry, the shared state contains $\proph(\tid,\_)$ along with the state of $\tid$, which is either $\linpending$ or $\lindone$.
$\linpending$ captures an uncommitted $\mcslsearch$, and
$\lindone$ describes the operation after it has been committed. \tw{Note that we omit the annotation of the pre and postcondition from $\atomicUpdate(\Phi)$ as it always refers to the specification of \code{search} in this proof.}

\begin{figure}[t]
  \centering
  \begin{lstlisting}[aboveskip=1pt,belowskip=0pt]
$\annotAtom{M.\; \mcslstate(r,M)}$
let $\mcslsearch$ $r$ $k$ =
  $\annot{\atomicUpdate(\Phi)}$ @\label{line-mcssearch'-proof-logatom-intro}@
  let $\tid$ = $\newproph$ in
  let $p$ = $\newproph$ in
  $\annot{\atomicUpdate(\Phi) * \proph(\tid, \_) * \proph(p, v_p)}$ @\label{line-mcssearch'-proof-case-split-pre1}@
  $\annot{\atomicUpdate(\Phi) * \proph(\tid, \_) * \proph(p, v_p) * \mcsstate(r,t,H_0)}$ @\label{line-mcssearch'-proof-case-split-pre2}@
  $\annot{\atomicUpdate(\Phi) * \proph(\tid, \_) * \proph(p, v_p) * \ghostState{s}{\authFrag\; H_0} * (v_0,t_0) \!=\! \maxH[H_0](k) * \mcscont(k, v_0, t_0)}$ @\label{line-mcssearch'-proof-case-split-pre3}@
  (* Case analysis on $v_p = v_0$, $v_p \neq v_0$: only showing $v_p \neq v_0$ *) @\label{line-mcssearch'-proof-case-split}@
  $\annot{\atomicUpdate(\Phi) * \proph(\tid, \_) * \proph(p, v_p) * (v_0,t_0) \!=\!\maxH[H_0](k) * \mcscont(k, v_0, t_0) * v_p \!\neq\! v_0 * \dots}$ @\label{line-mcssearch'-proof-case-split-post}@
    $\annot{\ldots * \atomicUpdate(\Phi) * \proph(\tid, \_) * (\forall t.\; (k,(v_p,t)) \in H_0 \Rightarrow t < t_0)}$ @\label{line-mcssearch'-proof-call-search-pre-frame}@
    $\annot{\ldots * \proph(\tid, \_) * \linpending(H_0,k,vp,t_0,\Phi) * \token}$ @\label{line-mcssearch'-proof-call-search-pre1}@
    $\annot{\ldots * \ghostState{r}{\authAuth\; R} * \tid \notin R
      * \helpingstate(H_0, k, v_p, t_0, \Phi, \token) * \token}$ @\label{line-mcssearch'-proof-call-search-pre2}@
  (* Ghost update: $\ghostState{r}{\authAuth\; R} \vsR \ghostState{r}{\authAuth\; R \cup \set{\tid}}$ *) @\label{line-mcssearch'-proof-call-search-pre3}@
  $\annot{\proph(p, v_p) * \token * \ghostState{r}{\authFrag\; \{tid\}}  * \mcscont(k, v_0, t_0) * v_p \!\neq\! v_0 * \mcsstate(r,t,H) }$ @\label{line-mcssearch'-proof-call-search-pre4}@
  let $v$ = search $r$ $k$ in @\label{line-mcssearch'-proof-call-search}@
  $\annot{\proph(p, v_p) * \token * \ghostState{r}{\authFrag\; \{tid\}} * v_p \!\neq\! v_0 * \mcsstate(r,t,H) * t_0 \le t' * (k, (v,t')) \in H}$ @\label{line-mcssearch'-proof-call-search-post}@
  $\resolve{p}{v}$; @\label{line-mcssearch'-proof-resolve-p}@
  $\annot{\token * \ghostState{r}{\authFrag\; \{\tid\}} * v \!\neq\! v_0 * \mcsstate(r,t,H) * t_0 \le t' * (k, (v,t')) \in H}$ @\label{line-mcssearch'-proof-resolve-p-post}@
  $\annot{\dots * \token * \ghostState{r}{\authFrag\; \{\tid\}} * v \!\neq\! v_0 * t_0 \le t' * (k, (v,t')) \in H * \helpingstate(H,k,v,t_0,\Phi,\token)}$
  $\annot{\dots * \token * \ghostState{r}{\authFrag\; \{\tid\}} * t_0 \le t' * (k, (v,t')) \in H * \lindone(H,k,v,t_0,\Phi,\token)}$
  $\annot{\dots * \token * \ghostState{r}{\authFrag\; \{\tid\}} * t_0 \le t' * (k, (v,t')) \in H * (\Phi(v) \lor \token)}$
  $\annot{\dots * \Phi(v) * \ghostState{r}{\authFrag\; \{\tid\}} * t_0 \le t' * (k, (v,t')) \in H * (\Phi(v) \lor \token) }$
  $\annot{\dots * \Phi(v) * \ghostState{r}{\authFrag\; \{\tid\}} * \lindone(H,k,v,t_0,\Phi,\token)}$
  $\annot{\Phi(v) * \mcsstate(r,t,H)}$ @\label{line-mcssearch'-proof-phi}@
  $\annot{\mcslstate(r, M) * M(k)=v}$ @\label{line-mcssearch'-proof-phi}@
  $v$
  $\annotAtom{\Ret v. \mcslstate(r,M) * M(k)=v}$
\end{lstlisting}

\caption{Outline for the proof of the client-level specification for $\mcslsearch$.}
\label{fig-multicopy-search'-proof}
\end{figure}

\tw{The proof outline for $\mcslsearch$ is shown in \refFig{fig-multicopy-search'-proof}. After creating the two prophecies $\tid$ and $p$, the proof case-splits on whether the thread requires helping or not (line~\ref{line-mcssearch'-proof-case-split}). We only consider the helping case (i.e., $(v_0,t_0) = \maxH[H_0](k) \land v_0 \neq v_p$), where $H_0$} is the initial upsert history and $v_p$ the prophesied return value). Here, the thread registers itself with the helping protocol by replacing $R$ with $R \cup \set{\tid}$ using rule \refRule{auth-set-upd} \tw{(line~\ref{line-mcssearch'-proof-call-search-pre3})}.
To do this, it first establishes \tw{$\linpending(H_0,k,v_p,t_0,\Phi)$} by transferring its obligation to linearize to the shared state\tw{, captured by the update token $\atomicUpdate(\Phi)$. The condition $\forall t. (k, (v_p,t)) \in H_0 \Rightarrow t < t_0$ follows from $v_0 \neq v_p$, the definition of $\maxH[H_0]$, and the invariant $\hunique(H_0)$.}
The thread also a creates a fresh non-duplicable token $\token$ that it will later trade in for the receipt $\Phi(v_p)$.

Let us briefly switch to the role played by the \code{upsert} that \tw{updates the logical value of $k$ to $v_p$ at some time $t \ge t_0$.}
When this \code{upsert} reaches its linearization point, our proof uses rule \refRule{auth-set-upd} to update the upsert history from $H$ to \tw{$H \cup \set{(k,(v_p,t))}$}, as required by the postcondition of~(\ref{eq:upsert-spec}), and also increments the global clock \tw{from $t$ to $t+1$}.
We must then show that \tw{$\mcsstate(r,t+1,H \cup \set{(k,(v_p,t))})$} holds after these ghost updates, which requires us to prove \tw{$\helpingproto(H \cup \set{(k,(v_p,t))})$} assuming $\helpingproto(H)$ was true before the update.
In particular, any $\mcslsearch$ thread that was in a $\linpending$ state \tw{$\helpingstate(H, k, v_p, t_0, \Phi, \token)$} and thus waiting to be helped by this \code{upsert} needs to be committed.
It can do this because the postcondition of these triples are satisfied after $H$ has been updated.
The proof then transfers the receipts $\Phi(v_p)$ back to the shared representation predicate, yielding new states \tw{$\lindone(H \cup \set{(k,(v_p,t))}, k, v_p, t_0, \Phi, \token)$} for each of these threads.

Coming back to the proof of the $\mcslsearch$ that needed helping,
after the call to $\code{search}$ on line~\ref{line-mcssearch'-proof-call-search}, we know from the postcondition of~(\ref{eq:search-spec}) that we must have \tw{$(k,(v,t')) \in H$ for some $t'$ such that $t' \ge t_0$ where $H$ is the new upsert history at this point}.
Moreover, after resolving the prophecy on line~\ref{line-mcssearch'-proof-resolve-p} we know \tw{$v_p=v$} and therefore \tw{$(k,(v_p,t')) \in H$}.
From the invariant, we can then conclude that the thread must be in a $\lindone$ state.
Since the thread owns the unique token $\token$, it trades it in to obtain $\Phi(v)$, which lets it complete the proof of its atomic triple specification~(\ref{eq:search-spec}).


\section{The LSM DAG Template}
\label{sec-template}

This section presents a general template for multicopy structures that generalizes the LSM (log-structured merge) tree discussed in \refSec{sec-lsm}. We prove linearizability of the template by verifying that all operations satisfy the template-level atomic triples (\refSec{sec-atomic-spec-search-recency}). The template and the proof parameterize over the implementation of the single-copy data structures used at the node-level. Instantiating the template for a specific implementation involves only sequential reasoning about the implementation-specific node-level operations.


We split the template into two parts. The first part is a template for \code{search} and \code{upsert} that works on general multicopy structures, i.e., arbitrary DAGs with locally disjoint edgesets. The second part (discussed in \refSec{sec-maintenance}) is a template for a maintenance operation that generalizes the compaction mechanism found in existing list-based LSM tree implementations to tree-like multicopy structures.

\refFig{fig-multicopy-algorithm} shows the code of the template for the core multicopy operations. The operations \code{search} and \code{upsert} closely follow the high-level description of these operations on the LSM tree (\refSec{sec-lsm}). The operations are defined in terms of implementation-specific helper functions \helperFn{findNext}, \helperFn{addContents}, and \helperFn{inContents}. \twout{The \code{upsert} additionally uses \code{readClock} and \code{incrementClock}, auxiliary functions that are ghost code. The ghost functions manipulate the ghost state that keeps track of the clock value, facilitating the proof while having no effect on program behavior.}


\begin{figure}
  \centering

\begin{minipage}[t]{.48\textwidth}
\begin{lstlisting}[aboveskip=0pt,belowskip=0pt]
let rec traverse $r$ $n$ $k$ =
  lockNode $n$;
  match |<inContents>| $r$ $n$ $k$ with @\label{line-mc-search-in-contents}@
  | Some $v$ -> unlockNode $n$; $v$ @\label{line-mc-search-return}@
  | None ->
    match |<findNext>| $r$ $n$ $k$ with
    | Some $n'$ -> @\label{line-mc-search-find-next-succ}@
      unlockNode $n$;
      traverse $r$ $n'$ $k$ @\label{line-mc-search-recurse}@
    | None -> unlockNode $n$; $\square$ @\label{line-mc-search-find-next-fail}@

let search $r$ $k$ = traverse $r$ $r$ $k$
\end{lstlisting}
\end{minipage}%
\begin{minipage}[t]{.48\textwidth}
\begin{lstlisting}[aboveskip=0pt,belowskip=0pt,firstnumber=last]
let rec upsert $r$ $k$ $v$ =
  lockNode $r$;
  let $\mathit{res}$ = |<addContents>| $r$ $k$ $v$ in
  if $\mathit{res}$ then 
    unlockNode $r$ @\label{line-mc-upsert-commit}@
  else begin
    unlockNode $r$;
    upsert $r$ $k$ $v$
  end
\end{lstlisting}
\end{minipage}
\caption{ The general template for multicopy operations \code{search} and \code{upsert}. The template can be instantiated by providing implementations of helper functions \helperFn{inContents}, \helperFn{findNext}, and \helperFn{addContents}. \helperFn{inContents}$\,r\,n\,k$ returns \tw{\code{Some$\,v$} if $(v,t')=C_n(k)$ for some $t'$}, and \code{None} otherwise. \helperFn{findNext}$\,r\,n\,k$ returns \code{Some$\,n'$} if $n'$ is the unique node such that $k \in \edgeset(n,n')$, and \code{None} otherwise. \helperFn{addContents}$\,r\,k\,v$ \tw{updates the contents of $r$ by setting the value associated with key $k$ to $v$}. The return value of \helperFn{addContents} is a Boolean which indicates whether the insertion was successful (e.g., if $r$ is full, insertion may fail leaving $r$'s contents unchanged). 
\label{fig-multicopy-algorithm}}
\end{figure}

The \code{search} operation calls the recursive function \code{traverse} on the root node.
\code{traverse}$\,r\,n\,k$ first locks the node $n$ and uses the helper function \helperFn{inContents}$\,r\,n\,k$ to check if a copy of key $k$ is contained in $n$.
If a copy of $k$ is found, then its \tw{associated value $v$} is returned after unlocking $n$.
Otherwise, \code{traverse} uses the helper function \helperFn{findNext} to determine the unique successor $n'$ of the given node $n$ and query key $k$ (i.e., the node $n'$ satisfying $k \in \edgeset(n,n')$). If such a successor $n'$ exists, \code{traverse} recurses on $n'$. Otherwise, \code{traverse} concludes that there is no copy of $k$ in the data structure and returns \tw{$\square$}.
Note that this algorithm uses fine-grained concurrency, as the thread executing the \code{search} holds at most one lock at any point (and no locks at the points when \code{traverse} is called recursively).

The \code{upsert}$\;r\;k\;v$ operation locks the root node and adds a new copy of the key $k$ with value $v$ to the contents of the root node using \helperFn{addContents}. \twout{In order to add a new copy of $k$, it must know the current clock value. This is accomplished by using the \code{readClock} function.} \code{addContents}$\;r\;k\;v$ adds the pair $(k,v)$ to the root node when it succeeds. \code{upsert} terminates by \twout{incrementing the clock value and} unlocking the root node. The \helperFn{addContents} function may however fail if the root node is full. In this case \code{upsert} calls itself recursively\footnote{For simplicity of presentation, we assume that a separate maintenance thread flushes the root if it is full to ensure that upserts eventually make progress.}.


\section{Verifying the Template}
\label{setc-template-proof}

We next discuss the correctness proof of the template operations. We will focus on the high-level proof ideas and key invariants and defer the detailed proof outline and encoding of the invariants in Iris to Appendix~\ref{sec-iris-proofs}.




\subsection{High-Level Proof Outline}
\label{sec-multicopy-search-proof}


\paragraph{Proof of \code{search}.}
We start with the proof of \code{search}. Recall that search recency is the affirmation that if $t_0$ is the logical timestamp of $k$ at the point when \code{search$\;r\,k$} is invoked, then the operation returns \tw{$v$ such that $(k, (v,t')) \in H$ and $t' \ge t_0$}. Since the \tw{value $v$} of $k$ retrieved by \code{search} comes from some node in the structure, we must examine the relationship between the upsert history $H$ of the data structure and the physical contents $C_n$ of the nodes $n$ visited as the search progresses. We do this by identifying the main invariants needed for proving search recency for arbitrary multicopy structures.

We refer to the \emph{spatial ordering} of the copies \tw{$(k, (v,t))$} stored in a multicopy structure as the ordering in which those copies are reached when traversing the data structure graph starting from the root node. Our first observation is that the spatial ordering is consistent with the temporal ordering in which the copies have been upserted. We referred to this property as the time-ordering invariant in our library analogy in \refSec{sec-library-analogy}: the farther from the root a search is, the older the copies it finds are. Therefore, if a \code{search$\;r\;k$} traverses the data structure without interference from other threads and returns the first copy of $k$ that it finds, then it is guaranteed to return the logical value of $k$ at the start of the search.

We formalize this observation in terms of the \emph{contents-in-reach} of a node. The contents-in-reach of a node $n$ is \tw{the partial function $\cir(n) \colon \KS \pto \VS \times \Nat$} defined recursively over the graph of the multicopy structure as follows:
\tw{
\begin{equation}
  \label{eq-cir}
  \cir(n)(k) \defeq \begin{cases}
    C_n(k) & \text{if } k \in \dom(C_n)\\
    \cir(n')(k) & \text{else if } \exists n'.\, k \in \edgeset(n, n')\\
    \bot & \text{otherwise}
  \end{cases}
\end{equation}}
Note that $\cir(n)$ is well-defined because the graph is acyclic and the edgesets labeling the outgoing edges of every node $n$ are disjoint.
\tw{We further define $\ts(\cir(n)(k))=t$ if $\cir(n)(k)=(\_,t)$ and $\ts(\cir(n)(k))=0$ if $k \notin \dom(\cir(n))$.}

For example, in the multicopy structure depicted in \refFig{fig-multicopy}, we have \tw{$\cir(r)=\{k_1 \rightarrowtail (\square,6), k_2 \rightarrowtail (d,7), k_3 \rightarrowtail (c,4)\}$ and $\cir(n_3)=C_{n_3}$.}


\makeatletter
\let\orgdescriptionlabel\descriptionlabel
\renewcommand*{\descriptionlabel}[1]{%
  \let\orglabel\label
  \let\label\@gobble
  \phantomsection
  \edef\@currentlabel{#1\unskip}%
  \let\label\orglabel
  \orgdescriptionlabel{#1}%
}
\makeatother

The observation that interference-free searches will find the current logical timestamp of their query key is then captured by the following invariant:
\begin{enumerate}[label=\textbf{Invariant~\arabic*},ref={\arabic*},itemindent=55pt,labelindent=0pt,leftmargin=0pt]
\item The logical contents of the multicopy structure is the contents-in-reach of its root node: $\maxH = \cir(r)$. \label{inv-mc1}
\end{enumerate}

In order to account for concurrent threads interfering with the search, we prove the condition $t_0 \leq t'$ for the timestamp $t'$ \tw{associated with the value} returned by the search. Intuitively, this is true because the contents-in-reach of a node $n$ can be affected only by upserts or maintenance operations, both of which only increase the timestamps associated with every key of any given node: upserts insert new copies into the root node and maintenance operations move recent copies down in the structure, possibly replacing older copies. This observation is formally captured by the following invariant:
\begin{enumerate}[resume*]
\item \label{inv-mc2} The contents-in-reach of every node only increases. That is, for every node $n$ and key $k$, if \tw{$\ts(\cir(n)(k))=t$} at some point in time and \tw{$\ts(\cir(n)(k))=t'$} at any later point in time, then $t \leq t'$.
\end{enumerate}

Finally, in order to prove the condition \tw{$(k,(v,t')) \in H$} of search recency, we need one additional property:
\begin{enumerate}[resume*]
\item \label{inv-mc3} All copies present in the multicopy structure have been upserted at some point in the past. That is, for all nodes $n$, $C_n \subseteq H$.
\end{enumerate}

Now let us consider an execution of \code{search} on a operation key $k$.
In addition to the above three general invariants, we need an inductive invariant for the traversal performed by the search: we require as a precondition for \code{traverse$\,r\,n\,k$} that \tw{$\ts(\cir(n)(k)) \ge t_0$} where $t_0$ is \tw{the timestamp of the logical value $v_0$} of $k$ at the point when \code{search} was invoked.
To see that this property holds initially for the call to \code{traverse$\,r\,r\,k$} in \code{search}, let $\maxH[H_0]$ be the logical contents at the time point when \code{search} was invoked.
The precondition \tw{$\mcscont(k,v_0,t_0)$} implies \tw{$\ts(\maxH[H_0](k)) \ge t_0$}, which, combined with Invariant~\ref{inv-mc1} implies that we must have had \tw{$\ts(\cir(r)(k)) \ge t_0$} at this point.
Since \tw{$\ts(\cir(r)(k))$} only increases over time because of Invariant~\ref{inv-mc2}, we can conclude that \tw{$\ts(\cir(r)(k)) \ge t_0$} when \code{traverse} is called. We next show that the traversal invariant is maintained by \code{traverse} and is sufficient to prove search recency.

Consider a call to \code{traverse$\,r\,n\,k$} such that \tw{$\ts(\cir(n)(k)) \ge t_0$} holds initially. We must show that the call returns \tw{$v$ such that $(k,(v,t')) \in H$ and $t' \ge t_0$ for some $t'$}. We know that the call to \helperFn{inContents} on line~\ref{line-mc-search-in-contents} returns either \tw{\code{Some$\,v$} such that $(v,t')=C_n(k)$} or \code{None} if $C_n(k) = \bot$. Let us first consider the case where \helperFn{inContents} returns \code{Some$\,v$}. In this case, \code{traverse} returns $v$ on line~\ref{line-mc-search-return}. By definition of $\cir(n)$ we have $\cir(n)(k)=C_n(k)$. Hence, we have $\ts(\cir(n)(k)) = t'$ and the precondition \tw{$\ts(\cir(n)(k)) \ge t_0$}, together with Invariant~\ref{inv-mc2}, implies $t' \ge t_0$. Moreover, Invariant~\ref{inv-mc3} guarantees \tw{$(k, (v,t')) \in H$}.

Now consider the case where \helperFn{inContents} returns \code{None}. Here, $k \notin \dom{C_n(k)}$, indicating that no copy has been found for $k$ in $n$. In this case, \code{traverse} calls \helperFn{findNext} to obtain the successor node of $n$ and $k$. In the case where the successor $n'$ exists (line~\ref{line-mc-search-find-next-succ}), we know that $k \in \edgeset(n,n')$ must hold. Hence, by definition of contents-in-reach we must have $\cir(n)(k)=\cir(n')(k)$. From \tw{$\ts{\cir(n)(k)} \ge t_0$} and Invariant~\ref{inv-mc2}, we can then conclude \tw{$\ts(\cir(n')(k)) \ge t_0$}, i.e. that the precondition for the recursive call to \code{traverse} on line~\ref{line-mc-search-recurse} is satisfied and search recency follows by induction.

On the other hand, if $n$ does not have any next node, then \code{traverse} returns $\square$ (line~\ref{line-mc-search-find-next-fail}), indicating that $k$ has not yet been upserted at all so far (i.e., has never appeared in the structure). In this case, by definition of contents-in-reach we must have $\cir(n)(k) = \bot$. Invariant~\ref{inv-mc2} then guarantees $\ts(\cir{n}{k}) = 0 = t_0$. \tw{The invariant $\init(H)$ on the upsert history then gives us $(k,(\square,0)) \in H$. Hence, search recency holds in this case for $t'=0$.}

\paragraph{Proof of \code{upsert}.}

In order to prove the logically atomic specification (\ref{eq:upsert-spec-revised}) of \code{upsert}, we must identify an atomic step where the clock $t$ is incremented and the upsert history $H$ is updated. Intuitively, this atomic step is when the lock on the root node is released (line~\ref{line-mc-upsert-commit} in \refFig{fig-multicopy-algorithm}) after \helperFn{addContents} succeeds. Note that in this case \helperFn{addContents} changes the contents of the root node from $C_r$ to \tw{$C_r' = C_r[k \rightarrowtail (v,t)]$}. Hence, in the proof we need to update the ghost state for the upsert history from $H$ to \tw{$H'=H \cup \set{(k,(v,t))}$}, reflecting that a new copy of $k$ has been upserted. It then remains to show that the three key high-level invariants of multicopy structures identified above are preserved by these updates.

First, observe that Invariant~\ref{inv-mc3}, which states $\forall n.\; C_n \subseteq H$, is trivially maintained: only $C_r$ is affected by the upsert and the new copy \tw{$(k,(v,t))$} is included in $H'$. Similarly, we can easily show that Invariant~\ref{inv-mc2} is maintained: $\cir(n)$ remains the same for all nodes $n \neq r$ and for the root node it increases, provided Invariant~\ref{inv-mc1} is also maintained.

Thus, the interesting case is Invariant~\ref{inv-mc1}. Proving that this invariant is maintained amounts to showing that \tw{$\maxH[H'](k)=(v,t)$}. This step critically relies on the following additional observation:
\begin{enumerate}[resume*]
\item \label{inv-mc4} All timestamps in $H$ are smaller than the current time of the global clock $t$.
\end{enumerate}
This invariant implies that $\maxH[H'](k) = \max(\maxH(k),(v,t))=(v,t)$, which proves the desired property. We note that Invariant~\ref{inv-mc4} is maintained because the global clock is incremented when $H$ is updated to $H'$, and, as we describe below, while $r$ is locked.

\twout{
In the proof of Invariant~\ref{inv-mc1} we have silently assumed that the timestamp $t$, which was obtained by reading the global clock at line~\ref{line-mc-upsert-read-clock}, is still the value of the global clock at the linearization point when the clock is incremented at line~\ref{line-mc-upsert-incr-clock}. This step in the proof relies on the observation that only \code{upsert} changes the global clock and it  does so only while the clock is protected by the root node's lock. Hence, for lock-based implementations of multicopy structures, we additionally require the following invariant: While a thread holds the lock on the root node, no other thread will change the value of the global clock.
}

In the remainder of this section, we discuss the key technical issue when formalizing the above proof in a separation logic like Iris.

\subsection{Iris Invariant}
\label{sec-mcs-invariant}

The Iris proof must capture the key invariants identified in the proof outline given above in terms of appropriate ghost state constructions.
We start by addressing the key technical issue that arises when formalizing the above proof in a separation logic like Iris: contents-in-reach is a recursive function defined over an arbitrary DAG of unbounded size.
This makes it difficult to obtain a simple local proof that involves reasoning only about the bounded number of modified nodes in the graph. The recursive and global nature of contents-in-reach mean that modifying even a single edge in the graph can potentially change the contents-in-reach of an unbounded number of nodes (for example, deleting an edge $(n_1, n_2)$ can change $\cir(n)$ for all $n$ that can reach $n_1$).
A straightforward attempt to prove that a template algorithm preserves Invariant~\ref{inv-mc2} would thus need to reason about the entire graph after every modification (for example, by performing an explicit induction over the full graph). We solve this challenge using the flow framework~\cite{DBLP:conf/esop/KrishnaSW20}.

\paragraph{Encoding Contents-in-Reach using Flows.}

The flow framework enables separation-logic-style reasoning about recursive functions on graphs. Certain restrictions apply. The function must be of the form $\flow \colon N \to \mDom$ where $N$ is the set of nodes of the graph and $(\mDom,\mOp,\mZero)$ is a commutative cancellative monoid, called the \emph{flow domain}. Further, $\flow$ must satisfy the \emph{flow equation}:
\begin{equation}
  \label{eqn-flow-equation}
    \forall n \in N.\; \flow(n) = \inflow(n) \mOp \kern-3pt \mBigOp_{n' \in N} \edgeFn(n', n)(\flow(n'))
    \tag{\textsf{FlowEqn}}
\end{equation}
Intuitively, this equation states that $\flow$ can be computed by assigning every node an initial value according to the \emph{inflow function} $\inflow \colon N \to \mDom$ and then propagating these values along the edges of the graph using the \emph{edge function} $\edgeFn \colon N \times N \to \mDom \to \mDom$ to reach a fixpoint. At each node $n$, the values propagated from predecessor nodes $n'$ are aggregated using the monoid operation $\mOp$. A function $\flow$ that satisfies the flow equation is called a \emph{flow} and a graph equipped with a flow is a \emph{flow graph}. The flow framework then enables us to reason compositionally about invariants of flow graphs expressed as node-local conditions that depend on a node's flow.

If we can define the contents-in-reach in terms of a flow, then we can use the notion of a \emph{flow interface} to prove locally that an update to the graph does not change the flow of any nodes outside the modified region.
The flow interface of a region consists of its outflow and inflow, maps that intuitively capture the contribution of this region to the flow of the rest of the world and the contribution of the outside world to this region's flow, respectively.
If the interface of a modified region is preserved, then the framework guarantees that the flow of the rest of the graph is unchanged.
Thus, our proofs need to  prove only that Invariant~\ref{inv-mc2} is preserved for a bounded set of affected nodes.

Technically, this kind of reasoning is enabled by the separation algebra structure of flow graphs (in particular the definition of flow graph composition), which extends the composition of partial graphs in standard separation logic so that the frame rule also preserves flow values of nodes in the frame.
Instead of performing an explicit induction over the entire graph structure to prove that contents-in-reach values continue to satisfy desired invariants, the necessary induction is hidden away inside the definition of flow graph composition (for more details see~\cite{DBLP:conf/esop/KrishnaSW20}).
Note that since \code{search} does not modify the multicopy structure, it trivially maintains the flow interface of the nodes it operates on, and hence any flow-based invariants.

Equation~(\ref{eq-cir}) defines contents-in-reach in a bottom-up fashion, starting from the leaves of the multicopy structure graph.  That is, the computation proceeds \emph{backwards} with respect to the direction of the graph's edges. This makes a direct encoding of contents-in-reach in terms of a flow difficult because the flow equation describes computations that proceed in the forward direction.

We side-step this problem by tracking auxiliary ghost information in the data structure invariant for each node $n$ in the form of a function \tw{$Q_n\colon \KS \pto \VS \times \Nat$}. If these ghost values satisfy
\begin{equation}
  \label{eq:cir-flow-1}
  Q_n = \lambda k.
  \begin{cases}
    \cir(n')(k) & \text{if } \exists n'.\; k \in \edgeset(n, n')\\
    \bot & \text{otherwise}
  \end{cases}
\end{equation}
and we additionally define
\begin{equation*}
  \label{eq-cir-flow-2}
  \Cir{n} \defeq \lambda k.\, \ite{k \in \dom(C_n(k))}{C_n(k)}{Q_n(k)}
\end{equation*}
then $\cir(n) = \Cir{n}$.
The idea is that each node stores $Q_n$ so that node-local invariants can use it to talk about $\cir(n)$. We then use a flow to propagate the purported values $Q_n$ forward in the graph to ensure that they indeed satisfy (\ref{eq:cir-flow-1}). Note that while an \code{upsert} or maintenance operations on $n$ may change $\Cir{n}$, it preserves $Q_n$. That is, operations do not affect the contents-in-reach of downstream nodes, allowing local reasoning about the modification of the contents of $n$.

In what follows, let us fix a multicopy structure over nodes $N$ and some valuations of the \tw{partial} functions $Q_n$.
The flow domain $\mDom$ for our encoding of contents-in-reach consists of multisets of \tw{key/value-timestamp} pairs \tw{$\mDom \defeq \KS \times (\VS \times \Nat) \to \Nat$} with multiset union as the monoid operation. The edge function induced by the multicopy structure is defined as follows:
\begin{equation}
  \label{eq-cir-edgeFn}
  \edgeFn(n,n')(\_) \defeq \chi(\setcomp{(k,Q_n(k))}{k \in \edgeset(n,n') \land k \in \dom(Q_n)})
\end{equation}
Here, $\chi$ takes a set to its corresponding multiset. Additionally, we let the function $\inflow$ map every node to the empty multiset. With the definitions of $\edgeFn$ and $\inflow$ in place, there exists a unique flow $\flow$ that satisfies (\ref{eqn-flow-equation}). Now, if every node $n$ in the resulting flow graph satisfies the following two predicates
\begin{align}
  \phi_1(n) \defeq {} & \forall k.\; Q_n(k) = \bot \lor (\exists n'.\,k \in \edgeset(n,n'))\\
  \phi_2(n) \defeq {} & \forall k\,p.\; \flow(n)(k,p) > 0 \Rightarrow B_n(k)=p
\end{align}
then $\Cir{n}=\cir(n)$. Note that the predicates $\phi_1$ and $\phi_2$  depend only on $n$'s own flow and its local ghost state (i.e., $Q_n$, $C_n$ and the outgoing edgesets $\edgeset(n,\_)$).

\paragraph{Encoding the Invariants in Iris.}
\label{sec-lsm-dag-invariant}
We can now define the template-specific invariant $\mcsinv(r,t,H)$ for the LSM DAG template, which is assumed by the representation predicate $\mcsstate(\mcsinv,\proto)(r,t,H)$ defined in \refFig{fig-helping-protocol-inv}. We denote this invariant by $\lsminv$ and it is defined as follows:
\begin{align*}
    \lsminv(r,t,H) \defeq {} & \exists N.\;
    \globalInv(r,t,H,N)
    *  \Sep_{n \in N}
          \arraycolsep=2pt
          \begin{array}[t]{rl}
            \exists b_n\, C_n\, Q_n. &
            \lockR(b_n,n, \nodePred(r, n, C_n, Q_n))\\
            {} * {} & \nodeShar(r, n, C_n, Q_n, H)
          \end{array}
\end{align*}
The existentially quantified variable $N$ denotes the set of nodes of the multicopy structure. The invariant itself consists of two parts. The predicate $\globalInv(r,t,H,N)$ states certain invariants about its parameters and contains \emph{global} ghost resources storing the values $t$ and $N$. The second part is an iterated separating conjunction stating ownership of the node-local resources associated with every node $n \in N$.

The resources associated with each node $n$ are split between two predicates.
The predicate $\nodeShar(r, n, C_n, Q_n, H)$ holds those resources associated with $n$ that can be accessed by any thread operating on the data structure regardless of whether $n$ is locked or not. In particular, it contains the two predicates $\phi_1(n)$ and $\phi_2(n)$ needed for our encoding of contents-in-reach. The second predicate $\nodePred(r, n, C_n, Q_n)$ contains all resources that are accessible only to a thread that currently holds the lock on $n$. Ownership of node-local ghost state such as $Q_n$ is shared between the two predicates. This ensures that a thread may update the values of these resources only when it holds $n$'s lock. Moreover, every thread can assume that the constraints imposed on these values by $\nodeShar$ are true, even at times when the thread does not hold the lock.

$\nodePred(r, n, C_n, Q_n)$ includes \tw{$\heapRep(r, n, \edgeset(n,\cdot), \valof(C_n))$}, which is a predicate that encapsulates all resources specific to the implementation of the node-specific data structure abstracted by node $n$. In particular, this predicate owns the resources associated with the physical representation of the data structure and ties them to the abstract ghost state representing the high-level multicopy structure: the node's \tw{physical contents $\valof(C_n)$ (i.e., $C_n$ without timestamps)} and the edgesets of its outgoing edges $\edgeset(n,\cdot)$. Our template proof is parametric in the definition of $\heapRep$ and  depends only on the following two assumptions that each implementation used to instantiate the template must satisfy.
First, we require that $\heapRep$ is not duplicable:
    \[\heapRep(r, n, \mathit{es}, V_n) * \heapRep(r', n, \mathit{es}', V_n') \vdash \FALSE\]
Moreover, $\heapRep$ must guarantee disjoint edgesets:
\[\heapRep(r, n, \mathit{es}, V_n) \vdash \forall n_1\, n_2.\, n_1 = n_2 \lor \mathit{es}(n_1) \cap \mathit{es}(n_2) = \emptyset\]

The predicate $\lockR(b,n,R_n)$ captures the abstract state of $n$'s lock and is used to specify the protocol providing exclusive access to the resource $R_n$ protected by the lock via the helper functions $\code{lockNode}$ and $\code{unlockNode}$. The Boolean $b$ indicates whether the lock is (un)locked. \tw{The specifications of the helper functions used by \code{search} and \code{upsert}, given in terms of the predicates $\lockR(b,n,R_n)$ and $\heapRep(r, n, \mathit{es}, V_n)$ are shown in \refFig{fig-multicopy-helper-specs}.}
We discuss further details in \refApp{sec-iris-proofs}.



\begin{figure}[t]
  \centering
  \begin{lstlisting}[aboveskip=1pt,belowskip=0pt]
$\annotAtom{b\,R.\; \lockR(b,n, R)}$ |<lockNode>| $n$ $\annotAtom{ \lockR(\true, n, R) * R}$@\label{multicopy-search-proof-lockNode-spec}@
$\annotAtom{R.\; \lockR(\true,n, R) * R}$ |<unlockNode>| $n$ $\annotAtom{\lockR(\false, n, R)}$@\label{multicopy-search-proof-unlockNode-spec}@
$\annot{\heapRep(r, n, \mathit{es}, V_n)}$ |<inContents>| $n$ $k$ $\annot{\Ret x. \heapRep(r, n, \mathit{es}, V_n) * x = \ite{k \in \dom(V_n)}{\some(V_n(k))}{\none}}$@\label{multicopy-inContent}@
$\annot{\heapRep(r, n, \mathit{es}, V_n)}$ |<findNext>| $n$ $k$ $\annot{\Ret x. \heapRep(r, n, \mathit{es}, V_n) * x = \ite{\exists n'.\; k \in \mathit{es}(n')}{\some(n')}{\none}}$
$\annot{\heapRep(r,r,\mathit{es}, V_r)}$ |<addContents>| $r$ $k$ $v$ $\annot{\Ret b. \heapRep(r,r,\mathit{es}, V_r') * V_r' = \ite{b}{V_r[k \rightarrowtail v]}{V_r}}$
\end{lstlisting}
\caption{\tw{Specifications of helper functions used by \code{search} and \code{upsert}.}}
\label{fig-multicopy-helper-specs}
\end{figure}

\section{Multicopy Maintenance Operations}
\label{sec-maintenance}

We next show that we can extend our multicopy structure template in \refSec{sec-template} with a generic maintenance operation without substantially increasing the proof complexity.
The basic idea of our proofs here is that for every timestamped copy of key $k$, denoted as the pair \tw{$(k,(v,t))$}, every maintenance operation either does not change the distance of \tw{$(k,(v,t))$} to the root or increases it while preserving an edgeset-guided path to \tw{$(k,(v,t))$}. Using these two facts, we can prove that all the structure invariants are also preserved.

\subsection{Maintenance template}

For the maintenance template, we consider a generalization  of the compaction operation found in LSM tree implementations such as LevelDB~\cite{level-db} and Apache Cassandra~\cite{apache-cassandra,apache-cassandra-compaction}.
  While those implementations work on lists for the high-level multicopy structure, our maintenance template supports arbitrary tree-like multicopy structures. 
  The code is shown in \refFig{fig-multicopy-algorithm-compaction}. The template uses the helper function \helperFn{atCapacity}$\,r\,n$ to test whether the size of $n$ (i.e., the number of non-$\bot$ entries in $n$'s contents) exceeds an implementation-specific threshold. If not, then the operation simply terminates. In case $n$ is at capacity, the function \helperFn{chooseNext} is used to determine the node to which the contents of $n$ can be merged. If the contents of $n$ can be merged to successor $m$ of $n$, then \helperFn{chooseNext} returns \code{Some$\,m$}. In case no such successor exists, then it returns \code{None}. If \helperFn{chooseNext} returns \code{Some$\,m$}, then the contents of $n$ are merged to $m$. By merge, we mean that some copies of keys are transferred from $n$ to $m$, possibly replacing older copies in $m$. The merge is performed by the helper function \helperFn{mergeContents}. It must ensure that all keys $k$ merged from $C_n$ to $C_m$ satisfy $k \in \edgeset(n,m)$.
  
  On the other hand, if \helperFn{chooseNext} returns \code{None}, then a new node is allocated using the function \helperFn{allocNode}. The new node is then added to the data structure using the helper function \helperFn{insertNode}. Here, the new edgeset $\edgeset(n,m)$ must be disjoint from all edgesets for the other successors $m'$ of $n$. Afterwards, the contents of $n$ are merged to $m$ as before.
  Note that the maintenance template never removes nodes from the structure. In practice, the depth of the structure is bounded by letting the capacity of nodes grow exponentially with the depth. The right hand side of \refFig{fig-multicopy-algorithm-compaction} shows the intermediate states of a potential execution of the \code{compact} operation.

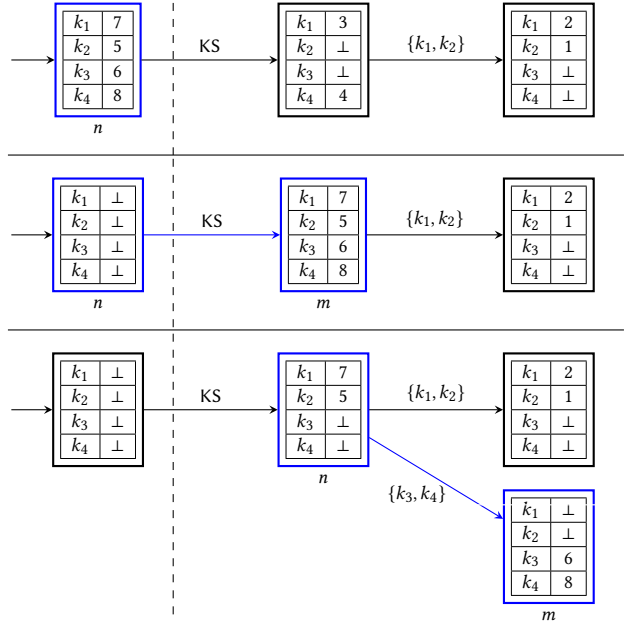
\begin{figure}[t]
  \centering
  \begin{minipage}{.37\textwidth}
    \centering
   \begin{lstlisting}[aboveskip=0pt,belowskip=0pt]
let rec compact $r$ $n$ =
  lockNode $n$;
  if |<atCapacity>| $r$ $n$ then begin
    match |<chooseNext>| $r$ $n$ with
    | Some $m$ -> 
      lockNode $m$;
      |<mergeContents>| $r$ $n$ $m$;@\label{line-mc-compaction-merge-contents}@
      unlockNode $n$;
      unlockNode $m$;
      compact $r$ $m$
    | None ->
      let $m$ = |<allocNode>| () in
      |<insertNode>| $r$ $n$ $m$;
      |<mergeContents>| $r$ $n$ $m$;
      unlockNode $n$;
      unlockNode $m$;
      compact $r$ $m$
  end
  else
    unlock $n$@\label{line-mc-compaction-final-unlock}@
  \end{lstlisting}
  \end{minipage}\hfill
  \begin{minipage}{.58\textwidth}
  \tikzset{
    unode/.style={circle, draw=black, thick, minimum size=4.5em, inner sep=2pt},
  }
  \begin{tikzpicture}[>=stealth,  every node/.style={scale=.8}, font=\small]

    \def\xsep{3.0}
    \def\ysep{-1.2}
    \def\xshift{2.5}
    \def\s{.5}
    \def\t{.8}

    \node[usnode, label = below:{$n$}, draw=blue] (r) {\mcbtable{k_1}{k_2}{k_3}{k_4}{7}{5}{6}{8}};
    \node[hnode, draw=white] (r') at ($(r) - (\xsep/2.5, 0)$) {};
    \node[usnode, label = below:{$$}] (n1) at ($(r) + (\xsep, 0) $) {\mcbtable{k_1}{k_2}{k_3}{k_4}{3}{\bot}{\bot}{4}};
    \node[usnode, label = below:{$$}] (n2) at ($(n1) + (\xsep, 0) $) {\mcbtable{k_1}{k_2}{k_3}{k_4}{2}{1}{\bot}{\bot}};

    \draw[edge] (r') to (r);
    \draw[edge] (r) to node[above] {$\KS$} (n1);
    \draw[edge] (n1) to node[above] {$\{k_1,k_2\}$} (n2);

    \draw[edge,-] ($(r) + (-\xsep/2.5,-1.25)$) to ($(n2) + (\xsep/3,-1.25)$);
    \draw[edge,dashed,-] ($(r) + (\xsep/3,.75)$) to ($(r) + (\xsep/3,-1.25)$);

 \end{tikzpicture}
 \begin{tikzpicture}[>=stealth,  every node/.style={scale=.8}, font=\small]

    \def\xsep{3.0}
    \def\ysep{-1.2}
    \def\xshift{2.5}
    \def\s{.5}
    \def\t{.8}

    \node[usnode, label = below:{$n$},draw=blue] (r) at (0,-8) {\mcbtable{k_1}{k_2}{k_3}{k_4}{\bot}{\bot}{\bot}{\bot}};
    \node[hnode, draw=white] (r') at ($(r) - (\xsep/2.5, 0)$) {};
    \node[usnode, label = below:{$m$},draw=blue] (n1) at ($(r) + (\xsep, 0) $) {\mcbtable{k_1}{k_2}{k_3}{k_4}{7}{5}{6}{8}};
    \node[usnode, label = below:{$ $}] (n2) at ($(n1) + (\xsep, 0) $) {\mcbtable{k_1}{k_2}{k_3}{k_4}{2}{1}{\bot}{\bot}};

    \draw[edge] (r') to (r);
    \draw[edge, draw=blue] (r) to node[above] {$\KS$} (n1);
    \draw[edge] (n1) to node[above] {$\{k_1,k_2\}$} (n2);

    \draw[edge,-] ($(r) + (-\xsep/2.5,-1.25)$) to ($(n2) + (\xsep/3,-1.25)$);
    \draw[edge,dashed,-] ($(r) + (\xsep/3,1)$) to ($(r) + (\xsep/3,-1.25)$);
  \end{tikzpicture}
  \begin{tikzpicture}[>=stealth,  every node/.style={scale=.8}, font=\small]

    \def\xsep{3.0}
    \def\ysep{-1.8}
    \def\xshift{2.5}
    \def\s{.5}
    \def\t{.8}

    \node[usnode, label = below:{$$}] (r) {\mcbtable{k_1}{k_2}{k_3}{k_4}{\bot}{\bot}{\bot}{\bot}};
    \node[hnode, draw=white] (r') at ($(r) - (\xsep/2.5, 0)$) {};
    \node[usnode, label = below:{$n$},draw=blue] (n1) at ($(r) + (\xsep, 0) $) {\mcbtable{k_1}{k_2}{k_3}{k_4}{7}{5}{\bot}{\bot}};
    \node[usnode, label = below:{$$}] (n2) at ($(n1) + (\xsep, 0) $) {\mcbtable{k_1}{k_2}{k_3}{k_4}{2}{1}{\bot}{\bot}};
    \node[usnode, label = below:{$m$},draw=blue] (m) at ($(n1) + (\xsep, \ysep)$) {\mcbtable{k_1}{k_2}{k_3}{k_4}{\bot}{\bot}{6}{8}};

    \draw[edge] (r') to (r);
    \draw[edge] (r) to node[above] {$\KS$} (n1);
    \draw[edge] (n1) to node[above] {$\{k_1,k_2\}$} (n2);
    \draw[edge,draw=blue] (n1) to node[below] {$\{k_3,k_4\}\quad\quad$} (m);

    \draw[edge,-,draw=white] ($(r) + (-\xsep/2.5,-1.25)$) to ($(n2) + (\xsep/3,-1.25)$);
    \draw[edge,dashed,-] ($(r) + (\xsep/3,1)$) to ($(r) + (\xsep/3,-2.75)$);
 \end{tikzpicture}
  \end{minipage}

  \caption{ Maintenance template for tree-like multicopy structures. The template can be instantiated by providing implementations of helper functions \helperFn{atCapacity}, \helperFn{chooseNext}, \helperFn{mergeContents}, \helperFn{allocNode}, and \helperFn{insertNode}.
    \helperFn{atCapacity}$\,r\,n$ returns a Boolean value indicating whether node $n$ has reached its capacity.
    The helper function \helperFn{chooseNext}$\,r\,n$ returns \code{Some$\;m$} if there exists a successor $m$ of $n$ in the data structure into which $n$ should be compacted, and \code{None} in case $n$ cannot be compacted into any of its successors.
    \helperFn{mergeContents}$\,r\,n\,m$ (partially) merges the contents of $n$ into $m$.
    Finally, \helperFn{allocNode} is used to allocate a new node and \helperFn{insertNode}$\,r\,n\,m$ inserts node $m$ into the data structure as a successor of $n$.
    The right hand side shows a possible execution of \code{compact}. Edges are labeled with their edgesets. The nodes $n$ and $m$ in each iteration are marked in blue. \tw{For simplicity, we here assume that the values are identical to their associated timestamps and only show the timestamps.}
\label{fig-multicopy-algorithm-compaction}}
\end{figure}

\subsection{High-level proof of \code{compact}}

The verification framework presented in \refSec{sec-mcs} can be easily extended to accommodate any maintenance operation on multicopy structures that does not change the data structure's abstract state. That is, we need to prove that \code{compact} satisfies the following atomic triple:
\begin{equation*}
   \atomicTriple{t\,H.\; \mcsstate(r,t, H)}{\code{compact}\;r}{\mcsstate(r,t, H)}
 \end{equation*}
 This specification says that \code{compact} logically takes effect in a single atomic step, and at this step the abstract state of the data structure does not change. \tw{We prove that \code{compact} satisfies this specification relative to the specifications of the implementation-specific helper functions shown in \refFig{fig-multicopy-helper-specs-compaction}. The postcondition of \helperFn{mergeContents} is given with respect to an (existentially quantified) set of keys $K$ that are merged from $V_n$ to $V_m$, resulting in new content sets $V_n'$ and $V_m'$. The new contents are determined by the functions $\mathit{mergeLeft}$ and $\mathit{mergeRight}$ which are defined as follows:
\begin{align*}
  \mathit{mergeLeft}(K, V_n, \mathit{Es}, V_m) \defeq {} &
  \lambda k.\, \ite{k \in K \cap \dom(V_n) \cap \mathit{Es}}{\bot}{V_n(k)}\\
  \mathit{mergeRight}(K, V_n, \mathit{Es}, V_m) \defeq {} &
  \lambda k.\, \ite{k \in K \cap \dom(V_n) \cap \mathit{Es}}{V_n(k)}{V_m(k)}
\end{align*}}

\begin{figure}[t]
  \centering
  \begin{lstlisting}[aboveskip=1pt,belowskip=0pt]
$\annot{\heapRep(r, n, \mathit{es}_n, V_n)}$ |<atCapacity $r$ $n$>| $\annot{\Ret b. \heapRep(r, n, \mathit{es}_n, V_n)}$

$\annot{\heapRep(r, n, \mathit{es}_n, V_n)}$ 
|<chooseNext>| $r$ $n$
$\annot{\Ret v. \heapRep(r, n, \mathit{es}_n, V_n) *
  (v = \some(m) * \mathit{es}_n(m) \neq \emptyset \lor v = \none * \code{needsNewNode}(r, n, \mathit{es}_n, V_n))}$

$\annot{\TRUE}$ |<allocNode>| $r$ $\annot{\Ret m. \heapRep(r,m, (\lambda n'.\,\emptyset), \emptyset)}$

$\annot{\heapRep(r,n,\mathit{es}_n, V_n) * \code{needsNewNode}(r, n, \mathit{es}_n, V_n) * \heapRep(r,m,  (\lambda n'.\,\emptyset), \emptyset)}$
|<insertNode>| $r$ $n$ $m$
$\annot{\heapRep(r,n,\mathit{es}_n', V_n) * \heapRep(r,m, (\lambda n'.\,\emptyset), \emptyset) *  \mathit{es}_n' = \mathit{es}_n[m \rightarrowtail \mathit{es}_n'(m)] * \mathit{es}_n'(m) \neq \emptyset}$

$\annot{\heapRep(r, n, \mathit{es}_n, V_n) * \heapRep(r, m, \mathit{es}_m, V_m) * \mathit{es}_n(m) \neq \emptyset}$ 
|<mergeContents>| $r$ $n$ $m$
$\annot{\heapRep(r, n, \mathit{es}_n, V_n') * \heapRep(r, m, \mathit{es}_m, V_m') * V_n' = \mathit{mergeLeft}(K, V_n, \mathit{Es}, V_m) * V_m' = \mathit{mergeRight}(K, V_n, \mathit{Es}, V_m)}$
\end{lstlisting}
\caption{\tw{Specifications of helper functions used by \code{compact}.}}
\label{fig-multicopy-helper-specs-compaction}
\end{figure}
 
Technically, the linearization point of the operation occurs when all locks are released, just before the function terminates. However, the interesting part of the proof is to show that the changes to the physical contents of nodes $n$ and $m$ performed by each call to \helperFn{mergeContents}  at line~\ref{line-mc-compaction-merge-contents} preserve the abstract state of the structure as well as the invariants. In particular, the changes to $C_n$ and $C_m$ also affect the contents-in-reach of $m$. We need to argue that this is a local effect that does not propagate further in the data structure, as we did in our proof of \code{upsert}.


\paragraph{Auxiliary invariants.}
When proving the correctness of \code{compact}, we face two technical challenges. The first challenge arises when establishing that \code{compact} changes the contents of the nodes involved in such a way that the high-level invariants are maintained. In particular, we must reestablish Invariant~\ref{inv-mc2}, which states that the contents-in-reach of each node can only increase over time. Compaction replaces downstream copies of keys with upstream copies. Thus, in order to maintain Invariant~\ref{inv-mc2}, we need the additional auxiliary invariant that the timestamps of keys in the contents of nodes can only decrease as we move away from the root:
\begin{enumerate}[resume*]
\item \label{inv-mc6} The (timestamp) contents of a node is not smaller than the contents-in-reach of its successor. That is, for all keys $k$ and nodes $n$ and $m$, if $k \in \edgeset(n,m)$ and $C_n(k) \neq \bot$ then \tw{$\ts(\cir(m)(k)) \le \ts(C_n(k))$}.
\end{enumerate}

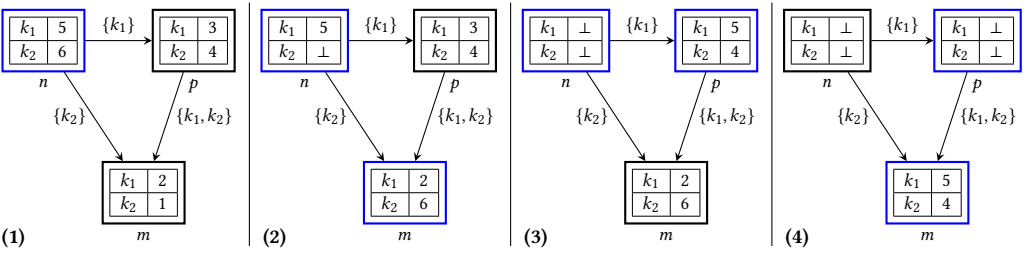
\begin{figure}[t]
  \centering
  \tikzset{
    unode/.style={circle, draw=black, thick, minimum size=4.5em, inner sep=2pt},
  }
  \begin{minipage}{.25\linewidth}
  \begin{tikzpicture}[>=stealth,  every node/.style={scale=.77}, font=\small]

    \def\xsep{2.0}
    \def\ysep{-2.0}
    \def\xshift{2.5}
    \def\s{.5}
    \def\t{.8}

    \node[usnode, label = below:{$n$}, draw=blue] (r) {\mctabletworows{k_1}{k_2}{5}{6}};
    \node[hnode, draw=white] (r') at ($(n2) + (\xsep/2.3, 0)$) {};
    \node[usnode, label = below:{$p$}] (n1) at ($(r) + (\xsep, 0) $) {\mctabletworows{k_1}{k_2}{3}{4}};
    \node[usnode, label = below:{$m$}] (n2) at ($(r) + (\xsep/1.5, \ysep)$) {\mctabletworows{k_1}{k_2}{2}{1}};

    \draw[edge] (r) to node[above] {$\{k_1\}$} (n1);
    \draw[edge] (n1) to node[right] {$\{k_1,k_2\}$} (n2);
    \draw[edge] (r) to node[left] {$\{k_2\}$} (n2);

    \draw[edge,-] ($(n1) + (\xsep/2.7,.5)$) to ($(n1) + (\xsep/2.7,-2.7)$);
    \node[hnode] (step) at ($(r) + (-\xsep/5,-2.6)$) {\bf(1)};

  \end{tikzpicture}
\end{minipage}%
\begin{minipage}{.25\linewidth}
\begin{tikzpicture}[>=stealth,  every node/.style={scale=.77}, font=\small]

    \def\xsep{2.0}
    \def\ysep{-2.0}
    \def\xshift{2.5}
    \def\s{.5}
    \def\t{.8}

    \node[usnode, label = below:{$n$}, draw=blue] (r) {\mctabletworows{k_1}{k_2}{5}{\bot}};
    \node[hnode, draw=white] (r') at ($(n2) + (\xsep/2.3, 0)$) {};
    \node[usnode, label = below:{$p$}] (n1) at ($(r) + (\xsep, 0) $) {\mctabletworows{k_1}{k_2}{3}{4}};
    \node[usnode, label = below:{$m$}, draw=blue] (n2) at ($(r) + (\xsep/1.5, \ysep)$) {\mctabletworows{k_1}{k_2}{2}{6}};

    \draw[edge] (r) to node[above] {$\{k_1\}$} (n1);
    \draw[edge] (n1) to node[right] {$\{k_1,k_2\}$} (n2);
    \draw[edge] (r) to node[left] {$\{k_2\}$} (n2);

    \draw[edge,-] ($(n1) + (\xsep/2.7,.5)$) to ($(n1) + (\xsep/2.7,-2.7)$);
    \node[hnode] (step) at ($(r) + (-\xsep/5,-2.6)$) {\bf(2)};

  \end{tikzpicture}
  \end{minipage}%
  \begin{minipage}{.25\linewidth}
  \begin{tikzpicture}[>=stealth,  every node/.style={scale=.77}, font=\small]

    \def\xsep{2.0}
    \def\ysep{-2.0}
    \def\xshift{2.5}
    \def\s{.5}
    \def\t{.8}

    \node[usnode, label = below:{$n$}, draw=blue] (r) {\mctabletworows{k_1}{k_2}{\bot}{\bot}};
    \node[hnode, draw=white] (r') at ($(n2) + (\xsep/2.3, 0)$) {};
    \node[usnode, label = below:{$p$}, draw=blue] (n1) at ($(r) + (\xsep, 0) $) {\mctabletworows{k_1}{k_2}{5}{4}};
    \node[usnode, label = below:{$m$}] (n2) at ($(r) + (\xsep/1.5, \ysep)$) {\mctabletworows{k_1}{k_2}{2}{6}};

    \draw[edge] (r) to node[above] {$\{k_1\}$} (n1);
    \draw[edge] (n1) to node[right] {$\{k_1,k_2\}$} (n2);
    \draw[edge] (r) to node[left] {$\{k_2\}$} (n2);

    \draw[edge,-] ($(n1) + (\xsep/2.7,.5)$) to ($(n1) + (\xsep/2.7,-2.7)$);

    \node[hnode] (step) at ($(r) + (-\xsep/5,-2.6)$) {\bf(3)};

  \end{tikzpicture}
\end{minipage}%
\begin{minipage}{.25\linewidth}
\begin{tikzpicture}[>=stealth,  every node/.style={scale=.77}, font=\small]

    \def\xsep{2.0}
    \def\ysep{-2.0}
    \def\xshift{2.5}
    \def\s{.5}
    \def\t{.8}

    \node[usnode, label = below:{$n$}] (r) {\mctabletworows{k_1}{k_2}{\bot}{\bot}};
    \node[usnode, label = below:{$p$}, draw=blue] (n1) at ($(r) + (\xsep, 0) $) {\mctabletworows{k_1}{k_2}{\bot}{\bot}};
    \node[usnode, label = below:{$m$}, draw=blue] (n2) at ($(r) + (\xsep/1.5, \ysep)$) {\mctabletworows{k_1}{k_2}{5}{4}};

    \draw[edge] (r) to node[above] {$\{k_1\}$} (n1);
    \draw[edge] (n1) to node[right] {$\{k_1,k_2\}$} (n2);
    \draw[edge] (r) to node[left] {$\{k_2\}$} (n2);

    \node[hnode] (step) at ($(r) + (-\xsep/5,-2.6)$) {\bf(4)};
    
    \draw[edge,-,draw=white] ($(n1) + (\xsep/.5,.5)$) to ($(n1) + (\xsep/.5,-2.7)$);

  \end{tikzpicture}
\end{minipage}%
\caption{Possible execution of the \code{compact} operation on a DAG. Edges are labeled with their edgesets. The nodes undergoing compaction in each iteration are marked in blue.\label{fig-multicopy-compaction-dag}}
\end{figure}

We can capture Invariant~\ref{inv-mc6} in our data structure invariant $\mcsstate(r,t,H)$ by adding the following predicate as an additional conjunct to the predicate $\nodeShar(r, n, C_n, \Cir{n})$:
\begin{align}
  \phi_3(n) \defeq {} & \forall k.\, \tw{\ts(Q_n(k)) \le \ts(\Cir{n}(k))}
\end{align}
  
The second challenge is that the maintenance template generates only tree-like structures. This implies that at any time there is at most one path from the root to each node in the structure. We will see that this invariant is critical for maintaining Invariant~\ref{inv-mc6}. However, the data structure invariant presented thus far allows for arbitrary DAGs.

To motivate this issue further, consider
the multicopy structure in step \textbf{(1)} of
\refFig{fig-multicopy-compaction-dag}. The logical contents of this
structure (i.e. the contents-in-reach of $n$) is
\tw{$\{k_1 \rightarrowtail (5,5), k_2 \rightarrowtail (6,6)\}$}.

The structure in step \textbf{(2)} shows the result obtained after executing \code{compact$\,r\,n$} to completion where $n$ has been considered to be at capacity and the successor $m$ has been chosen for the merge, resulting in \tw{$(k_2,(6,6))$} being moved from $n$ to $m$. Note that at this point the logical contents of the data structure is still \tw{$\{k_1 \rightarrowtail (5,5), k_2 \rightarrowtail (6,6)\}$} as in the original structure. However, the structure now violates Invariant~\ref{inv-mc6} for nodes $p$ and $m$ since \tw{$\ts(\Cir{m}(k_2)) > \ts(C_p(k_2))$}.

Suppose that now a new compaction starts at $n$ that still considers $n$ at capacity and chooses $p$ for the merge. The merge then moves the copy \tw{$(k_1,(5,5))$} from $n$ to $p$. The graph in step \textbf{(3)} depicts the resulting structure.
The compaction then continues with $p$, which is also determined to be at capacity. Node $m$ is chosen for the merge, resulting in \tw{$(k_1, (5,5))$} and \tw{$(k_2, (4,4))$} being moved from $p$ to $m$. At this point, the second compaction terminates. The final graph in step \textbf{(4)} shows the structure obtained at this point. Observe that the logical contents is now \tw{$\{k_1 \rightarrowtail (5,5), k_2 \rightarrowtail (4,4)\}$}. Thus, this execution violates the specification of \code{compact}, which states that the logical contents must be preserved. In fact, \tw{a timestamp in} the contents-in-reach of $n$ has decreased, which violates Invariant~\ref{inv-mc2}.


We observe that although \code{compact} will create only tree-like structures, we can prove its correctness using a weaker invariant that does not rule out non-tree DAGs, but instead focuses on how \code{compact} interferes with concurrent \code{search} operations. This weaker invariant relies on the fact that for every key $k$ in the contents of a node $n$, there exists a unique search path from the root $r$ to $n$ for $k$. That is, if we project the graph to only those nodes reachable from the root via edges $(n,m)$ that satisfy $k \in \edgeset(n,m)$, then this projected graph is a list. Using this weaker invariant we can capture implementations based on B-link trees or skip lists which are DAGs but have unique search paths.

To this end, we recall from~\cite{DBLP:journals/tods/ShashaG88} the notion of the \emph{inset} of a node $n$, $\inset(n)$, which is the set of keys $k$ such that there exists a (possibly empty) path from the root $r$ to $n$, and $k$ is in the edgeset of all edges along that path. That is, since a \code{search} for a key $k$  traverses only those edges $(n,m)$ in the graph that have $k$ in their edgeset, the \code{search}  traverses (and accesses the contents of) only those nodes $n$ such that $k \in \inset(n)$.
Now observe that \code{compact}, in turn,  moves new copies of a key $k$ downward in the graph only along edges that have $k$ in their edgeset. The following invariant is a consequence of these observations and the definition of contents-in-reach:

\begin{enumerate}[resume*]
\item \label{inv-mc7} A key is in the contents-in-reach of a node only if it is also in the node's inset. That is, \tw{$\dom(\cir(n)) \subseteq \inset(n)$}.
\end{enumerate}
This invariant rules out the problematic structure in step \textbf{(1)} of \refFig{fig-multicopy-compaction-dag} because we have \tw{$k_2 \in \dom(\cir(p))$} but $k_2 \notin \inset(p)=\set{k_1}$.

Invariant~\ref{inv-mc7} alone is not enough to ensure that Invariant~\ref{inv-mc6} is preserved. For example, consider the structure obtained from \textbf{(1)} of \refFig{fig-multicopy-compaction-dag} by changing the edgeset of the edge $(n,p)$ to $\set{k_1,k_2}$. This modified structure satisfies Invariant~\ref{inv-mc7} but allows the same problematic execution ending in the violation of Invariant~\ref{inv-mc6} that we outlined earlier.
However, observe that in the modified structure $k_2 \in \edgeset(n,p) \cap \edgeset(n,m)$, which violates the property that all edgesets leaving a node are disjoint.
We have already captured this property in our data structure invariant (as an assumption on the implementation-specific predicate $\heapRep(r,n,\mathit{es},C_n)$).
However, in our formal proof we need to rule out the possibility that a search for $k$ can reach a node $m$ via two \emph{incoming} edgesets $\edgeset(n,m)$ and $\edgeset(p,m)$.
Proving that disjoint \emph{outgoing} edgesets imply unique search paths involves global inductive reasoning about the paths in the multicopy structure.
To do this using only local reasoning, we will instead rely on an inductive consequence of locally disjoint outgoing edgesets, which we capture explicitly as an additional auxiliary invariant (and which we will enforce using flows):
\begin{enumerate}[resume*]
\item \label{inv-mc8} The distinct immediate predecessors of any node $n$ have disjoint insets. More precisely, for all distinct nodes $n$, $p$, $m$, and keys $k$, if $k \in \edgeset(n,m) \cap \edgeset(p,m)$ then $ k \notin \inset(n) \cap \inset(p)$.
\end{enumerate}
Note that changing the edgeset of $(n,p)$ in \refFig{fig-multicopy-compaction-dag} to $\set{k_1,k_2}$ would violate Invariant~\ref{inv-mc8} because the resulting structure would satisfy $k_2 \in \edgeset(n,m) \cap \edgeset(p,m)$ and $k_2 \in \inset(n) \cap \inset(p)$.

In order to capture invariants~\ref{inv-mc7} and~\ref{inv-mc8} in
$\mcsstate(r,t,H)$, we introduce an additional flow that we use to
encode the inset of each node.
The encoding of insets in terms of a flow
follows~\cite{DBLP:conf/pldi/KrishnaPSW20}. That is, the underlying
flow domain is multisets of keys $\mDom = \KS \to \Nat$ and the actual
calculation of the insets is captured by (\ref{eqn-flow-equation}) if
we define:
\begin{align*}
  \edgeFn(n,n') \defeq{} & \lambda m.\, m \cap \edgeset(n,n') &
  \inflow(n) \defeq {} & \chi \ite{n=r}{\KS}{\emptyset}
\end{align*}
If $\flow_\inset$ is a flow that satisfies (\ref{eqn-flow-equation}) for these definitions
of $\edgeFn$ and $\inflow$, then for any node $n$ that is reachable
from $r$, $\flow_\inset(n)(k) > 0$ iff $k \in \inset(n)$.
Invariants~\ref{inv-mc7} and~\ref{inv-mc8} are then captured by the following two predicates, which we add to $\nodeShar$:
 \begin{align*}
   \phi_4(n) \defeq {} & \forall k.\,
                         k \in \dom(\Cir{n}) \Rightarrow \flow_\inset(n)(k) > 0 &
   \phi_5(n) \defeq {} & \forall k.\, \flow_\inset(n)(k) \le 1
 \end{align*}
Note that $\phi_5$ captures Invariant~\ref{inv-mc8} as a property of each individual node $n$ by taking advantage of the fact that the multiset $\flow_\inset(n)$ explicitly represents all of the contributions made to the inset of $n$ by $n$'s predecessor nodes.

We briefly explain why we can still prove the correctness of
\code{search} and \code{upsert} with the updated data structure
invariant. First note that \code{search} does not modify the contents,
edgesets, or any other ghost resources of any node. So the additional
conjuncts in the invariant are trivially maintained.

Now let us consider the operation \tw{\code{upsert$\,r\,k\,v$}}. Since \code{upsert} does not change the edgesets of any nodes, the resources and constraints related to the inset flow are trivially maintained, with the exception of $\phi_4(r)$: after the \code{upsert} we have \tw{$k \in \dom(\Cir{r})$} which may not have been true before. However, from $\inflow(r)(k)=1$, the flow equation, and the fact that the flow domain is positive, it follows that we must have $\flow_\inset(r)(k)>0$ (i.e., $k \in \inset(r)=\KS$). Hence, $\phi_4(r)$ is preserved as well.




\section{Proof Mechanization}
\label{sec-evaluation}

\begin{table}[t]
  \caption[Summary of templates and instantiations verified.]{
    Summary of templates and instantiations verified in Iris/Coq and \grasshopper.
    For each algorithm or
    library, we show the number of lines of code, lines of proof
    annotation (including specification), total number of lines, and
    the proof-checking/verification time in seconds.}
  \label{tab-line-counts}
  \setlength{\tabcolsep}{5pt}
  \resizebox{.5\textwidth}{!}{
  \begin{tabular}[t]{l r r r r}
    \multicolumn{5}{l}{\bf Templates (Iris/Coq)}\\[2pt]
    \hline
    \bf Module & \bf Code & \bf Proof & \bf Total & \bf Time \\
    \hline
Flow Library            & 0     & 3757     & 3757  & 41\\
Lock Implementation     & 10     & {\the\numexpr 362-10}     & 362   & 11\\
Client-level Spec       & 7     & {\the\numexpr 938-7}    & 938   & 40\\
DF Template             & 19     & {\the\numexpr 933-19}    & 933   & 90\\
LSM DAG Template        & 39     & {\the\numexpr 3705-39}  & 3705  & 353\\
\bf Total               &  \bf {\the\numexpr 0+10+7+19+39}  &  \bf {\the\numexpr (3757+362+938+933+3705)-(0+10+7+19+39)} & \bf {\the\numexpr 3757+362+938+933+3705}  & \bf {\the\numexpr 41+11+40+90+353}\\ 
    \hline
  \end{tabular}
  }%
  \resizebox{.5\textwidth}{!}{
  \begin{tabular}[t]{l r r r r}
    \multicolumn{5}{l}{\bf Implementations (\grasshopper)}\\[2pt]
    \hline
    \bf Module & \bf Code & \bf Proof & \bf Total & \bf Time \\
    \hline
Array Library           & 191   & 440   & 631   & 11\\
LSM Implementation      & 207   & 246   & 453   & 51\\
                        &       &    &    & \\
                        &       &    &    & \\
                        &       &    &    & \\                        
\bf Total               & \bf {\the\numexpr 191+207}   & \bf {\the\numexpr 440+246}   & \bf {\the\numexpr 631+453}   & \bf {\the\numexpr 11+51}\\
    \hline
  \end{tabular}
}%
\end{table}

We illustrate the proof methodology presented in this paper by verifying that the multicopy template algorithm (\refSec{sec-template}, \refSec{setc-template-proof}, and \refSec{sec-maintenance}) satisfies search recency.
We then instantiate the template to an LSM-like implementation to demonstrate an application of the template.
Our proof effort (summarized in \refTab{tab-line-counts}) also contains a mechanically-checked proof that search recency refines the Map ADT specification (\refSec{sec-client-template-refinement}).
We further verify a two-node multicopy structure template that can be instantiated to differential file (DF) structure implementations~\cite{DBLP:journals/tods/SeveranceL76}.
We include this template in our artifact to demonstrate the reuse of the helping proof and because it has a simpler invariant. Though, we provide no implementation for the two-node template. \tw{The artifact is available as a VM image on Zenodo\footnote{\url{https://zenodo.org/record/5496104}} and as source code on GitHub\footnote{\url{https://github.com/nyu-acsys/template-proofs/multicopy}}.}
Verification time was measured
on a laptop with an Intel Core i7-8750H CPU and 16GB RAM.

The client-level and template-level proofs were performed in Iris and
mechanically verified by the Coq tool, and comprise the left half of
\refTab{tab-line-counts}. The flow library formalizes the meta theory of flow interface cameras used in the template proofs. It extends the development of~\cite{DBLP:conf/pldi/KrishnaPSW20} with a general theory of multiset-based flow domains (about 900 lines).

Our LSM implementation is verified in the SMT-based separation logic tool \grasshopper, and is described in the right half of the table.
The implementation uses an unsorted array to store key-timestamp pairs for the (in-memory) root node (with upserts adding to one end of the array), and a read-only sorted array (also known as a sorted string table~\cite{level-db}) for the other (on-disk) nodes. This array models the contents of a file. The implementation uses a library of utility functions and lemmas for arrays that represent partial maps from keys to values.

We verify both the helper functions for the core search structure operations (\refFig{fig-multicopy-algorithm}) as well as those needed by the maintenance template (\refFig{fig-multicopy-algorithm-compaction}). Each operation demuxes between the code for in-memory and on-disk nodes based on the reference to the operation node. For instance, in the case of $\helperFn{mergeContents}\,r\,n\,m$, if $r = n$ then the operation flushes the in-memory node $n$ to the on-disk node $m$. Otherwise, both $n$ and $m$ must be on-disk nodes, which are then compacted. Alternatively, one could use separate implementations of each helper function for the two types of nodes.
The polymorphism could then be resolved statically by unfolding the recursion in the template algorithms once, letting helper function calls in the unfolded iteration go to the in-memory versions and all remaining ones to the on-disk versions.

There are two gaps in the verification that would need to be bridged to obtain a complete end-to-end proof. First, there is currently no way to formally compose the proofs done in Iris/Coq and \grasshopper. However, the two proofs are linked by the node-level specifications of helper functions such as \code{findNext} at the representation level. As with prior work~\cite{DBLP:conf/pldi/KrishnaPSW20}, we split our verification across two tools in order to take advantage of SMT-based automated techniques for the sequential implementation proofs which are tedious but not technically challenging. Second, \grasshopper does not support reasoning about file access directly. We effectively model each file as a RAM disk whose contents is mapped into memory. This is consistent with the abstract interface for performing file accesses in the LSM tree implementation of LevelDB~\cite{level-db}.

\tw{
One specific technical challenge that we had to overcome in the Iris formalization is related to the decoupling of the generic client-template proof from the template-implementation proofs. At the linearization point of \code{upsert}, the proof needs to reestablish the invariant of the helping protocol. That is, each template-implementation proof needs to update the relevant ghost resources used for encoding this invariant. We have eliminated the dependency of the template-implementation proof on the concrete representation of the helping protocol invariant by parameterizing this part of the proof over all possible helping protocols that can be maintained by an \code{upsert}. We discuss this issue in more detail in \refApp{sec-decoupling}.

A more desirable solution would be to restrict all reasoning related to the helping protocol to the client-template proof so that the template-implementation proofs do not depend on the helping protocol at all. Essentially, the idea would be to do the relevant ghost state updates in the client-template level proof of \code{upsert} when the template-level atomic triple of \code{upsert} is committed. Unfortunately, this idea cannot be realized with Iris' current definition of atomic triples. Proving that the helping protocol invariant is maintained involves the elimination of a so-called \emph{later} modality. That is, one needs to show that a physical computation step is executed at the linearization point (e.g. a memory read or write). However, Iris' atomic triples $\atomicTriple{\vec{x}.\;P}{e}{\val.\; Q}$ are in some sense too abstract, as they do not capture whether $c$ performs a physical computation step.  More fine-grained notions of atomic triples are a promising direction for  future work.}


\section{Related Work}
\label{sec-related}

Most closely related to our work is the edgeset framework for verifying
single-copy structure
templates~\cite{DBLP:journals/tods/ShashaG88,DBLP:conf/pldi/KrishnaPSW20}.
The edgeset framework hinges on the notion of the \emph{keyset} of a node, which is the set of keys that are allowed in the node. That is, a node's contents must be a subset of its keyset. Moreover, the keysets of all nodes must be disjoint. The contribution of~\citet{DBLP:conf/pldi/KrishnaPSW20} is to capture these invariants by a resource algebra in Iris and to show how keysets can be related to the search structure graph using flows to enable local reasoning about template algorithms for single-copy structures. Note that this work~\cite{DBLP:journals/tods/ShashaG88,DBLP:conf/pldi/KrishnaPSW20} is limited to single-copy structures since the keyset invariants enforce that every key appears in at most one node. In multicopy structures, the same key may appear in multiple nodes with different associated values.

Relative to~\cite{DBLP:journals/tods/ShashaG88,DBLP:conf/pldi/KrishnaPSW20}, the main technical novelties are: (i) we identify a node-local quantity (contents-in-reach) for multicopy structures that plays a similar role to the keyset in the single-copy case. Both the invariants that the contents-in-reach must satisfy as well as how the contents-in-reach is encoded using flows is substantially different from the keyset. (ii) We capture the order-preservation aspect of linearizability for multicopy structures in the notion of search recency. (iii) We develop and verify new template algorithms for multicopy structures.

In data structures based on RCU synchronization such as the Citrus tree~\cite{DBLP:conf/podc/ArbelA14}, the same key may temporarily appear in multiple nodes. However, such structures are not necessarily multicopy structures. Notably, in a Citrus tree, all copies of a key have the same associated value even in the presence of concurrent updates. Moreover, searches have fixed linearization points. This structure can therefore be handled, in principle, using the single-copy framework of~\citet{DBLP:conf/pldi/KrishnaPSW20} (by building on the formalization of the RCU semantics developed in~\cite{DBLP:conf/esop/GotsmanRY13} and the high-level proof idea for the Citrus tree of~\citet{DBLP:journals/pacmpl/FeldmanKE0NRS20}).

Several other works present generic proof arguments for verifying concurrent traversals of search structures that involve dynamic linearization points~\cite{DBLP:conf/podc/OHearnRVYY10, DBLP:conf/wdag/FeldmanE0RS18, DBLP:conf/ppopp/Drachsler-Cohen18, DBLP:journals/pacmpl/FeldmanKE0NRS20}. However, these approaches focus on single-copy structures and rely on global reasoning based on graph reachability.

The idea of tracking auxiliary ghost state about a data structure's history to simplify its linearizability proof has been used in many prior works (e.g.~\cite{DBLP:conf/esop/SergeyNB15,DBLP:conf/ecoop/DelbiancoSNB17,DBLP:conf/cav/BouajjaniEEM17}). We build on these works and apply this idea to decouple the reasoning about the non-local linearization points of searches from the verification of any specific multicopy structure template.


We have formalized the verification of our template algorithms in Iris~\cite{DBLP:journals/jfp/JungKJBBD18}. Our formalization particularly benefits from Iris's support for user-definable resource algebras, which can capture nontrivial ghost state such as flow interfaces. However, there are a number of other formal proof systems that provide mechanisms for structuring complex linearizability proofs, including other concurrent separation logics~\cite{DBLP:conf/concur/FuLFSZ10,DBLP:conf/pldi/SergeyNB15,DBLP:conf/esop/RaadVG15,DBLP:conf/ecoop/PintoDG14,DBLP:conf/popl/Dinsdale-YoungBGPY13,DBLP:journals/entcs/GardnerRWW14} as well as systems based on classical logic~\cite{DBLP:conf/tacas/ElmasQSST10,DBLP:conf/cav/KraglQH20,DBLP:conf/cav/KraglQ18}. We also make use of Iris's support for logically atomic triples and prophecy variables to reason modularly about the non-local dynamic linearization points of searches. Specifically, the proof discussed in~\refSec{sec-client-template-refinement} builds on the prophecy-based Iris proof of the RDCSS data structure from~\cite{DBLP:journals/pacmpl/JungLPRTDJ20} and adapts it to a setting where an unbounded number of threads perform ``helping''. The idea of using prophecy variables to reason about non-fixed linearization points has also been explored in prior work building on logics other than Iris~\cite{DBLP:phd/ethos/Vafeiadis08, DBLP:conf/vstte/SezginTQ10, DBLP:conf/tamc/ZhangFFSL12}.

\tw{
Our proofs rely on both history and prophecy-based reasoning. However, we use the two ideas separately in the two parts of the proof (client-template vs. template-implementation). It does not seem possible to prove the client-template part without using prophecies. The reason is that we use an atomic triple to express the client-level specification. The atomic triple needs to be committed at the actual linearization point. If we were to use only history-based information in the proof, then we would determine at the point when $(k,(v,t'))$ is found that the linearization point already happened in the past. However, at that point, it is already too late to commit the atomic triple.

A proof that uses only history-based verification and does not rely on atomic triples is likely possible.} For instance, one alternative approaches to using atomic triples is to prove that the template-level atomic specification contextually refines the client-level atomic specification of multicopy structures using a relational program logic. A number of prior works have developed such refinement-based approaches~\cite{DBLP:conf/fsttcs/BanerjeeNN16, DBLP:conf/lics/FruminKB18,DBLP:journals/corr/abs-2006-13635}, including for settings that involve unbounded helping~\cite{DBLP:conf/pldi/LiangF13, DBLP:conf/popl/TuronTABD13}. An alternative approach to using prophecy variables for reasoning about non-fixed linearization points is to explicitly construct a partial order of events as the program executes, effectively representing all the possible linearizations that are consistent with the observations made so far~\cite{DBLP:conf/esop/KhyzhaDGP17}.

There has also been much work on obtaining fully automated proofs of linearizability by static analysis and model
checking~\cite{DBLP:conf/esop/AbdullaJT18, DBLP:conf/cav/AmitRRSY07,  DBLP:conf/esop/BouajjaniEEH13,  DBLP:conf/icalp/BouajjaniEEH15,  DBLP:conf/cav/BouajjaniEEM17,  DBLP:conf/concur/HenzingerSV13,  DBLP:conf/cav/ZhuPJ15, DBLP:conf/vmcai/Vafeiadis09, DBLP:conf/cav/CernyRZCA10, DBLP:conf/cav/DragoiGH13, DBLP:conf/cav/LesaniMP14, DBLP:conf/tacas/AbdullaHHJR13}.
The proof framework presented in this paper is capable of reasoning about implementations that are beyond the reach of current automatic techniques, via interactive (though still machine-checked) template proofs. We hope that this framework will help to inform the design of future automated static analyzers for concurrent programs.




\tw{
  Multicopy structures such as the LSM tree are often used in file and database systems to organize data that spans multiple storage media, e.g., RAM and hard disks. Several prior projects have considered the formal verification of file systems. SibyllFS~\cite{DBLP:conf/sosp/RidgeSTGMS15} provides formal specifications for POSIX-based file system implementations to enable systematic testing of existing implementations. FSCQ~\cite{DBLP:conf/sosp/ChenZCCKZ15}, Yggdrasil~\cite{DBLP:conf/asplos/BornholtKLKTW16, DBLP:conf/osdi/Sigurbjarnarson16}, and DFSCQ~\cite{DBLP:conf/sosp/ChenCKWICKZ17} provide formally verified file system implementations that also guarantee crash consistency. However, these implementations do not support concurrent execution of file system operations.
Our work provides a framework for reasoning about the in-memory concurrency aspects of multicopy structures. However, we mostly abstract from issues related to the interaction with the different storage media. Notably, in our verified LSM tree implementation, we do not model disk failure and hence do not address crash consistency.
}

Distributed key/value stores have to contend with copies of keys being present in multiple nodes at a time. Several works verify consistency of operations performed on such data structures~\cite{DBLP:conf/wdag/ChordiaRRRV13,DBLP:journals/pacmpl/KakiNNJ18,DBLP:conf/ecoop/XiongCRG19}, including linearizability~\cite{DBLP:conf/pldi/WangEMP19}. In the distributed context, the main technical challenge arises from data replication and the ensuing weakly consistent semantics of concurrent operations. As we consider lock-based templates, we can assume a sequentially consistent memory model for our verification.
For lock-free multicopy structures such as the Bw-tree~\cite{DBLP:conf/icde/LevandoskiLS13a}, weak memory consistency may be a concern. \tw{Lock-free multicopy structures also require the development of new template algorithms, which then need to be shown linearizable with respect to the template-level specification. However, once this is established, linearizability with respect to the client-level specification is obtained for free. We also believe that the high-level invariants from ~\refSec{setc-template-proof} are applicable towards proving the template-level specification. For instance, each lock-free node-local list of the Bw-tree behaves like a multicopy structure and satisfies the identified invariants.}


\section{Conclusion}

This paper and the accompanying verification effort have made the
following contributions: We presented a general framework for
verifying concurrent multicopy structures. The framework introduces an
intermediate abstraction layer that enables reasoning about concurrent
multicopy structures in terms of template algorithms that abstract
from the data structure representation. We constructed such a template
algorithm that generalizes the log-structured merge tree to DAGs and
proved its correctness. The proof is decomposed into two parts to
maximize proof reuse: (1) a general reduction of linearizability of
multicopy structures that eliminates the need to reason about
non-local linearization points; and (2) a generic proof of the
template algorithm that abstracts from the data structure's memory
representation in concrete implementations.  The full proof is
formalized in the concurrent separation logic Iris and mechanized in
Coq. We have also verified an instantiation of the template algorithm
to LSM trees, resulting in the first formally-verified \tw{concurrent} multicopy search structure.


\endgroup

\begin{acks}                            
This work is funded in parts by  the
\grantsponsor{GS100000001}{National Science
  Foundation}{http://dx.doi.org/10.13039/100000001} under
grants~\grantnum{GS100000001}{1925605},
 \grantnum{GS100000001}{1815633}, 1934388, 1840761, and 1339362. Further funding came from NYU WIRELESS and from
  the New York University Abu Dhabi
Center for Interacting Urban Networks
(CITIES). We thank Elizabeth Dietrich and Raphael Sofaer for their help on mechanizing the proofs of the differential file template.  We also extend our gratitude to the anonymous reviewers of OOPSLA'21 whose questions helped us clarify the presentation. We would also like to thank  the reviewers of the book ~\cite{KrishnaETAL21CSS}, specifically  Maurice Herlihy, Eddie Kohler, Robbert Krebbers, K.\ Rustan M.\ Leino, and Peter M\"uller, for their suggestions to improve the presentation.
\end{acks}
 
\bibliography{references,dblp}


\clearpage

\appendix
\section{Iris Formalization}
\label{sec-iris-proofs}

In this section we present a more detailed summary of the proofs we have formalized in Iris. The full development can be found in the supplementary materials. We start with a brief discussion of some additional Iris concepts that we use in our proofs before proceeding to the actual proof details.

\subsection{Atomic Triples}

Before we discuss the details of our encoding of the helping protocol in terms of Iris ghost state, it is instructive to understand the
basic structure of a proof of an atomic triple
$\atomicTriple{x.\;P}{e}{\Ret\val. Q}$.
  \begin{mathpar}
    \inferH{logatom-intro}
    {\All \Phi. \hoareTriple{\atomicUpdate_{x. P, Q}(\Phi)}{e}{\Ret\val. \Phi(\val)}}
    {\atomicTriple{x.\;P}{e}{\Ret\val. Q}}

    \inferH{logatom-atom}
    {\All x . \hoareTriple{P}{e}{\Ret\val. Q} \\
      e \text{ atomic}
    }
    {\atomicTriple{x.\; P}{e}{\Ret\val. Q}}

    \inferH{au-abort}
    {\atomicTriple{x.\; P * P'}{e}{\Ret\val. P * Q'}}
    {\hoareTriple{\atomicUpdate_{x. P, Q}(\Phi) * P'}{e}{\Ret\val. \atomicUpdate_{x. P, Q}(\Phi) * Q'}}

    \inferH{au-commit}
    {\atomicTriple{x.\; P * P'}{e}{\Ret\val. Q * Q'}}
    {\hoareTriple{\atomicUpdate_{x. P, Q}(\Phi) * P'}{e}{\Ret\val. \Phi(\val) * Q'}}
  \end{mathpar}

The proof proceeds by proving a standard Hoare triple of the form $\All \Phi. \hoareTriple{\atomicUpdate_{x. P, Q}(\Phi)}{e}{\Ret\val. \Phi(\val)}$. Here, $\atomicUpdate_{x. P, Q}(\Phi)$ is the \emph{atomic update token}, which gives us the \emph{right} to use the resources in the precondition $P$ when executing atomic instructions up to the linearization point. The token also records our \emph{obligation} to preserve $P$ up to the linearization point, where $P$ must be transformed to $Q$ in one atomic step. This step consumes the update token. The universally quantified proposition $\Phi$ can be thought of as the precondition for the continuation of the client of the atomic triple. At the linearization point, when the atomic update token is consumed, the corresponding proof rule produces $\Phi$ as a receipt that the obligation has been fulfilled. This receipt is necessary to complete the proof of the Hoare triple.

This can only be done at $e$'s linearization point which must transform $P$ into $Q$ and yields $\Phi(v)$ as a receipt. Up to the linearization point, the resources in $P$ can be accessed using the rule \refRule{au-abort}, which ensures that $P$ is preserved.

\subsection{Invariants in Iris}

\tw{In the proof of an atomic triple $\atomicTriple{x.\;P}{e}{\Ret\val. Q}$, we can  access  the resources represented by the precondition $P$ only up to the linearization point of $e$.
At the linearization point, the resources of $P$ are transferred to the postcondition $Q$, which then become inaccessible to the proof.
This restriction presents a minor technical challenge when proving the atomic triple for \code{search}.
The linearization point of \code{search} can be in a concurrent \code{upsert}.
The proof of \code{upsert} updates the registry of the helping protocol for all \code{search} threads that it linearizes.
In the corresponding case of the proof of \code{search}, we then need to access resources associated with the data structure invariant to determine the new state of the helping protocol, after the \code{search} has already been linearized.
If the atomic preconditions of \code{search}'s specification governs all the relevant
resources, then we can no longer access these resources after the linearization point.}

\tw{This issue can be avoided by using invariants.} An invariant in Iris is a formula of the form $\inv{P}$, where $P$ is an arbitrary Iris proposition. Invariants provide an orthogonal mechanism to atomic triples in order to reason about ownership of resources describing shared state that can be concurrently accessed by many threads. Intuitively, an invariant is a property that, once established, will remain true forever. It is therefore a duplicable resource and can be freely shared with any thread.
  
However, in order to ensure that the invariant indeed remains valid once it has been established, Iris' proof rules for invariants impose restrictions on how the resources contained in an invariant can be accessed and manipulated. At any point in time, a thread can \emph{open} an invariant $\inv{P}$ and gain ownership of the contained resources $P$. These resources can then be used in the proof of a single atomic step of the thread's execution. After the thread has performed an atomic step with an open invariant, the invariant must be \emph{closed}, which amounts to proving that $P$ has been reestablished. Otherwise, the proof cannot succeed. In this sense, invariants behave much like the preconditions of atomic triples before the atomic triple has been committed, as captured by the rule \refRule{au-abort}. However, unlike atomic preconditions, which become inaccessible after the commit point, an invariant is always accessible as long as it is reestablished after each atomic step. 

The $\invName$ in $\inv{P}$ refers to the \emph{namespace} of the invariant. Namespaces are part of the mechanism used in Iris to keep track of invariants that are currently open and need to be closed before the next atomic step. This is necessary to avoid issues of re-entrancy in case of nested invariants, which would lead to logical inconsistencies. In the following, we omit these namespace annotations for ease of notation. For a more in-depth discussion of Iris' invariant mechanism and the relevant proof rules, we refer the interested reader to~\cite{DBLP:journals/jfp/JungKJBBD18}.

\subsection{Decoupling the Helping and Template Proofs}
\label{sec-decoupling}

In our proofs, we parameterize the representation predicate $\mcslstate$ by two abstract predicates $\mcsinv(r,t,H)$ and $\proto(H)$. We do this to achieve a complete decoupling of the template-specific proofs from the helping proof, which relates the client-level and template-level specifications, and vice versa. The predicate $\mcsinv(r,t,H)$ abstracts from the resources needed for proving that a particular multicopy structure template satisfies the template-level specifications. In particular, this predicate will store the authoritative version of the global clock $t$. The predicate $\proto(H)$ abstracts from the resources used to track the state of the helping protocol. 

The fact that $\proto$ depends on $H$ creates an unfortunate entanglement between the proofs performed at the two abstraction levels: at the linearization point of \code{upsert}, the upsert history must be updated from $H$ to \tw{$H \cup \set{(k,(t,v))}$}. This step happens in the proof of each particular template algorithm for \code{upsert}. Therefore, these proofs will also have to carry out the ghost update that replaces $\proto(H)$ by \tw{$\proto(H \cup \set{(k,(v,t))})$}.

Fortunately, the template-specific proof does not need to know how $\proto$ is defined. It only needs to know that for any values of $H$, $k$, and $t$, $\proto(H)$ can be updated to \tw{$\proto(H \cup \set{(k,(v,t))})$}. We capture this assumption on $\proto$ formally using a predicate $\mcsprotoupdate(\proto)$ that we define using Iris's linear view shift modality:
\[\mcsprotoupdate(\proto) \defeq (\forall \; H\, k\, t\, v.\; \proto(H) \vsR \proto(H \cup \set{(k,(v,t))}))\]
In order to verify a particular implementation of \code{search} and \code{upsert} with respect to a particular template-specific invariant $\mcsinv(r,t,H)$, one then needs to prove validity of the following Iris propositions:
\begin{align}
  & \forall\; \proto\;r\;k\;v.\; \sinv{\mcslinv(\mcsinv,\proto)(r)} \magicwand \mcsprotoupdate(\proto) \magicwand {} \notag\\
  & \quad \atomicTriple{t\, H.\; \mcsstate(r,t, H)}{\code{upsert}\;r\;k\;v}{\mcsstate(r,t + 1, H \cup (k,(v,t)))}\label{eq:upsert-spec-revised}\\[.5em]
  & \forall\; \proto\;r\;k\;v_0\;t_0.\; \sinv{\mcslinv(\mcsinv,\proto)(r)} \magicwand {} \mcscont(k, v_0, t_0) \magicwand \notag\\
  & \quad \atomicTriple{t\, H.\; \mcsstate(r,t, H)}{\code{search}\;r\;k}{\Ret v. \exists t'.\; \mcsstate(r,t, H) * t_0\!\leq\! t' \,*\, (k,(v,t')) \in H } \label{eq:search-spec-revised}
\end{align}
These propositions abstract over any helping protocol predicate $\proto$ that is compatible with the ghost update of $H$ performed by \code{upsert}.
Note that an atomic triple guarded by an invariant can be interpreted as satisfying the atomic triple under the assumption that the shared state satisfies the invariant.

We show below that there exists a specific helping protocol invariant $\helpingproto$ such that the client-level atomic triples hold for any implementation of \code{upsert} that satisfies proposition~(\ref{eq:upsert-spec-revised}) and any implementations of \code{search} that satisfies proposition~(\ref{eq:search-spec-revised}). Formally, if we abbreviate proposition~(\ref{eq:upsert-spec-revised}) by $\mcsupsertspec(\code{upsert},\mcsinv)$ and proposition~(\ref{eq:search-spec-revised}) by $\mcssearchspec(\code{search},\mcsinv)$. Then we show validity of the following two propositions:
\begin{align}
  & \forall \; \code{upsert}\;\mcsinv\;r\;k\;v.\; \mcsupsertspec(\code{upsert},\mcsinv) \magicwand {} \notag\\
  & \;\; \atomicTriple{M.\,\mcslstate(\mcsinv,\helpingproto)(r,M)}
    {\code{upsert}\;r\;k\;v}
    {\mcslstate(\mcsinv,\helpingproto)(r, M[k \rightarrowtail v])} \label{eq:upsert-seq-spec-proph}\\[.5em]
  & \forall \; \code{search}\;\mcsinv\;r\;k.\; \mcssearchspec(\code{search},\mcsinv) \magicwand {} \notag\\
  & \;\; \atomicTriple{M.\,\mcslstate(\mcsinv,\helpingproto)(r, M)}
    {\mcslsearch\;r\;k}
    {\Ret v. \mcslstate(\mcsinv,\helpingproto)(r,M) * M(k)=v} \label{eq:search-seq-spec-proph}
\end{align}
These propositions abstract over the template-specific parameters \code{upsert}, \code{search}, and $\mcsinv$, but fix the helping protocol invariant $\proto$ to be $\helpingproto$. In order to obtain the overall correctness proof of a specific template $(\code{upsert}, \code{search}, \mcsinv)$, the proofs of these generic propositions then  need to be instantiated only with the proofs of $\mcsupsertspec(\code{upsert},\mcsinv)$ and $\mcssearchspec(\code{search},\mcsinv)$.

\subsection{The Full Helping Proof}
\label{sec-iris-rep-predicates}

This section presents the full proof of the helping protocol described in \refSec{sec-helping-protocol}.

The proof outline in \refFig{fig-multicopy-search'-proof} uses rule~\refRule{logatom-intro} to obtain the atomic update token $\atomicUpdate(\Phi)$ for the atomic triple in proposition~(\ref{eq:search-seq-spec-proph}) right at the start of $\mcslsearch$ (line~\ref{line-mcssearch'-proof-logatom-intro-full}). The proof transfers ownership of $\atomicUpdate(\Phi)$ from the thread-local context of the $\mcslsearch$ thread to the thread-local context of the \code{upsert} thread that will linearize the search. We do this via the shared representation predicate $\mcslstate$, or, to be more precise, an invariant that we store in $\mcslstate$. In the proof of \code{upsert}, when the search is linearized, the associated receipt \tw{$\Phi(v)$} is  transferred via the shared invariant back to the $\mcslsearch$ thread. We next explain the details of this part of the proof along with the definitions of the involved predicates.

\begin{figure}[t]
  \centering
  \begin{lstlisting}[aboveskip=1pt,belowskip=0pt]
$\annot{\mcssearchspec(\code{search},\mcsinv)} * \annotAtom{M.\; \mcsstate(\mcsinv,\helpingproto)(r, M)}$
let $\mcslsearch$ $r$ $k$ =
  (* Start application of @\refRule{logatom-intro}@ *)
  $\annot{\atomicUpdate(\Phi)}$ @\label{line-mcssearch'-proof-logatom-intro-full}@
  let $\tid$ = $\newproph$ in
  let $p$ = $\newproph$ in
  $\annot{\atomicUpdate(\Phi) * \proph(\tid, \_) * \proph(p, v_p)}$ @\label{line-mcssearch'-proof-case-split-pre1-full}@
  $\annot{\atomicUpdate(\Phi) * \proph(\tid, \_) * \proph(p, v_p) * \sinv{\mcslinv(\mcsinv,\helpingproto)(r)}}$ @\label{line-mcssearch'-proof-case-split-pre2-full}@
  $\annot{\atomicUpdate(\Phi) * \proph(\tid, \_) * \proph(p, v_p) * \ghostState{s}{\authFrag\; H_0} * (v_0,t_0) \!=\! \maxH[H_0](k) * \mcscont(k, v_0, t_0)}$ @\label{line-mcssearch'-proof-case-split-pre3-full}@
  (* Case analysis on $v_p = v_0$, $v_p \neq v_0$: only showing $v_p \neq v_0$ *) @\label{line-mcssearch'-proof-case-split-full}@
  $\annot{\proph(p, v_p) * (v_0,t_0) \!=\!\maxH[H_0](k) * \mcscont(k, v_0, t_0) * v_p \!\neq\! v_0 * \dots}$ @\label{line-mcssearch'-proof-case-split-post-full}@
    $\annot{\ldots * \atomicUpdate(\Phi) * \proph(\tid, \_)}$ @\label{line-mcssearch'-proof-call-search-pre-frame-full}@
    $\annot{\ldots * \proph(\tid, \_) * \linpending(H_0,k,v_p,t_0\Phi) * \ghostState{sy(\tid)}{\fracHalf\; H_0} * \ghostState{sy(\tid)}{\fracHalf\; H_0} * \token}$ @\label{line-mcssearch'-proof-call-search-pre1-full}@
    $\annot{\ldots * \ghostState{r}{\authAuth\; R} * \tid \notin R
      * \sinv{\helpingstate(\tid, k, v_p, t_0, \Phi, \token)} * \reg(\tid, H_0, k, v_p, t_0, \Phi, \token) * \token}$ @\label{line-mcssearch'-proof-call-search-pre2-full}@
    (* Ghost update: $\ghostState{r}{\authAuth\; R} \vsR \ghostState{r}{\authAuth\; R \cup \set{\tid}}$ *) @\label{line-mcssearch'-proof-call-search-pre3-full}@
  $\annot{\proph(p, v_p) * \token * \ghostState{r}{\authFrag\; \{tid\}}  * \sinv{\helpingstate(\tid, k, v_p, t_0, \Phi, \token)} * \mcscont(k, v_0, t_0) * \mcsstate(r,t,H)}$ @\label{line-mcssearch'-proof-call-search-pre4-full}@
  let $v$ = search $r$ $k$ in @\label{line-mcssearch'-proof-call-search-full}@
  $\annot{\proph(p, v_p) * \token * \ghostState{r}{\authFrag\; \{tid\}} * \mcsstate(r,t,H) * (k, (v,t')) \in H * t_0 \!\le\! t' * \ghostState{s}{\set{(k, t')}}}$ @\label{line-mcssearch'-proof-call-search-post-full}@
  $\resolve{p}{v}$; @\label{line-mcssearch'-proof-resolve-p-full}@
  $\annot{\token * \ghostState{r}{\authFrag\; \{\tid\}} * v_p \!=\! v * \ghostState{s}{\set{(k, v_p)}}}$ @\label{line-mcssearch'-proof-resolve-p-post-full}@
  $\annot{\Phi(v_p) * v_p \!=\! v}$ @\label{line-mcssearch'-proof-phi-full}@
  $\annot{\Phi(v)}$
  $v$
  (* End application of @\refRule{logatom-intro}@ *) @\label{line-mcssearch'-proof-end-full}@
  $\annotAtom{\Ret v. \mcsstate(\mcsinv,\helpingproto)(r,M) * M(k)=v}$
\end{lstlisting}

\caption{Outline for the proof of proposition~(\ref{eq:search-seq-spec-proph}).}
\label{fig-multicopy-search'-proof}
\end{figure}

\begin{figure}[t]

  \begin{align*}
    \mcslstate(\mcsinv,\proto)(r,t,M) \defeq {} & \exists\, t\, H.\;
    \mcsstate(r,t,H) \,*\,  (\forall k.\; (M(k),\_)\!=\!\maxH(k)) \,*\, \sinv{\mcslinv(\mcsinv,\proto)(r)} \\
    \mcslinv(\mcsinv,\proto)(r) \defeq {} &
     \exists\, t\,H.\; \mcsstate^{\authAuth}(r, t, H) \,*\,
       \ghostState{s}{\authAuth\; H} \,*\, \ghostState{t}{\authAuth\; t} \,*\, \mcsinv(t,H) \,*\, \proto(H) \\
     & {} \,*\, \init(H) \,*\, \hunique(H) \,*\, \maxTS(t,H) \\[.5em]
    \helpingproto(H) \defeq {} & \exists\, R.\; \ghostState{r}{\authAuth\; R}
     * \Sep_{\tid \in R}        
           \; \exists\, k\, v_p\, t_0, \Phi\, \token.\; \reg(\tid, H, k, v_p, t_0, \Phi, \token)\\
    \reg(\tid, H, k, v_p, t_0, \Phi, \token) \defeq {} &
           \proph(\tid, \_)
           * \ghostState{sy(\tid)}{\fracHalf\; H}
           * \sinv{\helpingstate(\tid, k, v_p, t_0, \Phi, \token)}
    \\[.5em]
    \helpingstate(\tid, k, v_p, t_0, \Phi, \token) \defeq {}
    & \exists\; H.\; \ghostState{sy(\tid)}{\fracHalf\; H} \\
    & * (\linpending(H, k, v_p, t_0, \Phi) \lor \lindone(H,k,v_p,t_0,\Phi, \token))
    \\[.5em]
    \linpending(H, k, v_p, t_0, \Phi) \defeq {}
    & 
    \atomicUpdate(\Phi)
    * (\forall t.\; (k, (v_p, t)) \in H \Rightarrow t < t_0)
    \\[.5em]
    \lindone(H,k,v_p,t_0,\Phi, \token) \defeq {}
    &  (\Phi(v_p)\; \lor\; \token) * (\exists t.\, (k, (v_p,t)) \in H \land t \ge t_0)\\[.5em]
    \mcscont(k,v,t) \defeq {} & \ghostState{s}{\authFrag \set{(k,(v,t))}}
  \end{align*}
  \caption{Full definition of client-level representation predicate and invariants of helping protocol.}
  \label{fig-helping-protocol-inv-revised}
\end{figure}

\refFig{fig-helping-protocol-inv-revised} shows the full definition of the representation predicate $\mcslstate$ and the invariant that encodes the helping protocol.
The predicate $\mcslstate(\mcsinv,\proto)(r,t,M)$ contains the predicate $\mcsstate(r,t,H)$, used in the template-level atomic triples, and then defines $M$ in terms of $H$ \tw{via $(\forall k.\; (M(k),\_)\!=\!\maxH(k))$}.
All remaining (ghost) resources associated with the data structure are owned by the predicate $\mcslinv(\mcsinv,\proto)(r)$. In particular, this predicate stores the ghost resource $\ghostState{s}{\authAuth H}$ whose type is the authoritative RA over sets \tw{$\KS \times (\VS \times \nat)$}. This resource holds the authoritative version of the current upsert history $H$. Notably, we can then define the predicate \tw{$\mcscont(k,v_0,t_0)$} as the fractional resource \tw{$\ghostState{s}{\authFrag \set{(k,(v_0,t_0))}}$}, which expresses the auxiliary precondition \tw{$(k,(v_0,t_0)) \in H$} needed for search recency. We track the current value $t$ of the global clock in an authoritative maxnat camera at ghost location $\gamma_t$. The camera ensures that the clock value can only increase. 

$\mcslinv$ also stores the abstract template-level invariant $\mcsinv(r,t,H)$, and the abstract predicate $\proto(H)$ used for the bookkeeping related to the helping protocol. In addition, $\mcslinv$ states the three invariants, \tw{$\init(H)$, $\hunique(H)$, and $\maxTS(t,H)$ that are needed to prove the atomic triple of $\mcslsearch$}. 

Note that we add $\mcslinv(\mcsinv,\proto)(r)$ to $\mcslstate$ as an Iris invariant (indicated by the box surrounding the predicate).
This provides more flexibility when proving that a template operation satisfies its template-level atomic triple. By storing all resources in an invariant, the resources can be accessed in each atomic step, regardless of whether the operation has already passed its linearization point or not.

As $\mcslinv$ is an invariant, we must ensure that it is preserved in each atomic step. However, $H$ and $t$ change with each \code{upsert}, which means that these values must be existentially quantified by $\mcslinv$. Nevertheless, we need to ensure that the values $t$ and $H$ exposed in the representation predicate $\mcsstate(r,t,H)$ of the template-level atomic triples agree with the values stored in the authoritative resources inside $\mcslinv$. We do this by introducing an additional predicate $\mcsstate^{\authAuth}(r,t,H)$ that we also add to the invariant. We can think of the predicates $\mcsstate^{\authAuth}(r,t,H)$ and $\mcsstate(r, t, H)$ as providing two complementary views at the abstract state of the data structure, one from the perspective of the data structure's implementation, and one from the perspective of the client of the template-level atomic triples. Together, they provide the following important properties:
\begin{mathpar}
  \inferH{view-upd}{\mcsstate^{\authAuth}(r,t, H) * \mcsstate(r,t, H)}{\mcsstate^{\authAuth}(r,t', H') * \mcsstate(r,t', H')}
  
  \inferH{view-sync}{\mcsstate^{\authAuth}(r,t, H) * \mcsstate(r,t', H') \vdash t = t' \wedge H = H'}
  {}
\end{mathpar}
The rule~\refRule{view-upd} says that both predicates are required in order to update the views of the data structure. This occurs in the proof of \code{upsert} when both the invariant as well as the precondition of the template-level atomic triple are accessed at the linearization point. The rule~\refRule{view-sync} ensures that both copies are always in sync. The predicates $\mcsstate^{\authAuth}(r,t, H)$ and $\mcsstate(r,t, H)$ are defined using a combination of authoritative and exclusive RAs. We omit the precise definitions here as they are unimportant for our discussion.

Let us now return to the proof outline in \refFig{fig-multicopy-search'-proof}.
After creating the two prophecies $\tid$ and $p$, the proof uses rule \refRule{au-abort} to obtain access to the precondition \tw{$\mcslstate(\mcsinv,\helpingproto)(r,M)$} and opens it to extract the invariant $\sinv{\mcslinv(\mcsinv,\helpingproto)(r)}$ (line~\ref{line-mcssearch'-proof-case-split-pre1-full}).
As the invariant is persistent, it can be assumed before each of the remaining steps of the proof, though we must also show that it is preserved by each step.
In preparation for the call to \code{search}, the proof proceeds on line~\ref{line-mcssearch'-proof-case-split-pre2-full} by opening the invariant $\mcslinv(\mcsinv,\helpingproto)(r)$ and using rule \refRule{auth-set-snap} to take a snapshot $\ghostState{s}{\authFrag H_0}$ of the authoritative upsert history, whose value at this point we denote by $H_0$.
The proof then lets \tw{$(v_0, t_0) \defeq \maxH[H_0](k)$}, which implies \tw{$(k,(v_0,t_0)) \in H_0$}.
This fact and rule \refRule{auth-set-frag} then yield \tw{$\ghostState{s}{\authFrag \set{(k,(v_0,t_0))}} = \mcscont(k,v_0,t_0)$} (line~\ref{line-mcssearch'-proof-case-split-pre3-full}).

The proof now case splits on whether the value $v_p$ prophesied by $p$ satisfies \tw{$v_0 \neq v_p$ or $v_0 = v_p$} (line~\ref{line-mcssearch'-proof-case-split-full}).
The case $v_0 = v_p$ where the call to \code{search} completes without interference from any upserts on $k$ is straightforward. It implies \tw{$(M(k),\_)=\maxH[H_0](k)=(v_p,\_)$}, which allows us to commit the atomic triple right away, i.e., without requiring any help from an \code{upsert} thread. We therefore  show only the case $v_0 \neq v_p$ in detail, which involves the helping protocol captured by the predicate $\helpingproto$.
  
The predicate $\helpingproto$ keeps track of the \code{search} threads that require helping from $\code{upsert}$ threads.
The IDs of these threads are stored in an authoritative set $R$ at ghost location $\gamma_r$.
In the case \tw{$v_0 \neq v_p$} of the proof, we must therefore register the thread ID $\tid$ with the invariant by replacing $R$ with $R \cup \set{\tid}$ using rule \refRule{auth-set-upd} (line~\ref{line-mcssearch'-proof-call-search-pre3-full}). 
The predicate $\helpingproto(H_0)$ needs to be preserved by this ghost update, because it is part of the invariant $\mcslinv$.
This forces the proof to transfer some of its thread-local resources, captured by the predicate $\reg(\tid, H_0, k, v_p, t_0,\Phi, \token)$, to the invariant. The lines in the proof preceding the update of $R$ establish that this predicate holds.
In particular, the proof transfers the predicate $\helpingstate(\tid, k, v_p, t_0, \Phi, \token)$ to $\helpingproto(H_0)$.
This predicate keeps track of the current state of thread $\tid$ in the helping protocol.
Initially, the thread is in state $\pendingstate(H_0, k, v_p, t_0, \Phi)$.
To establish this predicate, the proof needs to transfer $\atomicUpdate(\Phi)$ to the invariant, as explained earlier. It also needs to ensure \tw{$(\forall t.\, (k,(v_p,t)) \in H_0 \Rightarrow t \leq t_0$, which follows from $\maxH[H_0](k) = (v_0, t_0)$ and $\hunique(H_0)$}. Finally, the proof creates a fresh exclusive token $\token$ that it will later trade in for the receipt $\Phi(v_p)$ of the committed atomic triple.

A minor technicality at this point in the proof is that we must also turn the predicate $\helpingstate$ into an Iris invariant before we add it to \tw{$\reg(\tid, H_0, k, v_p, t_0, \Phi, \token)$}. The reason is that the $\mcslsearch$ proof needs to be able to conclude that the $\Phi$ it will receive back from the invariant later after the call to \code{search} returns, is the same as the one it registers with the helping protocol before calling \code{search}. By turning $\helpingstate$ into an invariant, the predicate becomes a duplicable resource. This allows the proof to keep one copy of the predicate in its thread-local proof context. An unfortunate side effect of this solution is that, as an invariant, the predicate $\helpingstate$ cannot expose the upsert history as a parameter. However, the definition of $\helpingstate$ still depends on the exact value of the upsert history that is stored in the authoritative resource in the invariant. We solve this minor technicality by additionally storing the upsert history at an auxiliary fractional resource at the ghost location $\gamma_{\mathit{sy}(\tid)}$. The invariant maintains one such ghost resource for each thread $\tid$ that is registered for helping. We split this resource half-way between the predicates $\reg$ and $\helpingstate$. This ensures that the upsert history $H$ referenced in $\helpingstate$ is indeed equal to the authoritative one.

Line~\ref{line-mcssearch'-proof-call-search-pre1-full} shows the proof state after \tw{$\pendingstate(H_0, k, v_p, t_0, \Phi)$} has been established and the new resource $\ghostState{{\mathit{sy}(\tid)}}{H_0}$ has been allocated and split into the two halves. These resources are then assembled into \tw{$\helpingstate(\tid,k,v_p,t_0,\Phi,\token)$} and then \tw{$\reg(\tid, H_0, k, v_p, t_0,\Phi, \token)$} as shown on line~\ref{line-mcssearch'-proof-call-search-pre2-full}. Assembling the latter also requires the proof to give up ownership of $\proph(\tid,\_)$. By storing these prophecy resources in $\helpingproto(H_0)$ for all threads in $R$, the proof can use the fact that prophecy resources are exclusive to conclude $\tid \notin R$ before adding $\tid$ to $R$. This is needed to show that the assembled \tw{$\reg(\tid, H_0, k, v_p, t_0, \Phi, \token)$} can indeed be transferred to the invariant during the ghost update.

After the ghost update of $R$, we arrive at line~\ref{line-mcssearch'-proof-call-search-pre4-full}, at which point we are ready to perform the call to \code{search}. After using the invariant and $\mcscont(k,v_0,t_0)$ to obtain the atomic triple for \code{search} from $\mcssearchspec(\code{search},\mcsinv)$, the proof opens the precondition $\mcslstate(r,M)$ of $\mcslsearch$ once again and extracts the precondition $\mcsstate(r,t,H)$ for the template-level atomic triple. Here, $t$ and $H$ refer to the new values of the global clock and upsert history at the linearization point of $\code{search}$. Next, the proof executes the call to \code{search} using the template-level atomic triple, which leaves us with the new proof state on line~\ref{line-mcssearch'-proof-call-search-post-full}. We can now open the invariant again to obtain a new snapshot $\ghostState{s}{H'}$ of the upsert history and use rule \refRule{view-sync} to conclude that the snapshot value $H'$ is the same as the $H$ in $\mcsstate(r,t,H)$. Together with \tw{$(k,(v,t')) \in H$}, we can then establish the persistent proposition \tw{$\ghostState{s}{\set{(k,(v,t'))}}$}. After resolving the prophecy $p$ on line~\ref{line-mcssearch'-proof-resolve-p-post-full}, we additionally establish \tw{$v_p = v$}.

To complete the proof, we now open $\mcslinv(\mcsinv,\helpingproto)$ and use the resource $\ghostState{r}{\authFrag \set{\tid}}$ to conclude that $\tid \in R$. That is, we have \tw{$\reg(\tid, H, k, v_p, t_0, \Phi, \token)$} and can now use \tw{$\ghostState{s}{\set{(k,(v_p,t))}}$} and the fractional resources at ghost location $\gamma_{\mathit{sy}(\tid)}$ to show that the thread must be in state \tw{$\lindone(H,k,v_p,t_0,\Phi, \token)$}. Since the thread still owns the unique token $\token$, it can be exchanged for $\Phi(v_p)$ in the invariant (line~\ref{line-mcssearch'-proof-phi-full}). Using $\Phi(v_p)$, the proof can then complete the initial application of the rule \refRule{logatom-intro} to show the atomic triple of $\mcslsearch$.

\paragraph{Proof of Proposition~(\ref{eq:upsert-spec-revised}).}
By comparison, proving the client-level specification of \code{upsert} from its template-level specification is relatively simple. Recall that the goal here is to prove
\[  \atomicTriple{M.\,\mcslstate(\mcsinv,\helpingproto)(r, M)}
  { \code{upsert}\;r\;k\;v}{\mcslstate(\mcsinv,\helpingproto)(r, M[k \rightarrowtail v])}\]
assuming $\mcsupsertspec(\code{upsert},\mcsinv)$. Let us for a moment assume that we have already established $\mcsprotoupdate(\helpingproto)$. We can then use this property to obtain the template-level atomic triple for \code{upsert} from $\mcsupsertspec(\code{upsert},\mcsinv)$. Before the call to \code{upsert}, we open the precondition \tw{$\mcslstate(\mcsinv,\helpingproto)(r, M)$} to obtain $\mcsstate(r,t,H)$. Moreover, we use the properties \tw{$(\forall k.\; (M(k),\_)\!=\!\maxH(k))$}  and $\maxTS(t,H)$ from the invariant to conclude that \tw{$t_k < t$ for all $t_k,v_k$ such that $(k, (v_k,t_k)) \in H$}. Applying the atomic triple gives the postcondition $\mcsstate(r,t+1,H \cup \set{(k,(v,t))})$. \tw{It then follows that we have for all $k'$:
$\maxH[H \cup \set{(k,(v,t))}](k') = \maxH{}[k \rightarrowtail (v,t)](k') = (M[k \rightarrowtail t](k'), \_)$}.
This allows us to establish the desired postcondition $\mcslstate(r,M[k \rightarrowtail v])$.

Finally, we need to show that $\mcsprotoupdate(\helpingproto)$ is valid at the point where the template-level atomic triple of \code{upsert} is applied. That is, we must show that we can reestablish \tw{$\helpingproto(H \cup \set{(k,(v,t))})$}, assuming $\helpingproto(H)$ is true. In this proof, we make use of the fact that $\maxTS(t,H)$ holds, which we obtain from the invariant.
The important point to note here is that any $\mcslsearch$ thread $\tid \in R$ that was in state \tw{$\pendingstate(H, k, v, t_0, \Phi)$} where $t \ge t_0$ and thus waiting to be helped by the \code{upsert} thread performing the ghost update of $H$, cannot remain in this state because \tw{$\pendingstate(H \cup \set{(k,(v,t))}, k, v, t_0, \Phi)$} is unsatisfiable. The ghost update is, thus, forced to commit the atomic triples of all these $\mcslsearch$ threads using their update tokens via rule~\refRule{au-commit}. It can do this because the postcondition of these triples are satisfied after $H$ has been updated. The reasoning here follows similar steps as the proof of the postcondition of \code{upsert} above. After committing the atomic triple, the receipt $\Phi(v)$ is transferred to \tw{$\helpingproto(H \cup \set{(k,(v,t))})$} as part of the new state \tw{$\lindone(H \cup \set{(k,(v,t))},k,v,t_0,\Phi, \token)$} for each of these threads.

\subsection{Invariant for LSM DAG Template}
\label{sec-iris-invariant}


We next discuss the details of the template-specific invariant $\lsminv(r,t,H)$ that we use for the proof of the LSM DAG template. It is shown in \refFig{fig-multicopy-invariant}.
For presentation purposes, we omit the additional flow interface for encoding the inset flow and the related predicates needed for the proof of the maintenance operation.
The full details of the Iris development can be found on GitHub.\footnote{\url{https://github.com/nyu-acsys/template-proofs/multicopy}}

\begin{figure}[t!]
  \centering
  \tw{
  \[\begin{array}{ll}
    \multicolumn{2}{l}{\lsminv(r,t,H) \defeq {} \exists\; \interface.}
    \\
    & \hspace{4mm} \globalInv(r,t,H,\interface)\\
    &{} * \Sep_{n \in \dom(\interface)}        
          \arraycolsep=2pt
          \begin{array}[t]{rl}
            \exists b_n\, C_n\, Q_n. &
            \lockR(b_n,n, \nodePred(r, n, C_n, Q_n))\\
            {} * {} & \nodeShar(r, n, C_n, Q_n, H)
          \end{array}
    \end{array}
  \]
  \[\arraycolsep=0pt\begin{array}{rl}
  \text{where} \quad
    \globalInv(r,t,H,\interface) \defeq {} &
        \ghostState{I}{\authAuth\; \interface}
      * \ghostState{f}{\authAuth\; \dom(\interface)}
      * \inflowOfInt{\interface} = \lambda_0
      * \inFootprint(r)
    \\[.5em]
    \inFootprint(n) \defeq {}
    & \ghostState{f}{\authFrag \set{n}} 
    \\[.5em]
    \nodePred(r, n, C_n, Q_n) \defeq {}
    &  \exists \mathit{es}. \,
        \heapRep(r, n, \mathit{es}, C_n) \\
      \multicolumn{2}{l}{\quad\qquad{\arraycolsep=0pt
      \begin{array}{l}
      \quad {}
       * \ghostState{e(n)}{\fracHalf \mathit{es}}
      * \ghostState{c(n)}{\fracHalf C_n}
      * \ghostState{q(n)}{\fracHalf Q_n}
      * \ghostState{s}{\authFrag\; C_n}
    \end{array}}}
    \\[1.5em]
    \nodeShar(r, n, C_n, Q_n, H) \defeq {}
    & \exists \mathit{es}, \interface_n. \\
      \multicolumn{2}{l}{\quad\qquad{\arraycolsep=0pt
       \begin{array}[t]{l} \quad {}  
      * \ghostState{e(n)}{\fracHalf \mathit{es}}
      * \ghostState{c(n)}{\fracHalf C_n}
      * \ghostState{q(n)}{\fracHalf Q_n}\\
     \quad {} 
      * \ghostState{\interface}{\authFrag \interface_n}
      * \dom(\interface_n) = \set{n}
      * \inFootprint(n)
      * \closed(\mathit{es})\\
     \quad {}
      * \outflowOfInt{\interface_n} = \lambda n'. \,
      \chi(\setcomp{(k,Q_n(k))}{k \in \mathit{es}(n') \land k \in \dom(Q_n)})\\
     \quad {}
      * \ite{n = r}{\Cir{n} = \maxH * \inflowOfInt{\interface_n} = \lambda_0}{\TRUE}\\
     \quad {}
      * \Sep_{k \in \KS} \ghostState{\mathit{cir}(n)(k)}{\authAuth\; \ts(\Cir{n}(k))}\\
     \quad {}
      *  \phi_1(n, Q_n, \mathit{es}) * \phi_2(n, C_n, Q_n, \interface_n)
    \end{array}}}\\
    \\
    \closed(\mathit{es}) \defeq {}
    & \forall n'.\; \mathit{es}(n') \neq \emptyset \Rightarrow \inFootprint(n')
    \\[.5em]
    \Cir{n} \defeq {} & \lambda k.\, \ite{k \in \dom(C_n)}{C_n(k)}{Q_n(k)}\\[.5em]
    \phi_1(n,Q_n,\mathit{es}) \defeq {} & \forall k.\; Q_n(k) = \bot \lor (\exists n'.\,k \in \mathit{es}(n'))\\[.5em]
    \phi_2(n,C_n,Q_n,\interface_n) \defeq {} & \forall k\,p.\; \inflowOfInt{\interface_n}(n)(k,p) > 0 \Rightarrow B_n(k)=p
    \end{array}\]}
  \caption{The invariant for the multicopy structure template.}
  \label{fig-multicopy-invariant}
\end{figure}

The parameters $t$ and $H$ are to be interpreted as the current clock value and the current upsert history. The existentially quantified variable $\interface$ is the current global flow interface used for encoding the contents-in-reach flow. As discussed in \refSec{sec-mcs-invariant}, the invariant consists of two parts: a \emph{global} part described by the predicate $\globalInv$ and a \emph{local} part that holds for every node in the data structure. The set of nodes $N$ of the data structure graph is implicitly captured by the domain $\dom(\interface)$ of the flow interface $\interface$.

We discuss each part of \refFig{fig-multicopy-invariant} in detail, starting with the predicate $\globalInv(r,t,H,\interface)$:
\begin{itemize}
\item  We use an authoritative camera of flow interfaces at ghost location $\gamma_I$ to keep track of the global interface $\interface$, which is composed of singleton interfaces $\interface_n$ for each node $n \in \dom(\interface)$. The $\interface_n$ are tied to the implementation-specific physical representation of the individual nodes via the predicates $\nodePred$ and $\nodeShar$ as explained below.
  
\item The ghost resource $\ghostState{f}{\authAuth \dom(\interface)}$ keeps track of the footprint of the data structure using the authoritative camera of sets of nodes. We use this resource to maintain the invariant of \code{traverse} that the currently visited node $n$ remains part of the data structure while $n$ is unlocked.

\item We require that the global flow interface $\interface$ has no inflow ($\inflowOfInt{\interface}=\lambda_0$), as required for our encoding of contents-in-reach. That is, $\lambda_0$ maps all nodes to the empty multiset, the unit $0$ of the flow domain.


\item The condition $\inFootprint(r)$ guarantees that $r$ is always in the domain of the data structure. 


\end{itemize}
  
The resources for a node $n$ are split between the predicates $\nodePred(r, n, C_n, Q_n)$ and $\nodeShar(r, n, C_n, Q_n, H)$. The latter is always owned by the invariant whereas the former is protected by $n$'s lock and transferred between the invariant and the thread's local state upon locking the node and vice versa upon unlocking, as usual. We next discuss $\nodePred(r, n, C_n, Q_n)$:
\begin{itemize}
\item The first conjunct is the implementation-specific node predicate \tw{$\heapRep(r, n, \mathit{es}, \valof(C_n))$}. For each specific implementation of the template, this predicate must tie the physical representation of the node $n$ to its physical contents \tw{$\valof(C_n):\KS \pto \VS$} and a function $es:\nodeDom \to \pset{\KS}$, which captures the edgesets of $n$'s outgoing edges. 

\item The fractional resources at ghost locations $\gamma_{e(n)}$, $\gamma_{c(n)}$, and $\gamma_{q(n)}$ ensure that the predicates $\nodePred$ and $\nodeShar$ agree on the values of $\mathit{es}$, $C_n$, and $Q_n$ even when $n$ is locked.

\item The ghost resource $\ghostState{s}{\authFrag\; C_n}$ when combined with $\ghostState{s}{\authAuth\; H}$ implies $C_n \subseteq H$, which captures Invariant~\ref{inv-mc3}.

\item The final conjunct of $\nodePred$ guarantees sole ownership of the global clock by a thread holding the lock on the root node.
\end{itemize}    
  

Moving on to $\nodeShar(r, n, C_n, Q_n, H)$, this predicate contains those resources of $n$ that are available to all threads at all times:
\begin{itemize}
  \item The resource $\ghostState{\interface}{\authFrag \interface_n}$ guarantees that all the singleton interfaces $\interface_n$ compose to the global interface $\interface$, thus, satisfying the flow equation. Similarly, the predicate $\inFootprint(n)$ guarantees that $n$ remains in the data structure at all times. The predicate $\closed(\mathit{es})$ ensures that the outgoing edges of $n$ point to nodes which are again in the data structure. Together with the condition $\inFootprint(r)$, this guarantees that all nodes reachable from $r$ must be in $\dom(\interface)$.

  \item The next conjunct of $\nodeShar$ defines the outflow of the singleton interface $\interface_n$ according to Equation~(\ref{eq-cir-edgeFn}) of our flow encoding of contents-in-reach. Note that, even though $\interface_n$ is not shared with the predicate $\nodePred$, only its inflow can change when $n$ is locked, because the outflow of the interface is determined by $Q_{n}$ and $\mathit{es}$, both of which are protected by $n$'s lock.
    
  \item The constraint $\Cir{n} = \maxH$ holds if $r=n$ and implies Invariant~\ref{inv-mc1}. We further require here that the interface of the root node has no inflow ($\inflowOfInt{\interface_n}=\lambda_0$), a property that we need in order to prove that \code{upsert} maintains the flow-related invariants. Moreover, we use for every key $k$ an authoritative maxnat resource at ghost location $\gamma_{\mathit{cir}(n)(k)}$ to capture Invariant~\ref{inv-mc2}.

  \item The last two conjuncts of $\nodeShar$ complete our encoding of contents-in-reach and ensure that we must indeed have $\Cir{n} = \cir(n)$ at all atomic steps.
    

\end{itemize}

Finally, we note that Invariant~\ref{inv-mc4}, which is needed for the proof of \code{upsert}, is already captured by the predicate $\maxTS(t,H)$ included in the full invariant $\mcslinv(\lsminv,\proto)(r)$.

\subsection{Detailed Proof of Template Operations}

We now have all the ingredients to proceed with the proof of the template operations.

\subsubsection{Proof of \code{search}.}
The proof relies on the specification of the implementation-specific helper functions provided in \refFig{fig-multicopy-helper-specs}.

\begin{figure}
  \centering
  \begin{lstlisting}[aboveskip=1pt,belowskip=0pt]
$\annot{\sinv{\mcslinv(\lsminv,\proto)(r)} * \mcscont(k,v_0,t_0)} * \annotAtom{t\; H.\; \mcsstate(r,t, H)}$@\label{multicopy-search-proof-search-start}@
let search $r$ $k$ =
  $\annot{\inFootprint(r) * \maxH_1(k) = \Cir{r}(k) * \ghostState{\ts(\mathit{cir}(r)(k))}{\authFrag\; \Cir{r}(k)} * t_0 \leq \ts(\maxH_1(k))}$@\label{multicopy-search-proof-search-traverse-pre1}@
  $\annot{\inFootprint(r) * \ghostState{\mathit{cir}(r)(k)}{\authFrag\; t_1} * t_0 \leq t_1} * \annotAtom{t\; H.\; \mcsstate(r,t, H)}$ @\label{multicopy-search-proof-search-traverse-pre2}@
  traverse $r$ $r$ $k$ @\label{multicopy-search-proof-search-traverse}@
$\annotAtom{\Ret v. \exists t'.\; \mcsstate(r, t, H)  * t_0 \le t' * (k,(v,t')) \in H}$

$\annot{\sinv{\mcslinv(\lsminv,\proto)(r)} * \inFootprint(n) * \ghostState{\mathit{cir}(n)(k)}{\authFrag\; t_1} * t_0 \leq t_1} * \annotAtom{t\; H.\; \mcsstate(r,t, H)}$@\label{multicopy-search-proof-traverse-start}@
let rec traverse $r$ $n$ $k$ =
  lockNode $n$;
  $\annot{\nodePred(r, n, C_n, Q_n) * \ghostState{\ts(\mathit{cir}(n)(k))}{\authFrag\; \Cir{n}(k)} * t_0 \leq \Cir{n}(k)}$@\label{multicopy-search-proof-traverse-lockNode}@
  match |<inContents>| $r$ $n$ $k$ with
  | Some $v$ ->
    $\annot{\nodePred(r, n, C_n, Q_n) * t_0 \leq \ts(\Cir{n}(k)) * k \in \dom(\valof(C_n)) * v = \valof(C_n)(k)}$ @\label{multicopy-search-proof-traverse-case-1}@
    $\annot{\nodePred(r, n, C_n, Q_n) * t_0 \leq \ts(\Cir{n}(k)) * (v,t') = C_n(k)}$ @\label{multicopy-search-proof-traverse-case-1}@
    (* Linearization point *)
    $\annot{\nodePred(r, n, C_n, Q_n) * t_0 \leq \ts(C_n(k)) * (v,t') = C_n(k)}$ @\label{multicopy-search-proof-traverse-case-1-commit-pre1}@
    $\annot{\nodePred(r, n, C_n, Q_n) * (k, (v,t')) \in H * t_0 \le t' * \mcsstate(r, t, H)}$ @\label{multicopy-search-proof-traverse-case-1-commit-pre2}@
    unlockNode $n$; $v$ @\label{multicopy-search-proof-traverse-case-1-commit}@
    $\annotAtom{\Ret v. \exists t'.\;\mcsstate(r,t, H) * (k,(v,t')) \in H * t_0 \le t'}$
  | None ->
    $\annot{\nodePred(r, n, C_n, Q_n) * t_0 \leq \ts(\Cir{n}(k)) * k \notin \dom(\valof(C_n))}$ 
    $\annot{\nodePred(r, n, C_n, Q_n) * t_0 \leq \ts(\Cir{n}(k)) * k \notin \dom(C_n)}$ @\label{multicopy-search-proof-traverse-case-2}@
    match |<findNext>| $r$ $n$ $k$ with
    | Some $n'$ ->
      $\annot{\heapRep(r, n, \mathit{es}, C_n) * \dots * t_0 \leq \ts(\Cir{n}(k)) * k \notin \dom(C_n) * k \in \mathit{es(n')} }$ @\label{multicopy-search-proof-traverse-case-2-a}@
      $\annot{\nodePred(r, n, C_n, Q_n) * \inFootprint(n') * \ghostState{\mathit{cir}(n')(k)}{\authFrag\; t_1} * t_0 \leq t_1}$@\label{multicopy-search-proof-traverse-case-2-a-unlock-pre}@ 
      unlockNode $n$;
      $\annot{\inFootprint(n') * \ghostState{\mathit{cir}(n')(k)}{\authFrag\; t_1} * t_0 \leq t_1} * \annotAtom{t\; H.\; \mcsstate(r,t, H)}$@\label{multicopy-search-proof-traverse-case-2-a-recurse-pre}@  
      traverse $r$ $n'$ $k$ @\label{multicopy-search-proof-traverse-case-2-a-recurse}@  
      $\annotAtom{\Ret v. \exists t'.\;\mcsstate(r,t, H) * (k,(v,t')) \in H * t_0 \le t'}$
    | None ->
      $\annot{\heapRep(r, n, \mathit{es}, C_n) * \dots * t_0 \leq \ts(\Cir{n}(k)) * k \notin \dom(\valof(C_n)) * \forall n'.\; k \notin \mathit{es}(n')}$@\label{multicopy-search-proof-traverse-case-2-b}@
      $\annot{\heapRep(r, n, \mathit{es}, C_n) * \dots * t_0 \leq \ts(\Cir{n}(k)) * k \notin \dom(\Cir{n}(k))}$
      $\annot{\nodePred(r, n, C_n, Q_n) * t_0 \leq \ts(\Cir{n}(k)) * \ts(\Cir{n}(k)) = 0}$@\label{multicopy-search-proof-traverse-case-2-b-commit-pre1}@
      (* Linearization point *)
      $\annot{\nodePred(r, n, C_n, Q_n) * t_0 = \ts(\Cir{n}(k)) = 0 * \mcsstate(r, t, H) }$ 
      $\annot{\nodePred(r, n, C_n, Q_n) * (k, (\square,0)) \in H * t_0 = 0 * \mcsstate(r, t, H) }$
      unlockNode $n$; $\square$
$\annotAtom{\Ret v. \exists t'.\;\mcsstate(r,t, H) * (k,(v,t')) \in H * t_0 \le t'}$
\end{lstlisting}

\caption{Proof of \code{search}.}
\label{fig-multicopy-search-proof}
\end{figure}

\refFig{fig-multicopy-search-proof} provides the outline of the proof of \code{search}. The intermediate assertions shown throughout the proof represent the relevant information from the proof context at the corresponding point. By convention, all the newly introduced variables are existentially quantified. Note that the condition \tw{$\mcscont(k,v_0,t_0)$} is persistent and, hence, holds throughout the proof. Moreover, the invariant $\sinv{\mcslinv(\lsminv,\proto)(r)}$ is maintained throughout the proof since \code{search} does not modify any shared resources. We therefore do not include these resources explicitly in the intermediate assertions.

As most of the actual work is done by the recursive function \code{traverse}, we start with its atomic specification:
\begin{align*}
  &\mkblue{\sinv{\mcslinv(\lsminv,\proto)(r)} \magicwand \inFootprint(n) \magicwand \ghostState{\mathit{cir}(n)(k)}{\authFrag\; t_1} \magicwand t_0 \leq t_1 \magicwand} \\
  &\qquad \atomicTriple{t\; H.\; \mcsstate(r, t, H)}{\code{traverse}\;r\;n\;k}
{\Ret v. \exists t'.\;\mcsstate(r, t, H) \,*\, (k,(v,t')) \in H \,*\, t_0\!\le\!t'}
\end{align*} 
Recall the traversal invariant \tw{$t_0 \le \ts(\cir(n)(k))$} that we used in our informal proof of search recency.
The ghost resource $\ghostState{\mathit{cir}(n)(k)}{\authFrag\; t_1}$, together with $t_0 \le t_1$ in the precondition of the above specification precisely capture this invariant.
In addition, \code{traverse} assumes the invariant $\mcslinv(\lsminv,\proto)(r)$ and requires that $n$ must be a node in the graph, expressed by the predicate $\inFootprint(n)$.
The operation then guarantees to return \tw{a value $v$} such that search recency holds.

Let us for now assume that \code{traverse} satisfies the above specification and focus on the proof of \code{search}.
The precondition of \code{search $r$ $k$} assumes the invariant $\mcslinv(\lsminv,\proto)(r)$ and the predicate \tw{$\mcscont(k, v_0, t_0)$} (line~\ref{multicopy-search-proof-search-start}).
We must, hence, use these to establish the precondition for the call \code{traverse $r$ $n$ $k$} on line~\ref{multicopy-search-proof-search-traverse}.
To this end, we open the invariant from which we can directly obtain $\inFootprint(r)$.
Next, we unfold the definition of \tw{$\mcscont(k,v_0,t_0)$} to obtain $\ghostState{s}{\authFrag\; \set{(k,(v_0,t_0)}}$. Snapshotting $\ghostState{s}{\authAuth\; H_1}$ in the invariant for the current upsert history $H_1$, we can conclude $(k,(v_0,t_0) \in H_1$ and therefore $t_0 \leq \ts(\maxH[H_1](k))$.
From $\nodeShar(r,r,C_r,Q_r,H_1)$ in the invariant we can further deduce $\maxH[H_1](k) = \Cir{r}(k)$ and \tw{$\ghostState{\mathit{cir}(r)(k)}{\authFrag\; \ts(\Cir{r}(k))}$}  (line~\ref{multicopy-search-proof-search-traverse-pre1}).
By substituting both \tw{$\ts(\maxH[H_1](k))$} and \tw{$\ts(\Cir{r}(k))$} with a fresh existentially quantified variable $t_1$ we obtain the precondition of \code{traverse} (line~\ref{multicopy-search-proof-search-traverse-pre2}).
We now commit the atomic triple of \code{search} on the call to \code{traverse} and immediately obtain the desired postcondition.

\paragraph{Proof of \code{traverse}.}
Finally, we prove the assumed specification of \code{traverse}. The proof starts at line~\ref{multicopy-search-proof-traverse-start}.
The thread first locks node $n$ which yields ownership of the predicate $\nodePred(r, n, C_n, Q_n)$.
At this point, we also open the invariant to take a fresh snapshot of the resource $\ghostState{\mathit{cir}(n)(k)}{\authAuth\; \ts(\Cir{n}(k))}$ to conclude $t_0 \leq \ts(\Cir{n}(k))$ from the precondition of \code{traverse} (line~\ref{multicopy-search-proof-traverse-lockNode}).
Next, the thread executes $\helperFn{inContents}\,r\,n\,k$. The precondition of this call, i.e. $\heapRep(r, n, \mathit{es}, \valof(C_n))$, is available to us as part of the predicate $\nodePred(r, n, C_n, Q_n)$.
Depending on the return value $v$ of \helperFn{inContents} we end up with two subcases.

In the case where \tw{$x = \some(v)$, we know $k \in \dom(\valof(C_n))$ and $v=\valof(C_n)(k)$ which lets us conclude $(v,t') = C_n(k)$ for some $t'$} (line~\ref{multicopy-search-proof-traverse-case-1}). The call to \code{unlockNode} on line~\ref{multicopy-search-proof-traverse-case-1-commit} will be the linearization point of this case.
To obtain the desired postcondition of the atomic triple, we first retrieve $\nodeShar(r, n, C_n, \Cir{n}, H)$ from the invariant and open its definition.
From $k \in \dom(C_n)$ and the definition of $\Cir{n}$ we first obtain
$\Cir{n}(k)=C_n(k)$. This leaves us with the proof context on line~\ref{multicopy-search-proof-traverse-case-1-commit-pre1}.
Now we access the precondition $\mcsstate(r,t, H)$ of the atomic triple and sync it with the view of $H$ and $t$ in the invariant.
Moreover, we use the resource $\ghostState{s}{\authFrag\; C_n}$ to infer $C_n \subseteq H$, which then implies \tw{$(k,(v,t')) \in H$} (line~\ref{multicopy-search-proof-traverse-case-1-commit-pre2}).
The call to \code{unlockNode} returns $\nodePred(r, n, C_n, Q_{n})$ to the invariant and commits the atomic triple, which concludes this case.

For the second case where the return value of \helperFn{inContents} is \tw{$x=\none$, we have $k \notin \dom(C_n)$} and the thread calls \helperFn{findNext}.
Here, we unfold $\nodePred(r, n, C_n, Q_{n})$ to retrieve $\heapRep(r,n,\mathit{es},\valof(C_n))$, which is needed to satisfy the precondition of \helperFn{findNext}.
We then end up again with two subcases: one where there is a successor node $n'$ such that $k \in \mathit{es}(n')$, and the other where no such node exists.
Let us consider the first subcase, which is captured by the proof context on line~\ref{multicopy-search-proof-traverse-case-2-a}.
Now, before the thread unlocks $n$, we have to reestablish the precondition of \code{traverse} for the recursive call on line~\ref{multicopy-search-proof-traverse-case-2-a-recurse}.
To do this, we first open the invariant and retrieve $\nodeShar(r, n, C_n, Q_n, H)$. From predicate $\closed(\mathit{es})$, we obtain the resource $\inFootprint(n')$ because $\mathit{es}(n') \neq \emptyset$. $\inFootprint(n')$ is the first piece for the precondition of \code{traverse}. To obtain the remaining pieces, we must retrieve $\nodeShar(r, n', C_{n'}, Q_n', H)$ from the invariant. This is possible because we can infer $n' \in \dom(\interface)$ using $\inFootprint(n')$.

We first observe that $k \notin \dom(C_n)$ implies $\Cir{n}(k)=Q_n(k)$ and which together with $k \in \mathit{es}(n')$ gives us
\[\outflowOfInt{\interface_n}(n')(k, \Cir{n}(k)) > 0.\]
From the fact that $\interface_n$ and $\interface_{n'}$ compose, we can conclude $\inflowOfInt{\interface_{n'}}(n')(k, \Cir{n}(k)) > 0$.
It then follows from the constraint $\phi_2(n',C_n,Q_n,\interface_{n'})$ that $\Cir{n'}(k)=\Cir{n}(k)$.
We can further take a fresh snapshot of the resource \tw{$\ghostState{\mathit{cir}(n')(k)}{\authAuth\; \ts(\Cir{n'}(k))}$} to obtain our final missing piece \tw{$\ghostState{\mathit{cir}(n')(k)}{\authFrag\; \ts(\Cir{n'}(k))}$} for the precondition of \code{traverse}.
Then substituting both $\Cir{n}(k)$ and $\Cir{n'}(k)$ by a fresh variable $t_1$ and folding predicate $\nodePred(r, n, C_n, Q_{n})$ we arrive at line~\ref{multicopy-search-proof-traverse-case-2-a-unlock-pre}.
Unlocking $n$ transfers ownership of $\nodePred(r, n, C_n, Q_{n})$ back to the invariant.
The resulting proof context satisfies the precondition of the recursive call to \code{traverse}, which we use to commit the atomic triple by applying the specification of \code{traverse} inductively.

We are left with the last subcase where $n$ has no successor for $k$ (line~\ref{multicopy-search-proof-traverse-case-2-b}).
Here, we proceed similarly to the first case above:
we conclude $\Cir{n}(k)=Q_n(k)$ from $k \notin \dom(C_n)$ and use $\forall n'.\; k \notin \mathit{es}(n')$ and $\phi_1(n,Q_n,\mathit{es})$ to conclude that $Q_{n}(k) = \Cir{n}(k)=C_n(k)=\bot$. This, in turn, implies $\ts(\Cir{n}(k)) = 0$.
After folding predicate $\nodePred(r, n, C_n, Q_n)$ we arrive on line~\ref{multicopy-search-proof-traverse-case-2-b-commit-pre1}.
The linearization point in this case is at the point when $n$ is unlocked, so we access the precondition $\mcsstate(r,t, H)$ of the atomic triple and sync it with the view of $H$ and $t$ in the invariant.
\tw{Using the property $\init(H)$ from the invariant, we can conclude $(k, (\square, 0)) \in H$.}
The call to \code{unlockNode} returns $\nodePred(r, n, C_n, Q_{n})$ to the invariant and commits the atomic triple.
\tw{The postcondition then follows for $t'=0$ and $v=\square$, which is the return value in this case.}

This completes the proof of \code{search}.

\subsubsection{Proof of \code{upsert}.}

We now focus on proving the correctness of the \code{upsert} operation.
\refFig{fig-two-node-upsert-proof} shows the proof outline.
The proof relies on the specification of the implementation-specific helper function \helperFn{addContents} provided in \refFig{fig-multicopy-helper-specs}.
The specification simply says that when the function succeeds, the copy of $k$ is updated to $v$ in $V_r=\valof(C_r)$. In case it fails, then no changes are made to $V_r$.


\begin{figure}[t!]
  \centering
  \begin{lstlisting}[aboveskip=1pt,belowskip=0pt]
$\annot{\heapRep(r,r,\mathit{es}, C_r)}$
  |<addContents>| $r$ $k$ $t$
$\annot{\Ret\val. \heapRep(r,r,\mathit{es}, C_r') * C_r' = \ite{v}{C_r[k \rightarrowtail t]}{C_r}}$ @\label{tn-upsert-proof-addContents-spec}@
$\annot{\ghostState{t}{q\; \authAuth\; t} * q > 0}$ readClock () $\annot{\Ret\val. \ghostState{t}{q\; \authAuth\ t} * \val = t }$ @\label{tn-upsert-proof-readClock-spec}@
$\annot{\ghostState{t}{\authAuth\ t}}$ incrementClock () $\annot{\ghostState{t}{\authAuth\ t + 1}}$ @\label{tn-upsert-proof-incrementClock-spec}@

$\annot{\sinv{\mcslinv(\lsminv,\proto)(r)} * (\forall \; H\, k\, t\, v.\; \proto(H) \vsR \proto(H \cup \set{(k,(v,t))}))} * \annotAtom{t\, H.\; \mcsstate(t, H)}$
let upsert $r$ $k$ $v$ =
  lockNode $r$;
  $\annot{\nodePred(r,r, C_r,Q_r)}$ @\label{tn-upsert-proof-lock-r-1}@
  $\annot{\heapRep(r,r,\mathit{es}, \valof(C_r)) * \dots * \ghostState{s}{\authFrag\; C_r} * \ghostState{c(r)}{\fracHalf C_r}}$ @\label{tn-upsert-proof-lock-r-2}@
  let $\result$ = |<addContents>| $r$ $k$ $v$ in @\label{tn-upsert-proof-add-content}@
  if $\result$ then
    $\annot{\heapRep(r,r,\mathit{es}, V_r') * \dots * \ghostState{s}{\authFrag\; C_r} * \ghostState{c(r)}{\fracHalf C_r} \,*\, V_r' = \valof(C_r)[k \rightarrowtail v]}$ @\label{tn-upsert-proof-add-content-succ}@
    (* Linearization point *)
    $\annot{\heapRep(r,r,\mathit{es}, V_r') * \dots * \ghostState{s}{\authFrag\; C_r} * \ghostState{c(r)}{\fracHalf C_r} * \ghostState{s}{\authAuth\; H} * \ghostState{t}{\authAuth t} * C_r' = C_r[k \rightarrowtail (v,t)] * V_r' = \valof(C_r')}$ @\label{tn-upsert-proof-commit-pre-1}@
    $\annot{\heapRep(r,r,\mathit{es}, \valof(C_r')) * \dots * \ghostState{s}{\authFrag\; C_r} * \ghostState{c(r)}{\fracHalf C_r} * \ghostState{s}{\authAuth\; H} * \ghostState{t}{\authAuth t} \\ {} * C_r' = C_r[k \rightarrowtail (v,t)] * H' = H \cup \{(k, (v,t))\} * \mcsstate(r,t,H)}$ @\label{tn-upsert-proof-commit-pre-2}@
    (* Ghost updates: $\ghostState{s}{\authAuth\; H} \vsR \ghostState{s}{\authAuth\; H'}$, $\ghostState{c(r)}{1 C_r} \vsR \ghostState{c(r)}{1 C_r'}$,
                        $\ghostState{t}{\authAuth t} \vsR \ghostState{t}{\authAuth t+1}$, $\ghostState{\mathit{cir}(n)(k)}{\authAuth\; \ts(\Cir{n}(k))} \vsR \ghostState{\mathit{cir}(n)(k)}{\authAuth\; t}$ *)
    $\annot{\heapRep(r,r,\mathit{es}, \valof(C_r')) * \dots  * \ghostState{s}{\authFrag\; C_r} * \ghostState{c(r)}{\fracHalf C_r'} * \ghostState{s}{\authAuth\; H'} * \ghostState{t}{\authAuth t+1} \\ {} * C'_r = C_r[k \rightarrowtail t] * H' = H \cup \{(k, (v,t))\} * \mcsstate(r,t, H) }$ @\label{tn-upsert-proof-commit-post-1}@
    $\annot{\nodePred(r,r, C_r',Q_r) * \ghostState{s}{\authAuth\; H'} * \ghostState{t}{\authAuth t} * C'_r = C_r[k \rightarrowtail t] * H' = H \cup \{(k, (v,t))\} * \mcsstate(r,t, H) }$ 
    unlockNode $r$
    $\annot{\ghostState{s}{\authAuth\; H'} * \ghostState{t}{\authAuth t+1} * C'_r = C_r[k \rightarrowtail t] * H' = H \cup \{(k, (v,t))\} * \mcsstate(r,t, H) }$ @\label{tn-upsert-proof-commit-post-2}@
    $\annotAtom{ \mcsstate(r, t+1, H \cup \{(k, (v,t))\}) }$ @\label{tn-upsert-proof-unlock-r}@
  else begin
    $\annot{\heapRep(r,r,\mathit{es}, V_r') * \dots * \ghostState{s}{\authFrag\; C_r} * \ghostState{c(r)}{\fracHalf C_r} \,*\, V_r' = \valof(C_r)}$ @\label{tn-upsert-proof-add-content-fail}@
    $\annot{\nodePred(r,r, C_r,Q_r)}$
    unlockNode $r$;
    upsert $r$ $k$
  end
$\annotAtom{ \mcsstate(t+1, H \cup \{(k, t)\}) }$
\end{lstlisting}
\caption{Proof of \code{upsert}.}
\label{fig-two-node-upsert-proof}
\end{figure}

With everything needed for the proof of \code{upsert} in place, let us now walk through the proof outline shown in \refFig{fig-two-node-upsert-proof}. We start with the invariant $\sinv{\mcslinv(\lsminv,\proto)(r)}$, the view shift assumption on $\proto$, and the atomic precondition $\annotAtom{t\, H.\; \mcsstate(r,t, H)}$. The invariant can be accessed at each atomic step, but must also be reestablished after each step.
Similarly, the atomic precondition is accessible at each atomic step, and must either be used to generate the postcondition of the atomic triple or the precondition must be reestablished. 

The thread first locks the root node, which transfers ownership of $\nodePred(r,r,\mathit{es},C_r)$ from the invariant to the thread (line~\ref{tn-upsert-proof-lock-r-1}). At this point, we unfold the definition of $\nodePred(r,r,\mathit{es},C_r)$ (line~\ref{tn-upsert-proof-lock-r-2}) as we will need the contained resources later in the proof.
The thread now calls \helperFn{addContents} to update $r$ with the new \tw{value $v$} for $k$. This leaves us with two possible scenarios depending on whether the return value $\result$ is $\true$ (line~\ref{tn-upsert-proof-add-content-succ}) or $\false$ (line~\ref{tn-upsert-proof-add-content-fail}).

In the case where \code{addContents} fails ($\result = \false$), no changes have been performed. Here, we simply fold $\nodePred(r,r,\mathit{es},C_r)$ again, unlock $r$ to transfer ownership of the node's resources back to the invariant, and commit the atomic triple on the recursive call to \code{upsert}.

In the case where \code{addContents} succeeds ($\result = \true$), we obtained $V_r' = \valof(C_r)[k \rightarrowtail v]$ from its postcondition, where $V_r'$ is the new physical contents of the root node. The thread will next call \helperFn{unlockNode} to unlock the root node $r$. This will be the linearization point of this branch of the conditional expression. Hence, we will also have to update all ghost resources to their new values at this point. \tw{To prepare committing the atomic triple, we first open the invariant $\sinv{\mcslinv(\lsminv,\proto)(r)}$ to retrieve $\ghostState{s}{\authAuth\; H}$ and $\ghostState{t}{\authAuth t}$. Then we define $C_r' = C_r[k \rightarrowtail (v,t)]$, which together with $V_r' = \valof(C_r)[k \rightarrowtail v]$ gives us $V_r' = \valof(C_r')$. We further define $H' = H \cup (k, (v,t))$. This leaves us in the proof state on line~\ref{tn-upsert-proof-commit-pre-1}. Next, we access the precondition of the atomic triple to obtain $\mcsstate(r,t_1,H_1)$ for some fresh variables $t_1$ and $H_1$.  We then obtain $\mcsstate^{\authAuth}(r,t, H)$ from the invariant and use rule~\refRule{view-sync} to conclude $H_1=H$ and $t_1=t$ (line~\ref{tn-upsert-proof-commit-pre-2}).}

\tw{
The actual commit of the atomic triple involves several steps. First, we update all relevant ghost resources where we use, in particular, the following rule for frame-preserving updates on fractional resources:
\begin{mathpar}
  \inferH{frac-upd}
  {}{(1, s) \mupd (1, s')}
\end{mathpar}

\begin{itemize}
\item We update the authoritative version of the upsert history at ghost location $\gamma_s$ in the invariant from $H$ to $H'$, using rule~\refRule{auth-set-upd}.
\item We use the pendent of the rule~\refRule{auth-set-upd} for the authoritative maxnat camera to update the global clock at ghost location $\gamma_t$ in the invariant from $t$ to $t+1$.
\item We use rule~\refRule{frac-upd} to update the resource holding the root's contents at location $\gamma_{c(r)}$ from $C_r$ to $C_r'$ by reassembling the full resource from the half owned by the invariant, respectively, the half owned by the local proof context. After the update, the resource is split again into two halves, with one half returned to the invariant. 
\item We use the pendent of the rule~\refRule{auth-set-upd} for the authoritative maxnat camera to update the resource holdings the roots contents-in-reach for $k$ at ghost location $\gamma_{\mathit{cir}(r)(k)}$ from $\ts(\Cir{r}(k))$ to $t$. This is possible because $\Cir{r} = \maxH$ and $\maxTS(t, H)$ hold according to the invariant, which together imply $\ts(\Cir{r}(k)) < t$.
\end{itemize}
This leaves us with the new proof context shown on line~\ref{tn-upsert-proof-commit-post-1}. We then reassemble the predicate $\heapRep(r,r,\mathit{es}, V_r')$ from the proof context at line~\ref{tn-upsert-proof-commit-post-1} to satisfy the precondition of \helperFn{unlockNode}. We have all the relevant pieces available, except for $\ghostState{s}{\authFrag\; C_r'}$. We obtain this remaining piece by observing that $\ghostState{s}{\authFrag\; C_r}$ implies $C_R \subseteq H'$ which in turn implies $C_r' \subseteq H'$ by definition of $C_r'$ and $H'$. Using rule~\refRule{auth-set-snap} we obtain $\ghostState{s}{\authFrag\; H'}$, which we can rewrite to $\ghostState{s}{\authFrag\; (H' \cup C_r')}$ using the previously derived equality $H' = H' \cup C_r'$. Applying rule~\refRule{auth-set-frag}, we can then infer $\ghostState{s}{\authFrag\; H'} * \ghostState{s}{\authFrag\; C_r'}$ and after throwing away the first conjunct, we are left with the desired missing piece.

The call to \helperFn{unlockNode} then returns $\heapRep(r,r,\mathit{es}, V_r')$ to the invariant, leaving us in the state shown on line~\ref{tn-upsert-proof-commit-post-2}. Finally, we use rule~\refRule{view-upd} to update the client's and invariant's views of the data structures state to $\mcsstate(r, t + 1, H')$ and $\mcsstate^{\authAuth}(r, t + 1, H')$, respectively. This commits the atomic triple, yielding the desired postcondition on line~\ref{tn-upsert-proof-unlock-r}.}

It remains to show that the updates of the ghost resources preserve the invariant.
That is, we need to prove that all constraints in the invariant involving $t$, $H$, $C_r$, and $\Cir{r}'$ are maintained if we replace them with $t+1$, $H'$, $C_r'$, and $\Cir{r}'=\lambda k'.\,\ite{k' \in \dom(C_r')}{C_r'(k')}{Q_r(k')}$, respectively.

First note that we can use the view shift assumption on $\proto$ to reestablish $\proto(H')$ from $\proto(H)$. Further note that by definition we have $\Cir{r}' = \Cir{r}[k \rightarrowtail (v,t)]$. Hence, the update to the ghost location $\mathit{cir}(r)(k)$ preserved the invariant. We can similarly see that the updates to $\gamma_t$, $\gamma_S$, and $\gamma_{c{r}}$ preserve the invariant.

Next, observe that $\maxTS(t+1,H')$ and $\hunique(H')$ follow directly from the definition of $H'$, $\maxTS(t, H)$, and $\hunique(H)$. We obtain the later two from the invariant prior to the ghost update of $t$ and $H$. Likewise, $\init(H')$ holds again because $H \subseteq H'$ and $\init(H)$ holds before the update.

To show $\maxH[H'] = \Cir{r}'$, we need to prove that for all keys $k'$
\[\maxH[H'](k') = \ite{k' \in \dom(C'_r)}{C'_r(k')}{Q_r(k')}\]
If $k' \neq k$, the equality follows directly from $\maxH=\Cir{r}$ and the definitions of $H'$ and $C_r'$. For the case where $k' = k$ observe that $\maxTS(t, H)$ implies $\maxH[H'](k) = (v,t)$. Moreover, we have $C_r'(k)=(v,t)$ by definition of $C_r'$.

Finally, observe that $\phi_2(r,C'_r, Q_r, \interface_r)$ holds trivially because $\inflowOfInt{\interface_r} = \lambda_0$. Hence, we conclude that the updates maintain the invariant, which also concludes the proof of \code{upsert}.

Our careful encoding of contents-in-reach ensured that the sets $Q_n$ and the node-local interfaces $\interface_n$ are not affected by the upsert for any node $n$, including $r$. This considerably simplified the prove that the invariant $\mcslinv(\lsminv,\proto)(r)$ is maintained at the linearization point.

\subsubsection{Proof of \code{compact}.}

As discussed in \refSec{sec-maintenance}, we need to extend the data structure invariant $\lsminv$ with ghost resources that track the inset of each node. We do this via
an additional flow interface. The new global interface, denoted $\ointerface$, is stored at a new ghost location $\gamma_\ointerface$ in $\lsminv$. \tw{Similar to the contents-in-reach flow interface $\interface$, the associated RA is authoritative flow interfaces over the flow domain of multisets of keys}. We add the constraint $\dom(\interface)=\dom(\ointerface)$ to ensure that $\interface$ and $\ointerface$ agree on which nodes belong to the graph. 

The actual calculation of the insets is captured by the following constraint on the outflow of the singleton interfaces $\ointerface_n$ which we add to the predicate $\nodeShar$:
\begin{equation}
  \label{eq-rint-edgeFn}
  \outflowOfInt{\ointerface_n} = \lambda n'\,k.\,\ite{k \in \mathit{es}_n(n')}{\inflowOfInt{\ointerface_n}(n)(k)}{0}
\end{equation}
Additionally, we add the following constraint to $\globalInv$, which requires that the global interface $\ointerface$ gives the full keyspace as inflow to the root node $r$ and no inflow to any other node:
\begin{equation*}
  \label{eq-rint-global-inflow}
  \inflowOfInt{\ointerface} = \lambda n\,k.\, \ite{n=r}{1}{0}
\end{equation*}
Together, these constraints guarantee that for any node $n$ that is reachable from $r$, $\inflowOfInt{\ointerface_n}(k) > 0$ iff $k \in \inset(n)$.

Finally, we add the following predicates to $\nodeShar$ in order to capture Invariants~\ref{inv-mc6}-\ref{inv-mc8}:
\begin{align*}
  \phi_3(n, C_n, Q_n) \defeq {} & \forall k.\, \ts(Q_n(k)) \leq \ts(\Cir{n}(k))\\
   \phi_4(n, C_n, Q_n, \ointerface_n) \defeq {} & \forall k.\,
                                                k \in dom(\Cir{n}) \Rightarrow \inflowOfInt{\ointerface_n}(n)(k) > 0\\
   \phi_5(n, \ointerface_n) \defeq {} & \forall k.\, \inflowOfInt{\ointerface_n}(n)(k) \leq 1
\end{align*}

The specifications of the implementation-specific helper functions assumed by \code{compact} are provided in \refFig{fig-multicopy-helper-specs-compaction}. A thread performing \code{compact$\,n$} starts by locking node $n$ and checking if node $n$ is at full capacity using the helper function \helperFn{atCapacity}. By locking node $n$, the thread receives the resources available in $\nodePred(r, n, C_n, Q_n)$, for some contents $C_n$ and successor contents-in-reach $Q_{n}$.  The precondition of \helperFn{atCapacity} requires the predicate $\heapRep(r, n, \mathit{es}_n, \valof(C_n))$, which is available to the thread as part of $\nodePred(r, n, C_n, Q_n)$. The return value of \helperFn{atCapacity\,$n$} is a boolean indicating whether node $n$ is full or not. The precise logic of how the implementation of \helperFn{atCapacity} determines whether a node is full is immaterial to the correctness of the template, so the specification of \helperFn{atCapacity} abstracts from this logic. If $n$ is not full, \code{compact} releases the lock on $n$, transferring ownership of $\nodePred(r, n, C_n, Q_{n})$ back to the invariant and then terminates. The call to \code{unlockNode} on line~\ref{line-mc-compaction-final-unlock} is the commit point of the atomic triple in the else branch of the conditional.

Thus, let us consider the other case, i.e. when $n$ is full. Here, the contents of node $n$ must be merged with the contents of some successor node $m$ of $n$. This node $m$ is determined by the helper function \helperFn{chooseNext}. \helperFn{chooseNext} either returns an existing successor $m$ of $n$ (i.e., $\mathit{es}'_n(m) \neq \emptyset$), or $n$ needs a new successor node, which we capture by the implementation-specific predicate $\code{needsNewNode}(r, n, \mathit{es}, \valof(C_n))$. In the former case, we can establish that $m$ is part of the data structure due to the fact that the edgeset of $n$ directs some keys to $m$. This follows from the property $\code{closed}(n)$ in the invariant $\mcslinv(\lsminv,\proto)(r)$.

In the latter case, a new node is allocated using the helper function \helperFn{allocNode} and inserted into the data structure as a successor of $n$ using the helper function \helperFn{insertNode}. After this call, $m$ becomes reachable from the root node $r$ from $n$. To ensure that the node-local invariant of $n$ is maintained, $m$ must be ``registered'' in $\mcslinv(\lsminv,\proto)(r)$.
To this end, we must extend the domain of the global flow interfaces tracked by $\lsminv$ with the new node $m$. This can be done using a frame-preserving update of the authoritative version of the interfaces at ghost locations $\gamma_\interface$ and $\gamma_\ointerface$. 
Showing that the invariant is preserved afer this update is easy because the postcondition of \helperFn{insertNext} together with the invariant guarantee that $m$ has no outgoing edges and can only be reached via $n$. In particular, the fact that $V_m = \emptyset$ and $\mathit{es}_m=\lambda n.\, \emptyset$ imply in this case that $\phi_4(m,C_m,Q_m,\ointerface_m)$ holds, where $C_m=\Cir{m}=\emptyset$. Additionally, the conjunct $\mathit{es}_n' = \mathit{es}_n[m \rightarrowtail \mathit{es}_n'(m)]$ ensures that the inset of any nodes other than $n$ and $m$ do not change. Hence, the flow interfaces at $\gamma_\interface$ and $\gamma_\ointerface$ can be contextually extended to include node $m$ in their domains.

Overall, once it has been established that node $m$ is in the domain
of the global flow interfaces $\interface$ and $\ointerface$ in the invariant $\mcslinv(\lsminv,\proto)(r)$, $m$ is locked and the contents of $n$ is (partially) merged into $m$ using the helper function \helperFn{mergeContents} (line~\ref{line-mc-compaction-merge-contents}).

Let us now examine the specification of \helperFn{mergeContent} in detail. \helperFn{mergeContents$\,n\,m$} merges the data from node $n$ into node $m$. By merge, we mean that copies of keys drawn from some set $K \subseteq \KS$ are transferred from $n$ to $m$, possibly replacing older copies in $m$.
In general, \helperFn{mergeContents} modifies the contents of the two nodes according to the specification given in \refFig{fig-multicopy-helper-specs-compaction}. The precondition demands ownership of the physical representation of the nodes' contents and that $m$ is a successor of $n$.
\tw{
The contents are modified to $V_n'$ and $V_m'$ respectively such that
\begin{align*}
  V_n' = {} &  \mathit{mergeLeft}(K, V_n, \mathit{Es}, V_m)\\
  V_m' = {} & \mathit{mergeRight}(K, V_n, \mathit{Es}, V_m)
\end{align*}
We can extend the definition of the functions $\mathit{mergeLeft}$ and $\mathit{mergeRight}$ to work on the contents extended with timestamps:
\begin{align*}
  \mathit{mergeLeft}(K, C_n, \mathit{Es}, C_m) \defeq {} &
  \lambda k.\, \ite{k \in K \cap \dom(C_n) \cap \mathit{Es}}{\bot}{C_n(k)}\\
  \mathit{mergeRight}(K, C_n, \mathit{Es}, C_m) \defeq {} &
  \lambda k.\, \ite{k \in K \cap \dom(C_n) \cap \mathit{Es}}{C_n(k)}{C_m(k)}
\end{align*}
We then define the new (timestamp) contents of $n$ and $m$ as:
\begin{align*}
  C_n' = {} & \mathit{mergeLeft}(K, C_n, \mathit{es}_n(m), C_m)\\
  C_m' = {} & \mathit{mergeRight}(K, C_n, \mathit{es}_n(m), V_m)
\end{align*}
Note that this gives us $V_n' = \valof(C_n')$ and $V_m' = \valof(C_m')$.

We next explain that, together, these constraints and definitions ensure that we can consistently update all relevant ghost resources in the invariant by locally recalculating the values of the ghost resources constrained by the contents of $n$ and $m$. In particular, the contents-in-reach of $m$ can only increase and the contents-in-reach of all other nodes, including $n$ remain unchanged.

We let the set $K' \defeq \dom(C_n) \setminus \dom(C_n')$ denote all keys whose copies are merged from $C_n$ into $C_m$. That is, we have $K' = K \cap \dom(C_n) \cap \mathit{es}_n(m)$ and for all keys $k \notin K'$, we have $C_n'(k)=C_n(k)$ and $C_m'(k)=C_m(k)$. We will use this observation freely in the remainder of the proof.}

Before we proceed with the rest of the proof, we fix instantiations for the existentially quantified variables in the invariant $\mcslinv(\lsminv,\proto)(r)$ when we assume it before the call to \helperFn{mergeContents}.
For a node $p$ in the structure, we denote by $\interface_p$ and $\ointerface_p$ the fragmental singleton flow interface of node $p$ at ghost locations $\gamma_\interface$ and $\gamma_\ointerface$, respectively. Moreover, let $Q_{p}$ be the set stored at ghost location $\gamma_{q(p)}$ (i.e., the successor contents-in-reach of $p$ before the call to \helperFn{mergeContents}). 

First, since the update of $C_m$ also affects the contents-in-reach of $m$, we need to update the resource storing $\Cir{m}$ appropriately. We do this by defining:
\[\Cir{m}' \defeq \lambda k.\, \ite{k \in K'}{C_n(k)}{\Cir{m}}\]
and then replace for each key $k$, $\ts(\Cir{m}(k))$ at ghost location $\gamma_{\cir(m)(k)}$ by $\ts(\Cir{m}'(k))$.
To do this, the authoritative maxnat RA requires us to show that $\ts(\Cir{m}(k)) \le \ts(\Cir{m}'(k))$. If $k \notin K'$, then $\Cir{m}(k)=\Cir{m}'(k)$ by definition. Hence consider the case where $k \in K'$. We then have $k \in \dom(C_n)$ and $\Cir{m}'(k)=C_n$. From this and $k \in \mathit{e}_n(m)$, it follows that $\outflowOfInt{\interface_n}(m)(k,Q_n(k)) > 0$, which, using the flow equation, enables us to conclude $\inflowOfInt{\ointerface_n}(m)(k, Q_n(k)) > 0$. We then infer from $\phi_2(m,C_m,Q_m,\interface_m)$ that $Q_n(k)=\Cir{m}{k}$. Moreover, it follows from $\phi_3(n,C_n,Q_n)$ that $\ts(Q_n(k)) \le \ts(\Cir{n}(k))$. Since $\Cir{n}(k)=C_n(k)=\Cir{m}'(k)$, we can conclude that $\ts(\Cir{m}(k)) \le \ts(\Cir{m}'(k))$ as desired.

  Second, we observe that since \helperFn{mergeContents$\,n\,m$} does not change the edgesets of any nodes, the insets of all nodes also remain unchanged. In particular, for all nodes $p$, the singleton interfaces $\ointerface_{p}$ are unaffected and hence still compose to the global interface $\rinterface$. Additionally, this means that $\phi_5(p,\ointerface_p)$ and the constraint~(\ref{eq-rint-edgeFn}) on $\outflowOfInt{\ointerface_p}$ are trivially maintained. To see that $\phi_4(m,C_m',Q_m,\ointerface_{m})$ is also preserved, we note that if $k \notin \dom(\Cir{m})$ and $k \in \dom(\Cir{m}')$ for some $k$, then $k \in K'$ and hence $C_n(k)=\Cir{n}(k)=\Cir{m}'(k)$. It then follows from $\phi_4(n,C_n,Q_n,\ointerface_{n})$ that $\inflowOfInt{\ointerface_n}(n)(k) > 0$. Moreover, $k \in K'$ implies $k \in \mathit{es}_n(m)$ and, hence, $\outflowOfInt{\ointerface_n}(m)(k) > 0$. It then follows from the flow equation that $\inflowOfInt{\ointerface_m}(m)(k) > 0$.

Finally, we need to reflect the change in the contents-in-reach of $m$ in the local invariant of $n$ by updating the ghost resource holding $Q_n$ to the new value
\[Q_n' = \lambda k.\, \ite{k \in K'}{C_n(k)}{Q_n(k)}]\]
In turn, this requires an update of the singleton interfaces $\interface_n$ and $\interface_m$ to $\interface_n'$ and $\interface_m'$ such that:
\begin{align*}
\inflowOfInt{\interface_n'} \defeq {} & \inflowOfInt{\interface_n}\\
\outflowOfInt{\interface_n'} \defeq {} & 
\lambda n'\, (k,p). \outflowOfInt{\interface_n}(n')(k, p) + \ite{n'=m \land k \in K' \land C_n(k)=p}{1}{0}\\
\inflowOfInt{\interface_m'} \defeq {} & \lambda n'\,(k,p). \inflowOfInt{\interface_m}(n')(k, p)  + \ite{n'=m \land k \in K' \land C_n(k)=p}{1}{0}\\
\outflowOfInt{\interface_m'} \defeq {} & \outflowOfInt{\interface_m}
\end{align*}
First, note that the changes to the inflows and outflows match up consistently. One can therefore easily verify that the old and new singleton interfaces compose to the same larger two-node interface:
\[\interface_n \intComp \interface_m  = \interface_n' \intComp \interface_m'\]
This means that we can simultaneously replace all old interfaces by their new ones using a frame-preserving update of ghost location $\gamma_\interface$.

It remains to show that the relevant constraints in the invariant that depend on $\interface_n$, $\interface_m$, and $Q_n$ are preserved. We defer this part of the proof to the Iris development.

\end{document}